\documentclass[10pt, conference, letterpaper]{IEEEtran}
\IEEEoverridecommandlockouts
\usepackage{booktabs} 
\usepackage{amsmath}
\usepackage[english]{babel}
\usepackage[latin1]{inputenc}
\usepackage{amssymb}
\usepackage[ruled]{algorithm2e}
\usepackage{algpseudocode}
\usepackage{bm}
\usepackage{amsmath}
\usepackage{graphicx}
\usepackage{epsfig}
\usepackage{url}
\usepackage{psfrag}
\usepackage{balance}
\usepackage{pstricks}
\usepackage{times}
\usepackage{array}
\usepackage{subfigure}
\usepackage{float}
\usepackage{multirow}
\usepackage{mathtools}

\newtheorem{theorem}{\textbf{Theorem}}

\newenvironment{pf}{\noindent{\textbf{Proof.} }}{\hfill $\square$\medskip}
\algnewcommand\algorithmicforeach{\textbf{for each}}

\begin{document}

\title{REPT: A Streaming Algorithm of Approximating Global and Local Triangle Counts in Parallel
\thanks{\textsuperscript{*}Peng Jia, Jing Tao and Xiaohong Guan are corresponding authors.}
}
\author{
      \fontsize{11}{11}\selectfont
      Pinghui Wang$^{1,2}$, Peng Jia$^{1}$, Yiyan Qi$^{1}$, Yu Sun$^{1}$, Jing Tao$^{1,2,3}$, and Xiaohong Guan$^{2,1,4}$\\
      $^{1}$MOE Key Laboratory for Intelligent Networks and Network Security, Xi'an Jiaotong University, China\\
      $^{2}$Shenzhen Research School, Xi'an Jiaotong University, Shenzhen, China\\
      $^{3}$Zhejiang Research Institute, Xi'an Jiaotong University, Hangzhou, China\\
      $^{4}$Department of Automation and NLIST Lab, Tsinghua University, Beijing, China\\
      Email: \{phwang, pengjia, jtao, xhguan\}@sei.xjtu.edu.cn, \{qiyiyan, sunyuxajd2013\}@stu.xjtu.edu.cn
}
\maketitle

\begin{abstract}
Recently, considerable efforts have been devoted to approximately computing the global and local (i.e., incident to each node) triangle counts of a large graph stream represented as a sequence of edges.
Existing approximate triangle counting algorithms rely on sampling techniques to reduce the computational cost.
However, their estimation errors are significantly determined by the covariance between sampled triangles.
Moreover, little attention has been paid to developing parallel one-pass streaming algorithms that can be used to fast and approximately count triangles on a multi-core machine or a cluster of machines.
To solve these problems, we develop a novel parallel method REPT to significantly reduce the covariance (even completely eliminate the covariance for some cases) between sampled triangles.
We theoretically prove that REPT is more accurate than parallelizing existing triangle count estimation algorithms in a direct manner.
In addition, we also conduct extensive experiments on a variety of real-world graphs,
and the results demonstrate that our method REPT is several times more accurate than state-of-the-art methods.
\end{abstract}


\begin{figure*}[htb]
\center
\subfigure[$\tau$ vs $\eta$]{
\includegraphics[width=0.45\textwidth]{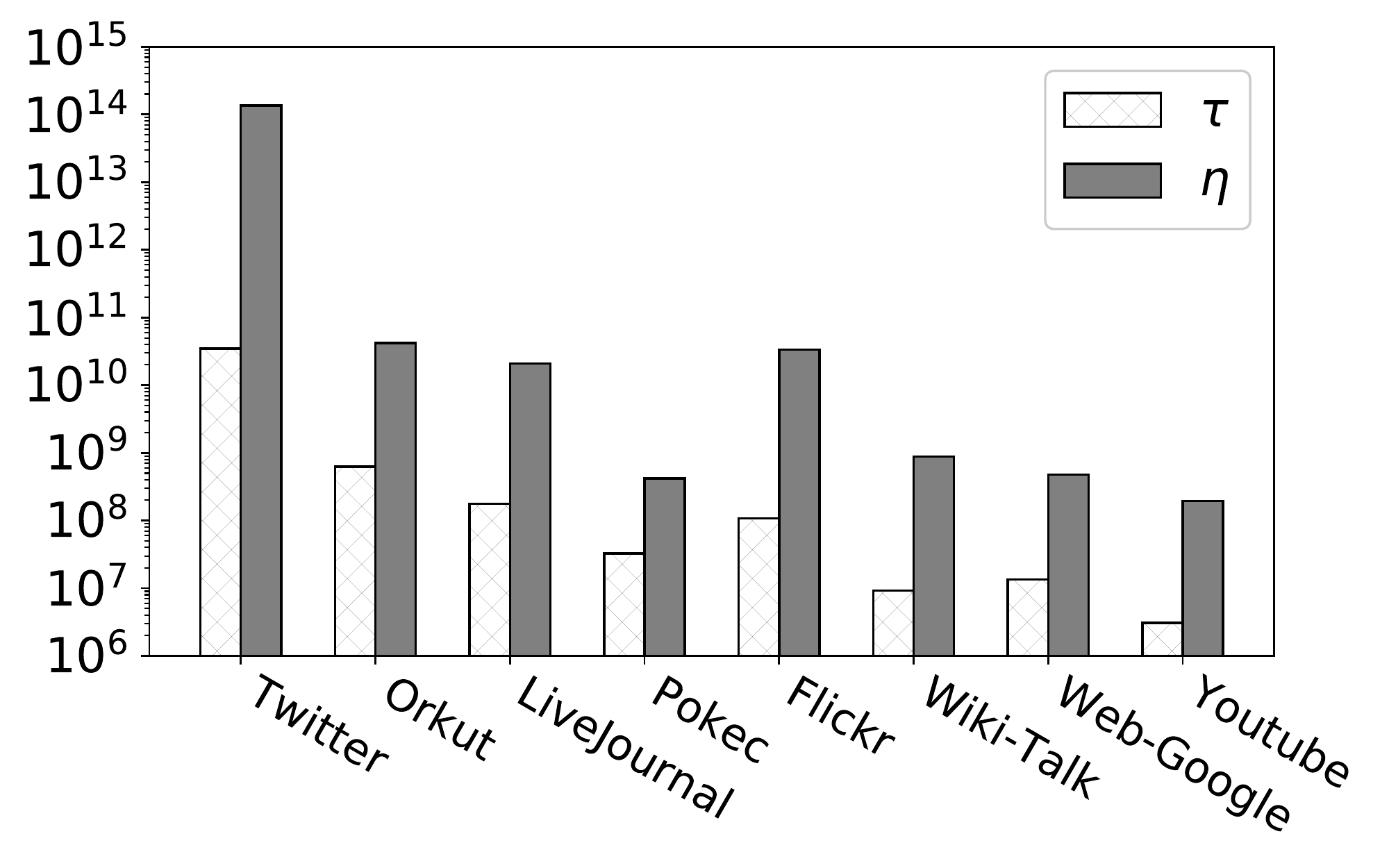}}
\subfigure[$\tau (p^{-2} - 1)$ vs $2\eta (p^{-1}- 1)$, $p=0.1$]{
\includegraphics[width=0.45\textwidth]{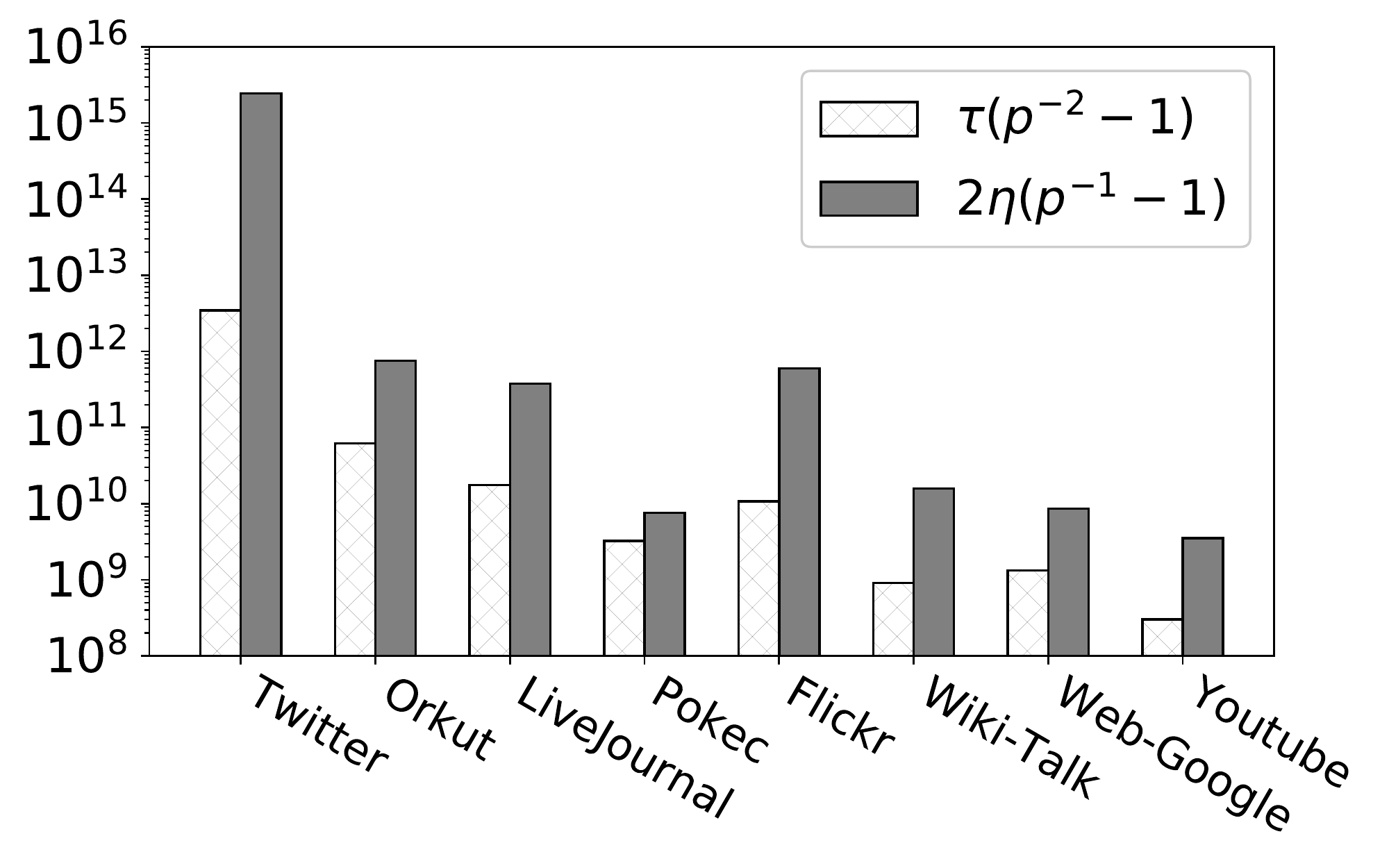}}
\subfigure[$\tau (p^{-2} - 1)$ vs $2\eta (p^{-1}- 1)$, $p=0.05$]{
\includegraphics[width=0.45\textwidth]{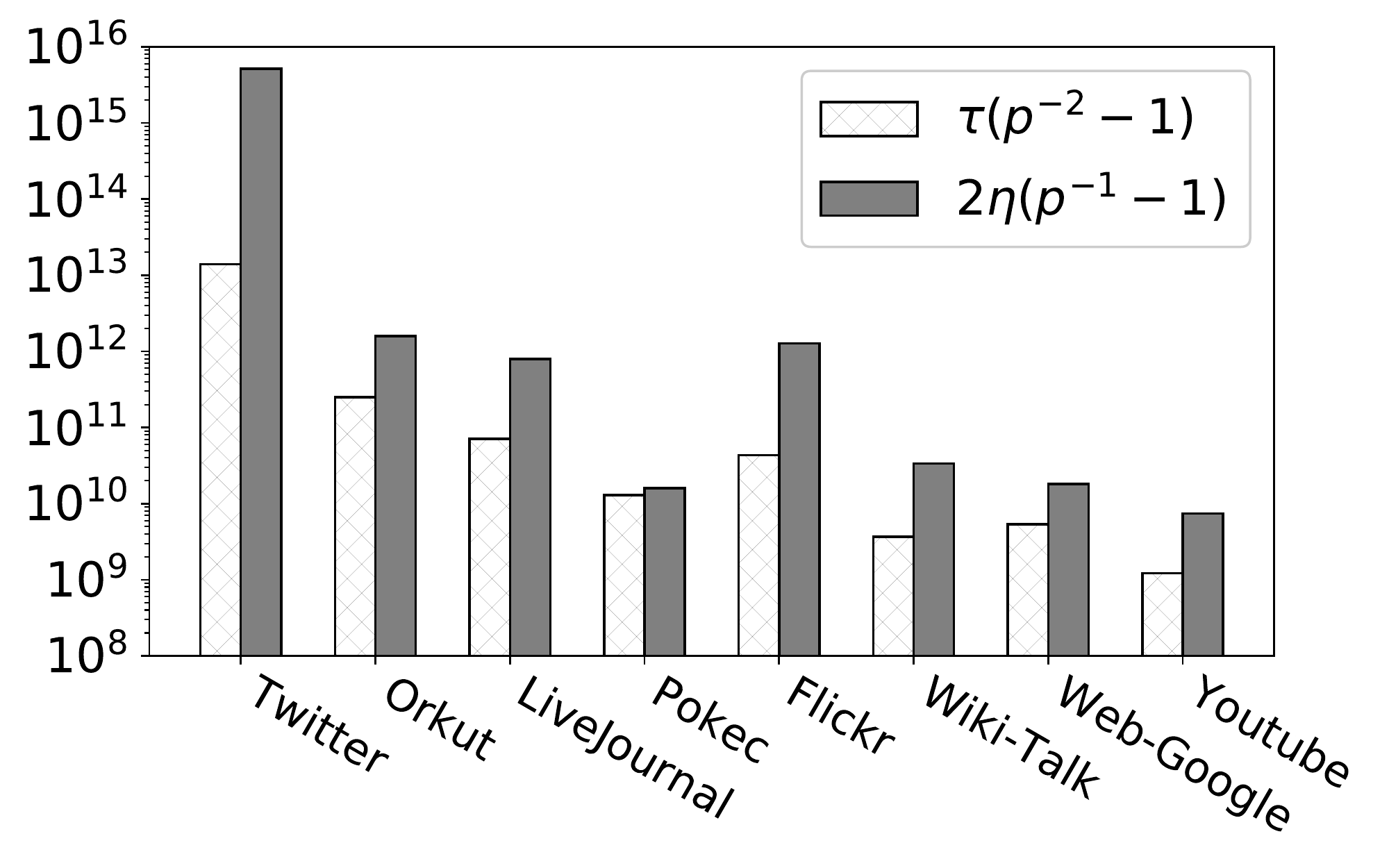}}
\subfigure[$\tau (p^{-2} - 1)$ vs $2\eta (p^{-1}- 1)$, $p=0.01$]{
\includegraphics[width=0.45\textwidth]{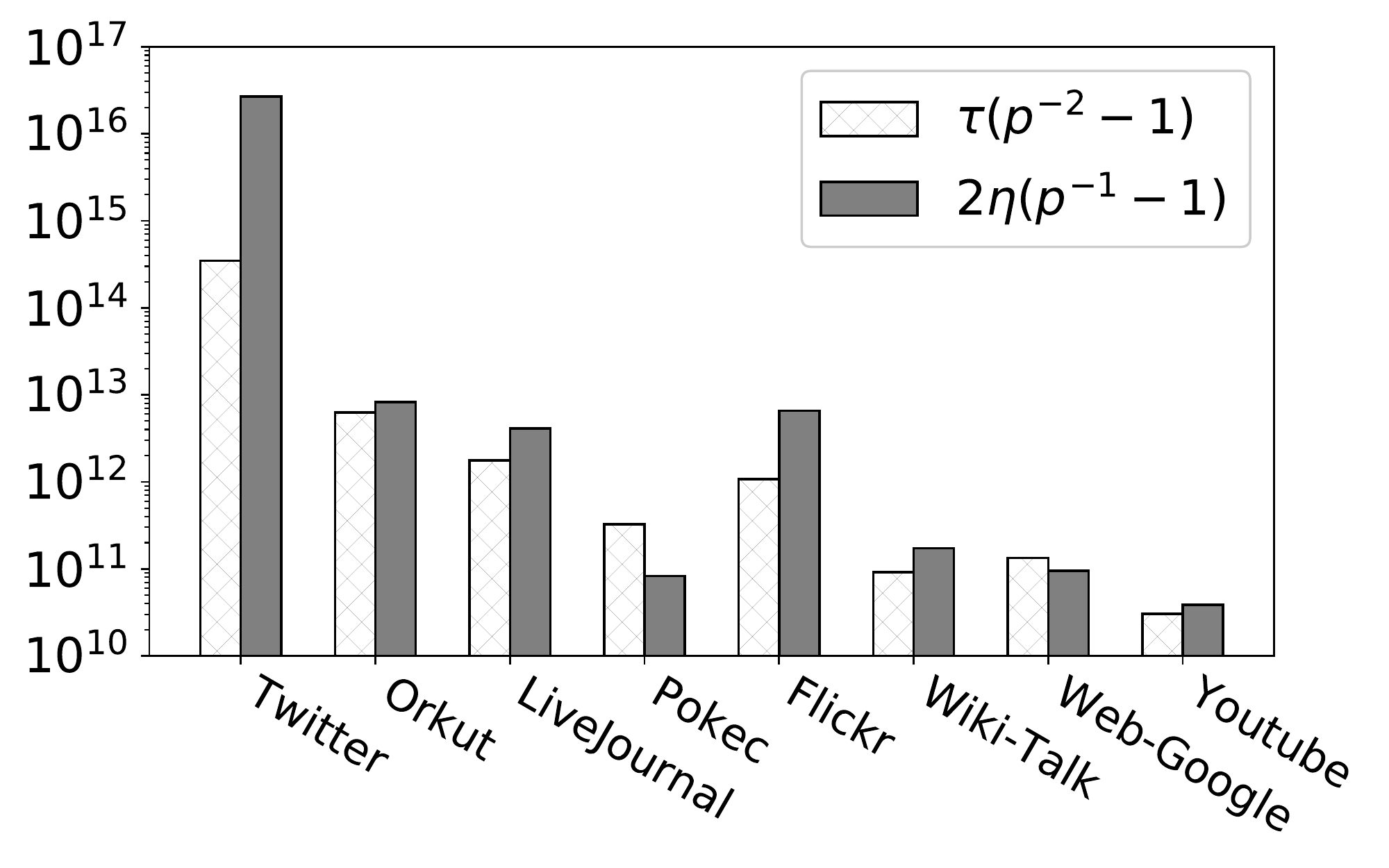}}
\caption{Terms in equation $\frac{\tau (p^{-2} - 1) + 2\eta (p^{-1} - 1)}{c}$, i.e., the variance of parallelizing MASCOT over $c$ processors, where the term $2\eta (p^{-1}- 1)$ is introduced by the covariance of sampled triangles. The statistics of used graphs are summarized in Table~\ref{tab:datasets} in Section~\ref{sec:results}.}
\label{fig:example}
\end{figure*}
\section{Introduction} \label{sec:introduction}
There is a growing interest to explore triangles in a large network,
which are basic local topology structures that form during the growth of the network,
and have been used for a variety of applications such as
spam webpage detection~\cite{BecchettiTKDD2010},
suspicious accounts detection on online social networks~\cite{YangTKDD2014,KangTKDE2014},
social role identification~\cite{Welser2007},
community detection~\cite{BerryPR2011},
topic mining~\cite{EckmannPNAS2002},
and motif detection~\cite{Milo2002,BerryPR2011}.
Recently, considerable attention has been paid to developing one-pass streaming algorithms for computing global and local (i.e., incident to each node) triangle counts of a large graph stream,
because it is critical for analyzing many real-world networks (e.g. mobile phone calling networks) that appear as a stream of edges.

Exactly counting the number of triangles in a large stream graph is a challenging \textbf{computational} task even using distributed and parallel processing frameworks such as MapReduce~\cite{TsourakakisKDD2009}.
To address this challenge,
existing algorithms~\cite{TsourakakisKDD2009,Bar-YossefSODA02,JowhariCOCOON2005,BuriolPODS2006,PavanyVLDB2013,JhaKDD2013,KutzkovWSDM2013,AhmedKDD2014,LimKDD2015,StefaniKDD16,WuTKDE16}
use sampling techniques (e.g., random edge sampling) to
quickly provide approximate estimates of global and local triangle counts,
which are sufficient for many types of analysis.
For instance, MASCOT~\cite{LimKDD2015} samples a fraction of edges from the edge stream $\Pi$ of interest on-the-fly,
which is generated by sampling each and every edge of stream $\Pi$ with a fixed probability $p$.
Tri{\`{e}}st~\cite{StefaniKDD16} extends revisor sampling techniques and samples edges of stream $\Pi$ with a fixed budget size,
where edges are inserted or deleted in an arbitrary order.
One can set a proper value of parameter $p$ for MASCOT (resp. budget size for Tri{\`{e}}st) to achieve desired time and space complexities for approximating estimates of global and local triangle counts.
Besides, both methods estimate global and local triangle counts of stream $\Pi$ based on the number of \textbf{\emph{semi-triangles}},
where a semi-triangle refers to a triangle whose first two edges on stream $\Pi$ are sampled no matter whether its last edge on stream $\Pi$ is sampled or not.
However, these algorithms are customized for the single core environment,
and it is unknown how to use a multi-core machine or a cluster of machines to improve their performance.

To compute global and local triangle counts in parallel, the straightforward method is to parallelize existing sampling algorithms in a direct manner.
Specifically,
one can conduct multiple independent trials and obtain a triangle count estimation by averaging results from these multiple independent trials,
where each trial is performed by a \emph{\textbf{processor}},
referring to either a thread on a multi-core machine or a machine in a distributed computing environment in this paper.
However, this cannot efficiently reduce estimation errors.
In detail, the variance of MASCOT is $\tau (p^{-2} - 1) + 2\eta (p^{-1}- 1)$ for estimating the global triangle count $\tau$,
where $\eta$ is the number of unordered pairs $(\sigma, \sigma^*)$ of distinct triangles that share an edge $g$ and are such that $g$ is neither the last edge of triangle $\sigma$
nor the last edge of triangle $\sigma^*$ on stream $\Pi$.
Usually, $\eta$ is larger than $\tau$ by several orders of magnitude.
For example, as shown in Figure~\ref{fig:example}(a), we can see that $\eta$ is about $11$ to $3,900$ times larger than $\tau$ for many real-world graphs including
Twitter, Orkut, LiveJournal, Pokec, Flickr, Wiki-Talk, Web-Google, and YouTube (the statistics of these graphs are summarized in Table~\ref{tab:datasets} in Section~\ref{sec:results}).
One can conduct MASCOT on available processors in parallel
to obtain independent estimates $\tilde\tau^{(1)}, \ldots, \tilde\tau^{(c)}$ and estimate $\tau=\frac{1}{c}\sum_{i=1}^c \tilde\tau^{(i)}$,
where $c$ is the number of processors used for estimating the global triangle count $\tau$.
In this case, the variance of the estimate given by this simple method of parallelizing MASCOT is $\frac{1}{c}(\tau (p^{-2} - 1) + 2\eta (p^{-1}- 1))$.
The term $2\eta (p^{-1}- 1)$ is introduced by the covariance of sampled semi-triangles,
and we easily observe that the estimation error is dominantly determined by this term.
For example, as shown in Figure~\ref{fig:example}(b), we can see that $2\eta (p^{-1}- 1)$ is 2 to 355 times larger than $\tau (p^{-2} - 1)$ for all graph datasets when $p=0.1$.
As $p$ decreases, as shown in Figures~\ref{fig:example}(b)-(d), the difference between two terms becomes smaller.
When $p=0.01$, however, the term $2\eta (p^{-1}- 1)$ is still 2 to 35 times larger than $\tau (p^{-2} - 1)$ for graph datasets Twitter, LiveJournal, Pokec, Flickr, and Wiki-Talk.
The method of parallelizing Tri{\`{e}}st~\cite{StefaniKDD16} suffers from the same issue,
i.e., its estimation error is significantly dominated by the covariance between sampled semi-triangles.

To solve this problem, we develop a novel parallel method REPT (random edge partition and triangle counting).
REPT randomly distributes edges of stream $\Pi$ into different processors and approximately computes the number of triangles in parallel.
Similar to parallel MASCOT, it samples a fraction of edges from stream $\Pi$ and computes the number of semi-triangles on each processor.
On average, each processor samples and stores $p\times 100\%$ of all edges of stream $\Pi$ at any time.
Unlike parallel MASCOT, \emph{REPT does not generate each processor's sampled edge set independently}.
We develop a novel method to generate all processors' sampled edge sets and utilize their dependencies to significantly reduce or even completely eliminate the estimation error introduced by covariances of sampled semi-triangles.
For example, when $p=\frac{1}{m}$ and $c_1 m$ processors are available where $m=\{2, 3, \ldots\}$ and $c_1\in \{1, 2, \ldots\}$,
our method REPT reduces the variance of global triangle count estimates given by parallel MASCOT from $\frac{\tau (m - 1/m) + 2\eta (1- 1/m)}{c_1}$ to $\frac{\tau (m - 1)}{c_1}$.
We conduct extensive experiments on a variety of real-world large graphs to demonstrate the performance of our method,
and the experimental results demonstrate that our method REPT is several times more accurate than state-of-the-art methods.

The rest of this paper is organized as follows.
The problem is formulated in Section~\ref{sec:problem}.
Section~\ref{sec:method} presents our method REPT for approximately counting triangles in graph streams in parallel.
The performance evaluation and testing results are presented in Section~\ref{sec:results}.
Section~\ref{sec:related} summarizes related work.
Concluding remarks then follow.

\section{Formulated Problem}\label{sec:problem}
To formally define our problem,
we first introduce some notations.
Denote $\Pi$ as the undirected graph stream of interest,
which represents a sequence of undirected edges.
For any discrete time $1\le t \le t_\text{max}$,
let $e^{(t)}=(u^{(t)}, v^{(t)})$ denote the $t^\text{th}$ edge of stream $\Pi$,
where $u^{(t)}$ and $v^{(t)}$ are the edge's two endpoints and $t_\text{max}$ is the size of stream $\Pi$.
Let $G=(V, E)$ be the undirected graph consisting of all edges occurring in stream $\Pi$,
where $V$ and $E$ are the node and edge sets respectively.
Denote $\Delta$ as the set of triangles in graph $G$.
For any node $v\in V$, let $\Delta_v\subseteq \Delta$ denote the set of triangles that include node $v$.
Let $\tau = |\Delta|$ (i.e., the cardinality of set $\Delta$) denote the global triangle count of stream $\Pi$
and $\tau_v = |\Delta_v|$ denote the local triangle count of node $v$.
In this paper, we focus on designing a parallel algorithm for fast and accurately estimating $\tau$ and $(\tau_v)_{v\in V}$.
It is useful for time interval based applications such as network traffic anomaly detection.
For example, $\Pi$ is a network packet stream collected on a router in a time interval (e.g., one hour in a day),
and one wants to compute global and local triangle counts for each interval, i.e., $\tau$ and $(\tau_v)_{v\in V}$ for each interval.
For ease of reading, we list notations used throughout the paper in Table~\ref{tab:notations}.
\begin{table}[tb]
\begin{center}
\caption{Table of notations.\label{tab:notations}}
\begin{tabular}{|c|l|} \hline
$G=(V, E)$&undirected graph consists of all edges in $\Pi$\\ \hline
$\Delta$&the set of triangles in $G$\\ \hline
$\Delta_v, v\in V$&the set of triangles in $G$ including node $v$\\ \hline
$\tau = |\Delta|$&the number of triangles\\ \hline
$\tau_v = |\Delta_v|, v\in V$&the number of triangles including node $v$\\ \hline
\multirow{4}{*}{$\eta$}&the number of unordered pairs $(\sigma, \sigma^*)$ of\\
&distinct triangles in $\Delta$ that share an edge $g$\\
&and are such that $g$ is neither the last edge\\
&of $\sigma$ nor the last edge of $\sigma^*$ on $\Pi$\\ \hline
\multirow{4}{*}{$\eta_v$}&the number of unordered pairs $(\sigma, \sigma^*)$ of\\
&distinct triangles in $\Delta_v$ that share an edge\\
&$g$ and are such that $g$ is neither the last\\
&edge of $\sigma$ nor the last edge of $\sigma^*$ on $\Pi$\\ \hline
$p=1/m, m\in \{2, 3, \ldots\}$&the sampling probability\\ \hline
$c$&the number of available processors\\ \hline
\multirow{2}{*}{$E^{(i)}, 1\le i\le c$}&the set of edges stored by processor $i$,\\
&which is built on-the-fly\\ \hline
\multirow{3}{*}{$N_u^{(i)}, 1\le i\le c$}&the set of neighbors of node $u$\\
&in the graph consisting of edges in $E^{(i)}$\\
&which varies over time\\ \hline
$\Delta^{(i)}, 1\le i\le c$&the set of semi-triangles in $E^{(i)}$\\ \hline
\multirow{2}{*}{$\Delta_v^{(i)}, 1\le i\le c$}&the set of semi-triangles in $E^{(i)}$ including\\
&node $v$\\ \hline
$\tau^{(i)} = |\Delta^{(i)}|, 1\le i\le c$&the number of semi-triangles in $E^{(i)}$\\ \hline
\multirow{2}{*}{$\tau_v^{(i)} = |\Delta_v^{(i)}|, 1\le i\le c$}&the number of semi-triangles in $E^{(i)}$\\
&including node $v$\\ \hline
$h(\cdot)$&hash function used for REPT when $c\le m$\\ \hline
\multirow{2}{*}{$(h_1(\cdot), h_2(\cdot), \ldots)$}&a series of hash functions used for REPT\\
&when $c> m$\\ \hline
\end{tabular}
\end{center}
\end{table}

\begin{algorithm}[tb]
\SetKwFunction{insert}{insert}
\SetKwFunction{delete}{delete}
\SetKwFunction{continue}{continue}
\SetKwFunction{UpdateTriangleCNT}{UpdateTriangleCNT}
\SetKwFunction{rand}{rand}
\SetKwInOut{Input}{input}
\SetKwInOut{Output}{output}
\Input{edge stream $\Pi$.}
\Output{$\hat\tau$, $\hat\tau_v$, $v\in V$.}
\BlankLine

\ForEach {processor $i\in {1,\ldots,c}$}{
$E^{(i)}\gets \emptyset$, $\tau^{(i)}\gets 0$, $\tau_v^{(i)}\gets 0$, $v\in V$\;
\ForEach {$(u, v)\in \Pi$}{
        $\UpdateTriangleCNT(i, (u,v))$\;
        \If {$h(u,v)==i$}{
            $E^{(i)}\gets E^{(i)} \cup \{(u,v)\}$\;
        }
    }
}
$\hat \tau \gets \frac{m^2}{c} \sum_{i=1}^c \tau^{(i)}$\;
\ForEach {$v\in V$}{
    $\hat \tau_v \gets \frac{m^2}{c} \sum_{i=1}^c \tau_v^{(i)}$\;
}
\BlankLine
\textbf{Function} $\UpdateTriangleCNT(i, (u,v))$
$N_{u,v}^{(i)}\gets N_u^{(i)}\cap N_v^{(i)}$\;
$\tau^{(i)}\gets \tau^{(i)} + |N_{u,v}^{(i)}|$\;
$\tau_u^{(i)}\gets \tau_u^{(i)} + |N_{u,v}^{(i)}|$\;
$\tau_v^{(i)}\gets \tau_v^{(i)} + |N_{u,v}^{(i)}|$\;
\ForEach {$w\in N_{u,v}^{(i)}$}{
    $\tau_w^{(i)}\gets \tau_w^{(i)} + 1$\;
}
\caption{REPT($p=\frac{1}{m}$, $c \leq m$).}\label{alg:reptsame}
\end{algorithm}
\section{Our Method}\label{sec:method}
When our algorithm is applied on a cluster of machines,
we assume that each machine in the cluster has enough memory space to store $p\times 100\%$ of edges,
where we set a proper value of $p$ to achieve desired time and space complexities for approximating estimates of global and local triangle counts,
which is similar to~\cite{LimKDD2015}.
When our algorithm is applied on a multi-core machine,
we only use $c^* = \min(c, \left\lfloor\frac{M}{p|E|}\right\rfloor)$ cores, where $M$ is the available memory on the multi-core machine.
Perez et al.~\cite{Perez2015Ringo} reveal that big-memory and multi-core machines have become more affordable and widely available,
and the memory of one such big-memory machine can comfortably handle most real-world graphs being analyzed today.
Therefore, we assume $c^*=c$. The basic idea behind our algorithm is summarized as:
For each processor $i=1, \ldots, c$, we generate a set of edges $E^{(i)}$ from stream $\Pi$ on-the-fly.
At any time,
on average set $E^{(i)}$ consists of $p\times 100\%$ of occurred edges.
For simplicity, in this paper we set $p=\frac{1}{m}$, where  $m\in \{2, 3, \ldots\}$.
The method of generating $E^{(i)}$ will be discussed in detail later.
Let $\Delta^{(i)}$ denote the set of \textbf{\emph{semi-triangles}} whose first two edges on stream $\Pi$ are in set $E^{(i)}$ no matter whether their last edges on stream $\Pi$ are in set $E^{(i)}$ or not.
Denote $\tau^{(i)} = |\Delta^{(i)}|$.
Note that $\tau^{(i)}$ may be larger than the number of triangles consisting of three edges in $E^{(i)}$.
In this section, we introduce a method that uses $c$ processors to compute $\tau^{(1)}, \ldots, \tau^{(c)}$ in parallel
and then estimates $\tau$ based on all $\tau^{(1)}, \ldots, \tau^{(c)}$.
Let $\Delta_v^{(i)}$ denote the number of semi-triangles in set $\Delta^{(i)}$ that include node $v$.
Let $\tau_v^{(i)} = |\Delta_v^{(i)}|$.
Similarly, we estimate $\tau_v$ based on $\tau_v^{(1)}, \ldots, \tau_v^{(c)}$.
Next, we introduce our algorithms $\text{REPT}(\frac{1}{m}, c \leq m)$ and $\text{REPT}(\frac{1}{m}, c>m)$  for two different cases $c \leq m$ and $c>m$ respectively.

\subsection{Algorithm for Case $c \leq m$}\label{sec:same}
The pseudo code of $\text{REPT}(\frac{1}{m}, c \leq m)$ is shown in Algorithm~\ref{alg:reptsame}.
When $c \leq m$, we use processor $i$ to collect edges in set $E^{(i)}$ and keep track of $\tau^{(i)}$ as:
Let $N_u^{(i)}$ denote the set of neighbors of node $u$ in the graph consisting of all edges in set $E^{(i)}$.
Note that sets $E^{(i)}$ and $N_u^{(i)}$ are initialized to be empty and change over time.
Let $N_{u,v}^{(i)} = N_u^{(i)} \cap N_v^{(i)}$.
For each coming edge $(u, v)$ occurring in stream $\Pi$, we compute $|N_{u,v}^{(i)}|$, i.e., the number of semi-triangles in set $\Delta^{(i)}$ of which the last edge on $\Pi$ is $(u, v)$,
and then update counters $\tau^{(i)}$, $\tau_u^{(i)}$, $\tau_v^{(i)}$, and $(\tau_w^{(i)})_{w\in N_{u,v}^{(i)}}$ as:
$\tau^{(i)} \gets \tau^{(i)} + |N_u^{(i)} \cap N_v^{(i)}|$, $\tau_u^{(i)}\gets \tau_u^{(i)} + |N_{u,v}^{(i)}|$, $\tau_v^{(i)}\gets \tau_v^{(i)} + |N_{u,v}^{(i)}|$,
and $\tau_w^{(i)}\gets \tau_w^{(i)} + 1$.
All these counters are initialized to zero.
Let $h(u, v)$ be a hash function that uniformly and independently maps each edge $(u, v)$ to an integer in $\{1,...,m\}$ at random,
i.e., $P(h(u,v)=i)=\frac{1}{m}$ and $P(h(u,v)=i \wedge h(u',v')=i')=\frac{1}{m^2}$ when $(u,v)\ne (u',v')$, $i,i' \in \{1,...,m\}$.
Edge $(u, v)$ is inserted to set $E^{(i)}$ when $h(u, v)$ equals $i$.
Next, we derive the sampling probabilistic model of $\text{REPT}(\frac{1}{m}, c \leq m)$,
which is critical for computing global and local triangle count estimations and analyzing their errors.
\begin{theorem}\label{theorem:REDprobability}
For any $r$ edges of stream $\Pi$,
the probability of function $h$ distributing all these edges into the same set among $E^{(1)}, \ldots, E^{(c)}$ is $p_{r,c}=\frac{c}{m^{r}}$.
\end{theorem}
\begin{pf}
For a specific set $E^{(i)}$, $1\le i\le m$, the probability of function $h$ mapping all these $r$ edges into it is $\frac{1}{m^r}$.
Thus, we have
$p_{r,c}=\binom{c}{1}\times \frac{1}{m^r}=\frac{c}{m^{r}}$.
\end{pf}

For a triangle $\sigma \in \Delta$,
let $\zeta_\sigma$ be a random variable that equals 1 when triangle $\sigma$ occurs as a semi-triangle on a processor (i.e., the first two edges of triangle $\sigma$ on stream $\Pi$ are sampled by the processor no matter whether the last edge of triangle $\sigma$ on stream $\Pi$ is sampled by the processor or not)  and 0 otherwise.
We say triangle $\sigma$ is ``\emph{\textbf{sampled}}" by REPT if and only if $\zeta_\sigma = 1$.
From~Theorem~\ref{theorem:REDprobability}, then we easily have
\begin{theorem}\label{theorem:REDIprobability}
Each triangle $\sigma$ is ``sampled" by algorithm $\text{REPT}(\frac{1}{m}, c \leq m)$ with the same probability $p_{2,c}$,
i.e., $P(\zeta_\sigma=1) = p_{2,c} = \frac{c}{m^2}$.
\end{theorem}

\begin{figure*}[htb]
\center
\includegraphics[width=0.9\textwidth]{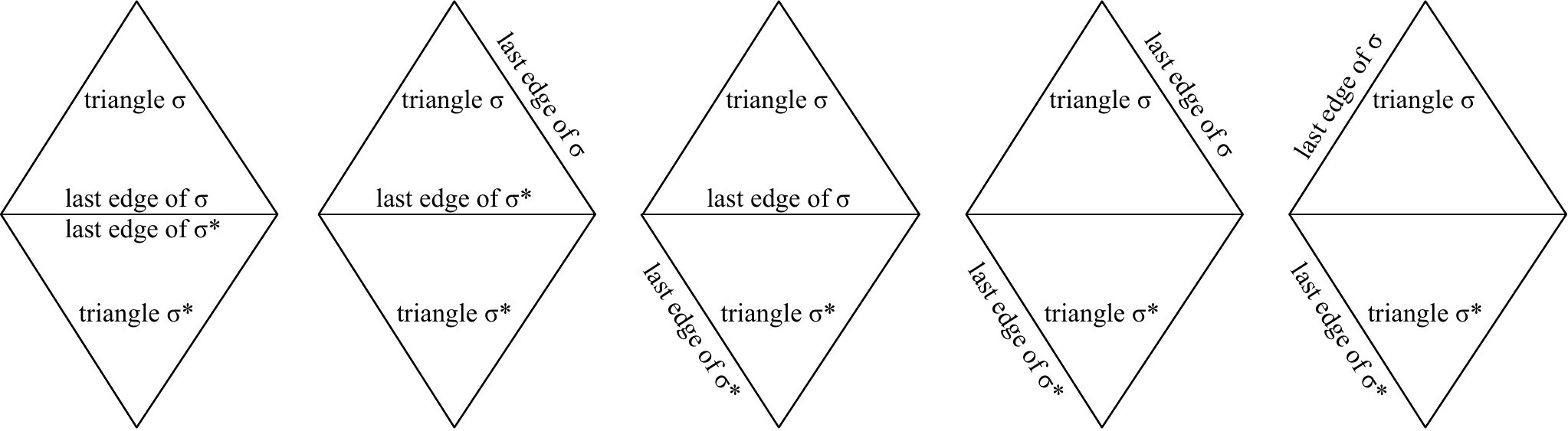}
\caption{Different cases of two distinct triangles $\sigma$ and $\sigma^*$ sharing an edge occurring in stream $\Pi$.}
\label{fig:covariance}
\end{figure*}

Based on the above Theorem, we estimate $\tau$ and $\tau_v$ as
\[
\hat \tau =\frac{\sum_{i=1}^c \tau^{(i)}}{p_{2,c}} = \frac{m^2}{c}\sum_{i=1}^c \tau^{(i)},
\]
\[
\hat \tau_v =\frac{\sum_{i=1}^c \tau_v^{(i)}}{p_{2,c}} = \frac{m^2}{c} \sum_{i=1}^c \tau_v^{(i)}, \quad v\in V.
\]
\begin{theorem}\label{theorem:lessvariance}
When $c \leq m$, the expectations of $\hat \tau$ and $\hat \tau_v$ given by REPT are
\[
\mathbb{E}(\hat \tau) = \tau, \quad \mathbb{E}(\hat \tau_v) = \tau_v.
\]
Let $\eta$ denote the number of unordered pairs $(\sigma, \sigma^*)$ of distinct triangles in set $\Delta$ that share an edge $g$ and are such that $g$ is neither the last edge of triangle $\sigma$
nor the last edge of triangle $\sigma^*$ on stream $\Pi$.
Similarly, let $\eta_v$ denote the number of unordered pairs $(\sigma, \sigma^*)$ of distinct triangles in set $\Delta_v$ that share an edge $g$ and are such that $g$ is neither the last edge of triangle $\sigma$
nor the last edge of triangle $\sigma^*$ on stream $\Pi$.
Then, the variances of $\hat \tau$ and $\hat \tau_v$ are
\[
\text{Var}(\hat \tau) =  \frac{\tau (m^2 - c) + 2\eta(m - c)}{c},
\]
\[
\text{Var}(\hat \tau_v)  =  \frac{\tau_v (m^2 - c) + 2\eta_v (m - c)}{c}.
\]
\end{theorem}
\noindent \textbf{Proof}.
From Theorem~\ref{theorem:REDprobability}, we have
\[
 \mathbb{E}(\hat \tau ) = \sum_{\sigma\in \Delta } \frac{\mathbb{E}(\zeta_\sigma)}{p_{2, c}} = |\Delta| = \tau.
\]
We compute the variance of $\hat \tau$ as
\begin{equation*}
\begin{split}
 &\text{Var}(\hat \tau)= \frac{\sum_{\sigma\in \Delta} \sum_{\sigma^*\in \Delta} \text{Cov}(\zeta_\sigma, \zeta_{\sigma^*})}{p_{2, c}^2}\\
 &=\frac{\sum_{\sigma\in \Delta} \text{Var}(\zeta_\sigma) + \sum_{\sigma, \sigma^*\in \Delta, \sigma\ne \sigma^*} \text{Cov}(\zeta_\sigma, \zeta_{\sigma^*})}{p_{2, c}^2}.
 \end{split}
\end{equation*}
From Theorem~\ref{theorem:REDIprobability}, we have $\text{Var}(\zeta_\sigma) = p_{2, c} - p_{2, c}^2.$
$\text{REPT}(\frac{1}{m}, c \leq m)$ samples triangles in an uncorrelated manner, when two triangles $\sigma$ and $\sigma^*$ share no edge, we easily have $\text{Cov}(\zeta_\sigma, \zeta_{\sigma^*})=\mathbb{E}(\zeta_\sigma\zeta_{\sigma^*}) - \mathbb{E}(\zeta_\sigma)\mathbb{E}(\zeta_{\sigma^*})=p_{2, c}^2 - p_{2, c}^2=0$.

Next, we compute $P(\zeta_\sigma=1 \wedge \zeta_{\sigma^*}=1)$ when triangles $\sigma$ and $\sigma^*$ share an edge.
Equation $\zeta_\sigma=1$ (resp. $\zeta_{\sigma^*}=1$) holds when the first two edges of triangle $\sigma$ (resp. $\sigma^*$) on stream $\Pi$ are mapped into the same set among $E^{(1)}, \ldots, E^{(c)}$ by function $h$. As shown in Figure~\ref{fig:covariance}, we observe:\\
1) when the shared edge is the last edge of triangle $\sigma$ or triangle $\sigma^*$ on stream $\Pi$ (e.g., the first three cases in Figure~\ref{fig:covariance}),
the first two edges of triangle $\sigma$ are different from the first two edges of triangle $\sigma^*$ on stream $\Pi$.
Then, we easily find that $P(\zeta_\sigma=1 \wedge \zeta_{\sigma^*}=1) = p_{2, c}^2$, therefore we have $\text{Cov}(\zeta_\sigma, \zeta_{\sigma^*})=p_{2, c}^2 - p_{2, c}^2 = 0$;\\
2) when the shared edge is not the last edge of triangle $\sigma$ or triangle $\sigma^*$ on stream $\Pi$ (e.g., the last two cases in Figure~\ref{fig:covariance}),
both equations $\zeta_\sigma=1$ and $\zeta_{\sigma^*}=1$ hold only and if only the first three edges of the triangle pair $(\sigma, \sigma^*)$ (i.e., the union of the first two edges of triangles $\sigma$ and $\sigma^*$) on stream $\Pi$ are mapped into the same set among $E^{(1)}, \ldots, E^{(c)}$ by function $h$.
According to Theorem~\ref{theorem:REDprobability}, then we easily find that $P(\zeta_\sigma=1 \wedge \zeta_{\sigma^*}=1) = \frac{c}{m^3}$,
therefore we have $\text{Cov}(\zeta_\sigma, \zeta_{\sigma^*})=p_{3, c} - p_{2, c}^2 = \frac{c}{m^3} - \frac{c^2}{m^4}$.

Based on the above observations, we easily have
\[
\text{Var}(\hat \tau) =  \frac{\tau (m^2 - c) + 2\eta(m - c)}{c}.
\]
Similarly, we have
\[
 \mathbb{E}(\hat \tau_v) = \sum_{\sigma\in \Delta_v } \frac{\mathbb{E}(\zeta_\sigma)}{p_{2, c}} = |\Delta_v| = \tau_v,
\]
\begin{equation*}
\begin{split}
 &\text{Var}(\hat \tau_v)= \frac{\sum_{\sigma\in \Delta_v} \sum_{\sigma^*\in \Delta_v} \text{Cov}(\zeta_\sigma, \zeta_{\sigma^*})}{p_{2, c}^2}\\
 &=\frac{\sum_{\sigma\in \Delta_v} \text{Var}(\zeta_\sigma) + \sum_{\sigma, \sigma^*\in \Delta_v, \sigma\ne \sigma^*} \text{Cov}(\zeta_\sigma, \zeta_{\sigma^*})}{p_{2, c}^2}\\
 &=\frac{\tau_v (m^2 - c) + 2\eta_v (m - c)}{c}. \qquad\qquad\qquad\qquad \hfill \square\medskip
 \end{split}
\end{equation*}

We can easily find that our method significantly reduces the estimation error caused by the covariance of sampled semi-triangles.
Especially, when $c=m$, the variances of $\hat \tau$ and $\hat \tau_v$  are $\text{Var}(\hat \tau) =  \tau (m - 1)$ and $\text{Var}(\hat \tau_v) =  \tau_v (m - 1)$.

\subsection{Algorithm for Case $c>m$}
When $c>m$, define $c_1 = \lfloor \frac{c}{m}\rfloor$ and $c_2 = c\%m$,
i.e., $c = c_1 m + c_2$, where $c_1\ge 1$ and $0\le c_2< m$.
We divide $c$ processors into $c_1 + 1$ groups:
Each of the first $c_1$ groups consists of $m$ processors and the last group consists of $c_2$ processors.
For each group, we apply the method REPT($p=\frac{1}{m}$, $c \leq m$) in Section~\ref{sec:same}.
Let $h_k$ denote the hash function used for generating the edge sets stored on the processors of the $k^\text{th}$ group, $1\le k\le c_1 + 1$.
We let $h_1, \ldots, h_{c_1 +1}$ independent with each other.
Therefore, the triangle counts given by these $c_1 + 1$ groups of processors are also independent.
Next, we introduce our algorithms for two different cases respectively.

\noindent\textbf{1) Algorithm for $c_2= 0$.}
We estimate $\tau$ and $\tau_v$ as
\[
\hat \tau = \frac{m}{c_1} \sum_{i=1}^{c_1 m} \tau^{(i)},
\]
\[
\hat \tau_v = \frac{m}{c_1} \sum_{i=1}^{c_1 m} \tau_v^{(i)}, \quad v\in V.
\]
Similar to the case $c=m$ mentioned in Section~\ref{sec:same}, we easily have $\text{Var}(\hat \tau)= \frac{\tau (m - 1)}{c_1}$ and $\text{Var}(\hat \tau_v)= \frac{\tau_v (m - 1)}{c_1}$.

\noindent\textbf{2) Algorithm for $c_2\neq 0$.}
In addition to the above estimate  of $\tau$ given by the first $c_1$ groups of processors, i.e.,
\[
\hat \tau^{(1)} = \frac{m}{c_1} \sum_{i=1}^{c_1 m} \tau^{(i)},
\]
with variance
\begin{equation}\label{eq:variance1}
\text{Var}(\hat \tau^{(1)}) = \frac{\tau (m-1)}{c_1},
\end{equation}
we also estimate $\tau$ based on the total number of semi-triangles occurring on the last group of $c_2$ processors as
\[
\hat \tau^{(2)} = \frac{m^2}{c_2} \sum_{i=c_1 m + 1}^c \tau^{(i)}.
\]
From Theorem~\ref{theorem:lessvariance},
we easily have
\begin{equation}\label{eq:variance2}
\text{Var}(\hat \tau^{(2)}) = \frac{\tau (m^2 - c_2) + 2\eta(m - c_2)}{c_2}.
\end{equation}
According to~\cite{Franklin1959}, we approximate $\tau$ by optimally combining these two independent and unbiased estimates $\hat \tau^{(1)}$ and $\hat \tau^{(2)}$
as
\[
\hat \tau = \frac{\text{Var}(\hat \tau^{(2)}) \hat \tau^{(1)} + \text{Var}(\hat \tau^{(1)}) \hat \tau^{(2)} }{\text{Var}(\hat \tau^{(1)}) + \text{Var}(\hat \tau^{(2)})}.
\]
The variance of $\hat \tau$ is
\[
\text{Var}(\hat \tau)= \frac{\text{Var}(\hat \tau^{(1)}) \text{Var}(\hat \tau^{(2)}) }{\text{Var}(\hat \tau^{(1)}) + \text{Var}(\hat \tau^{(2)})}.
\]
To compute $\text{Var}(\hat \tau^{(1)})$ and $\text{Var}(\hat \tau^{(2)})$,
we substitute $\tau$ with $\hat \tau^{(1)}$ in equations~(\ref{eq:variance1}) and~(\ref{eq:variance2}) because $\hat \tau^{(1)}$ has a smaller variance than $\hat \tau^{(2)}$,
and substitute $\eta$ with an estimate $\hat \eta$ obtained as
\[
\hat\eta = \sum_{i=1}^c \frac{m^3 \eta^{(i)}}{c},
\]
where $\eta^{(i)}$ is the number of unordered pairs $(\sigma, \sigma^*)$ of distinct triangles in set $\Delta^{(i)}$ that share an edge $g$ and are such that $g$ is neither the last edge of triangle $\sigma$ nor the last edge of triangle $\sigma^*$ on stream $\Pi$.
From the proof of Theorem~\ref{theorem:lessvariance}, we easily have $\mathbb{E}(\eta^{(i)}) = \frac{\eta}{m^3}$, therefore we obtain $\mathbb{E}(\hat\eta) = \eta$.
The method of computing $\eta^{(i)}$ will be discussed in detail later.

Similarly, we estimate the local triangle count $\tau_v$ as
\[
\hat \tau_v = \frac{\text{Var}(\hat \tau_v^{(2)}) \hat \tau_v^{(1)} + \text{Var}(\hat \tau_v^{(1)}) \hat \tau_v^{(2)} }{\text{Var}(\hat \tau_v^{(1)}) + \text{Var}(\hat \tau_v^{(2)})}, \quad v\in V,
\]
where $\hat \tau_v^{(1)}$ and $\hat \tau_v^{(2)}$ are defined as
\[
\hat \tau_v^{(1)} = \frac{m}{c_1} \sum_{i=1}^{c_1 m} \tau_v^{(i)}, \quad \hat \tau_v^{(2)} = \frac{m^2}{c_2} \sum_{i=c_1 m + 1}^c \tau_v^{(i)}.
\]
The variance of $\hat \tau_v$ is
\[
\text{Var}(\hat \tau_v)= \frac{\text{Var}(\hat \tau_v^{(1)}) \text{Var}(\hat \tau_v^{(2)}) }{\text{Var}(\hat \tau_v^{(1)}) + \text{Var}(\hat \tau_v^{(2)})}.
\]
We approximate $\text{Var}(\hat \tau_v^{(1)})$ as $\hat{\text{Var}}(\hat \tau_v^{(1)}) = \frac{\hat \tau_v^{(1)} (m-1)}{c_1}.$
Let $\eta_v^{(i)}$ denote the number of unordered pairs $(\sigma, \sigma^*)$ of distinct triangles in set $\Delta_v^{(i)}$ that share an edge $g$ and are such that $g$ is neither the last edge of triangle $\sigma$ nor the last edge of triangle $\sigma^*$ on stream $\Pi$.
We compute $\eta_v^{(i)}$ similarly to $\eta^{(i)}$, which will be discussed in detail later.
Similar to $\hat\eta$,  we estimate $\eta_v$ as $\hat\eta_v = \sum_{i=1}^c \frac{m^3 \eta_v^{(i)}}{c}$.
Then, we approximate $\text{Var}(\hat \tau_v^{(2)})$ as $\hat{\text{Var}}(\hat \tau_v^{(2)}) = \frac{\hat \tau_v^{(1)} (m^2 - c_2) + 2\hat\eta_v(m - c_2)}{c_2}.$
The pseudo code of $\text{REPT}(\frac{1}{m}, c>m\wedge c_2\neq 0)$ is shown in Algorithm~\ref{alg:reptmore}.

\noindent\textbf{Our method of computing $\eta^{(i)}$ and $\eta_v^{(i)}$.}
We use a counter $\tau_{(u, v)}^{(i)}$ to keep track of the number of triangles in set $\Delta^{(i)}$ that include edge $(u, v)$.
When a new edge $(u, v)$ occurring in $\Pi$ is inserted into $E^{(i)}$, we set $\tau_{(u, v)}^{(i)}= |N_{u,v}^{(i)}|$,
where $N_{u,v}^{(i)} = N_u^{(i)} \cap N_v^{(i)}$ records the set of common neighbors in the graph consisting of all edges in set $E^{(i)}$.
Note that $\tau_{(u, v)}^{(i)} = \tau_{(v, u)}^{(i)}$.
For each $w\in N_{u,v}^{(i)}$,
at any time,
we can easily find that $\tau_{(u, w)}^{(i)}$ also equals: 1) the number of unordered pairs $(\sigma, \sigma^*)$ of distinct triangles in set $\Delta_u^{(i)}$ that share an edge $(u, w)$ and are such that $(u, v)$ is the last edge among the five edges of the triangle pair $(\sigma, \sigma^*)$ on stream $\Pi$;
and 2) the number of unordered pairs $(\sigma, \sigma^*)$ of distinct triangles in set $\Delta_w^{(i)}$ that share an edge $(u, w)$ and are such that $(u, v)$ is the last edge among the five edges of the triangle pair $(\sigma, \sigma^*)$ on stream $\Pi$.
For each coming edge $(u, v)$ and each node $w\in N_{u,v}^{(i)}$,
therefore, we update counters $\eta^{(i)}$, $\eta_w^{(i)}$, $\eta_u^{(i)}$, $\eta_v^{(i)}$, $\tau_{(u, w)}^{(i)}$, and $\tau_{(v, w)}^{(i)}$ as
\[
\eta^{(i)}\gets \eta^{(i)} + \tau_{(u, w)}^{(i)} + \tau_{(v, w)}^{(i)},
\]
\[
\eta_w^{(i)}\gets \eta_w^{(i)} + \tau_{(u, w)}^{(i)} + \tau_{(v, w)}^{(i)},
\]
\[
\eta_u^{(i)}\gets \eta_u^{(i)} + \tau_{(u, w)}^{(i)},
\]
\[
\eta_v^{(i)}\gets \eta_v^{(i)} + \tau_{(v, w)}^{(i)},
\]
\[
\tau_{(u, w)}^{(i)}\gets \tau_{(u, w)}^{(i)} + 1,
\]
\[
\tau_{(v, w)}^{(i)}\gets \tau_{(v, w)}^{(i)} + 1.
\]

\begin{algorithm}
\SetKwFunction{insert}{insert}
\SetKwFunction{delete}{delete}
\SetKwFunction{continue}{continue}
\SetKwFunction{UpdateTrianglePairCNT}{UpdateTrianglePairCNT}
\SetKwFunction{rand}{rand}
\SetKwInOut{Input}{input}
\SetKwInOut{Output}{output}
\Input{edge stream $\Pi$.}
\Output{$\hat\tau$, $\hat\tau_v$, $v\in V$.}
\BlankLine

\ForEach {processor $i\in {1,\ldots, c}$}{
$E^{(i)}\gets \emptyset$, $\tau^{(i)}\gets 0$, $\eta^{(i)}\gets 0$\;
$\tau_v^{(i)}\gets 0$, $v\in V$\;
$\eta_v^{(i)}\gets 0$, $v\in V$\;
\ForEach {$(u, v)\in \Pi$}{
        $\UpdateTrianglePairCNT(i, (u,v))$\;
        $i_1 = \lfloor \frac{i}{m}\rfloor$\;
        $i_2 = i\%m$\;
        \If {$h_{i_1}(u,v)==i_2$}{
            $E^{(i)}\gets E^{(i)} \cup \{(u,v)\}$\;
            $\tau_{(u, v)}^{(i)}\gets |N_{u,v}^{(i)}|$\;
        }
    }
}
$c_1 = \lfloor \frac{c}{m}\rfloor$\;
$c_2 = c\%m$\;
$\hat \tau^{(1)} \gets \frac{m}{c_1} \sum_{i=1}^{c_1 m} \tau^{(i)}$\;
$\hat \tau^{(2)} \gets \frac{m^2}{c_2} \sum_{i=c_1 m + 1}^c \tau^{(i)}$\;
$\hat \eta\gets \sum_{i=1}^c \frac{m^3 \eta^{(i)}}{c}$\;
$w^{(1)} \gets \frac{\hat \tau^{(1)} (m-1)}{c_1}$\;
$w^{(2)} \gets \frac{\hat \tau^{(1)} (m^2 - c_2) + 2\hat\eta(m - c_2)}{c_2}$\;
$\hat \tau \gets \frac{w^{(2)} \hat \tau^{(1)} + w^{(1)} \hat \tau^{(2)} }{w^{(1)} + w^{(2)}}$\;
\ForEach {$v\in V$}{
    $\hat \tau_v^{(1)} \gets \frac{m}{c_1} \sum_{i=1}^{c_1 m} \tau_v^{(i)}$\;
    $\hat \tau_v^{(2)} \gets \frac{m^2}{c_2} \sum_{i=c_1 m + 1}^c \tau_v^{(i)}$\;
    $\hat \eta_v\gets \sum_{i=1}^c \frac{m^3 \eta_v^{(i)}}{c}$\;
    $w_v^{(1)} \gets \frac{\hat \tau_v^{(1)} (m-1)}{c_1}$\;
    $w_v^{(2)} \gets \frac{\hat \tau_v^{(1)} (m^2 - c_2) + 2\hat\eta_v(m - c_2)}{c_2}$\;
    $\hat \tau_v \gets \frac{w_v^{(2)} \hat \tau_v^{(1)} + w_v^{(1)} \hat \tau_v^{(2)} }{w_v^{(1)} + w_v^{(2)}}$\;
}
\BlankLine
\textbf{Function} $\UpdateTrianglePairCNT(i, (u,v))$
$N_{u,v}^{(i)}\gets N_u^{(i)}\cap N_v^{(i)}$\;
$\tau^{(i)}\gets \tau^{(i)} + |N_{u,v}^{(i)}|$\;
$\tau_u^{(i)}\gets \tau_u^{(i)} + |N_{u,v}^{(i)}|$\;
$\tau_v^{(i)}\gets \tau_v^{(i)} + |N_{u,v}^{(i)}|$\;
\ForEach {$w\in N_{u,v}^{(i)}$}{
    $\tau_w^{(i)}\gets \tau_w^{(i)} + 1$\;
    $\eta^{(i)}\gets \eta^{(i)} + \tau_{(u, w)}^{(i)} + \tau_{(v, w)}^{(i)}$\;
    $\eta_w^{(i)}\gets \eta_w^{(i)} + \tau_{(u, w)}^{(i)} + \tau_{(v, w)}^{(i)}$\;
    $\eta_u^{(i)}\gets \eta_u^{(i)} + \tau_{(u, w)}^{(i)}$\;
    $\eta_v^{(i)}\gets \eta_v^{(i)} + \tau_{(v, w)}^{(i)}$\;
    $\tau_{(u, w)}^{(i)}\gets \tau_{(u, w)}^{(i)} + 1$\;
    $\tau_{(v, w)}^{(i)}\gets \tau_{(v, w)}^{(i)} + 1$\;
}
\caption{REPT($p=\frac{1}{m}$, $c> m\wedge c\%m \neq 0$).}\label{alg:reptmore}
\end{algorithm}

\subsection{REPT vs Parallel MASCOT and Tri{\`{e}}st}
\noindent\textbf{Complexity comparison.} De Stefan et al.~\cite{StefaniKDD16} reveal that Tri{\`{e}}st almost has the same accuracy as MASCOT~\cite{LimKDD2015} for estimating global and local triangle counts at the end of stream $\Pi$,
which is consistent with our experimental results in Section~\ref{sec:results}.
Therefore, here we only theoretically compare the performance of our method REPT with the method of parallelizing MASCOT,
i.e., conducting MASCOT with the same edge sampling probability $p=\frac{1}{m}$ on $c$ processors in parallel
to obtain $c$ independent estimates $\tilde\tau^{(1)}, \ldots, \tilde\tau^{(c)}$ of the global triangle count $\tau$.
Similar to parallel MASCOT, each processor of REPT requires $O(p|E|)$ memory space,
and the time to process each edge $(u, v)$ of stream $\Pi$ is dominated by the computation of the shared
neighbors of nodes $u$ and $v$.
Later in our experiments we observe that REPT and parallel MASCOT almost have the same computational cost.

\noindent\textbf{Accuracy comparison.} From Lemma 6 in~\cite{LimKDD2015},
we easily derive the variance of estimate $\frac{1}{c}\sum_{i=1}^c \tilde\tau^{(i)}$ as
\[
\text{Var}(\frac{1}{c}\sum_{i=1}^c \tilde\tau^{(i)})=\frac{\tau (m^2 - 1) + 2\eta (m - 1)}{c}.
\]
Clearly, $\text{Var}(\frac{1}{c}\sum_{i=1}^c \tilde\tau^{(i)})$ is significantly larger than the variance of our method REPT especially for the case $c=\{m, 2m, \ldots\}$,
because $\eta$ is usually larger than $\tau$ by several orders of magnitude,
which is shown in Figure~\ref{fig:example}.
Similarly, we observe that our method PEPT outperforms parallel MASCOT for estimating local triangle counts.

\subsection{Scope and Limitations of REPT}
Our method REPT is developed for streaming graphs but not non-streaming graphs.
When the graph of interest is static and is stored in the memory,
one can easily parallelize the wedge sampling method~\cite{SeshadhriSADM2014}
to estimate the triangle count,
which could provide more accurate estimations than our method REPT under the same computational time.
When the graph of interest is given in the adjacency list format stored on disk,
one can use multi-core algorithms PATRIC~\cite{ArifuzzamanCIKM2013} and TC-Approx~\cite{ShunICDE2015} to exactly/approximately compute the triangle count,
which are also more accurate than our method REPT under the same computational time.
However, our method REPT may be faster than PATRIC and TC-Approx when the graph file is not given in the adjacency list format,
because both PATRIC and TC-Approx need to transform the original graph into the adjacency list format,
which may take a long period of time (e.g., $1,500$ seconds for the transformation in graph Twitter~\cite{Kwak2010}).

\section{Evaluation} \label{sec:results}
\subsection{Datasets}
We evaluate the performance of our method REPT on a variety of publicly available real-world graph datasets with up to a billion edges,
which are summarized in Table~\ref{tab:datasets}.
The algorithms are implemented in C++, and run on a computer with a Quad-Core Intel(R) Xeon(R) CPU E5-2690 v4 CPU 2.60GHz processor.
\begin{table}[htb]
\centering
\caption{Graph datasets used in our experiments.\label{tab:datasets}}
\begin{tabular}{|c|c|c|c|c|}
\hline
{\bf Graph}&{\bf nodes}&{\bf edges}&{\bf triangles}\\
\hline
Twitter~\cite{Kwak2010}&41,652,231&1,202,513,046&34,824,916,864\\
com-Orkut~\cite{YangICDM2012}&3,072,441&117,185,803&627,584,181\\
LiveJournal~\cite{YangICDM2012}&5,189,809&48,688,097&177,820,130\\
Pokec~\cite{Takac2012}&1,632,803&22,301,964&32,557,458\\
Flickr~\cite{Mcauley2012}&105,938&2,316,948&107,987,357\\
Wiki-Talk~\cite{Leskovec2010}&2,394,385&4,659,565&9,203,519\\
Web-Google~\cite{GoogleProgrammingContest2002}&875,713&	4,322,051&13,391,903\\
YouTube~\cite{YangICDM2012}&1,138,499&2,990,443&3,056,386\\
\hline
\end{tabular}
\end{table}

\subsection{Baselines}
Algorithms MASCOT~\cite{LimKDD2015} and Tri{\`{e}}st~\cite{StefaniKDD16} are the state-of-the-art one-pass streaming algorithms developed for estimating global and local triangle counts.
They both have several variants
and in our experiments we only study their improved variants (e.g. Tri{\`{e}}st-IMPR in~\cite{StefaniKDD16}).
We parallelize algorithm MASCOT on $c$ processors as: Each processor independently samples each and every edge of stream $\Pi$ with a fixed probability $p$
and then computes estimates of global and local triangle counts based on sampled edges.
Finally, we approximate global and local triangles by averaging estimates given by $c$ processors.
Similarly, we parallelize algorithm Tri{\`{e}}st on $c$ processors.
Tri{\`{e}}st needs to set the sampling budget (i.e., the number of maximum sampled edges) in advance.
In this paper, we set its sample budget to $p|E|$ for each processor,
where $|E|$ is the number of all edges of stream $\Pi$.
In addition, Ahmed et al.~\cite{ahmed2017sampling} present a new \emph{order-based reservoir sampling} framework \emph{GPS} (graph priority sampling)
which can be used for estimating global triangle counts.
For an edge arriving on the stream at time $t$, GPS assigns it a sampling weight,
which is computed on-the-fly depending on the set of sampled edges at time $t$.
GPS samples edges of highest priority according to their sampling weights.
It has two variants Post-Stream and In-Stream,
and we only study its improved variant In-Stream with lower variance.
In our experiments, we parallelize GPS on $c$ processors and the sample budget is set to $p|E|$ for each processor.
Because the sampled edges and their corresponding sampling weights all cost memory usage, each processor samples $\frac{p|E|}{2}$ edges for GPS.

\subsection{Error Metric}
For global and local triangle count estimations, we use the metric \emph{normalized root mean square error} (NRMSE) to evaluate the error of an estimation $\hat \mu$ with respect to its true value $\mu$.
Formally, NRMSE is defined as
\[
\text{NRMSE}(\hat \mu) = \sqrt{\text{MSE}(\hat\mu)}/\mu,
\]
where $\text{MSE}(\hat\mu)=\mathbb{E}((\hat\mu-\mu)^2)=\text{Var}(\hat\mu)+\left(\mathbb{E}(\hat\mu)-\mu\right)^2$.

\begin{figure*}[!t]
\centering
\subfigure[Twitter]{\includegraphics[width=0.232\textwidth]{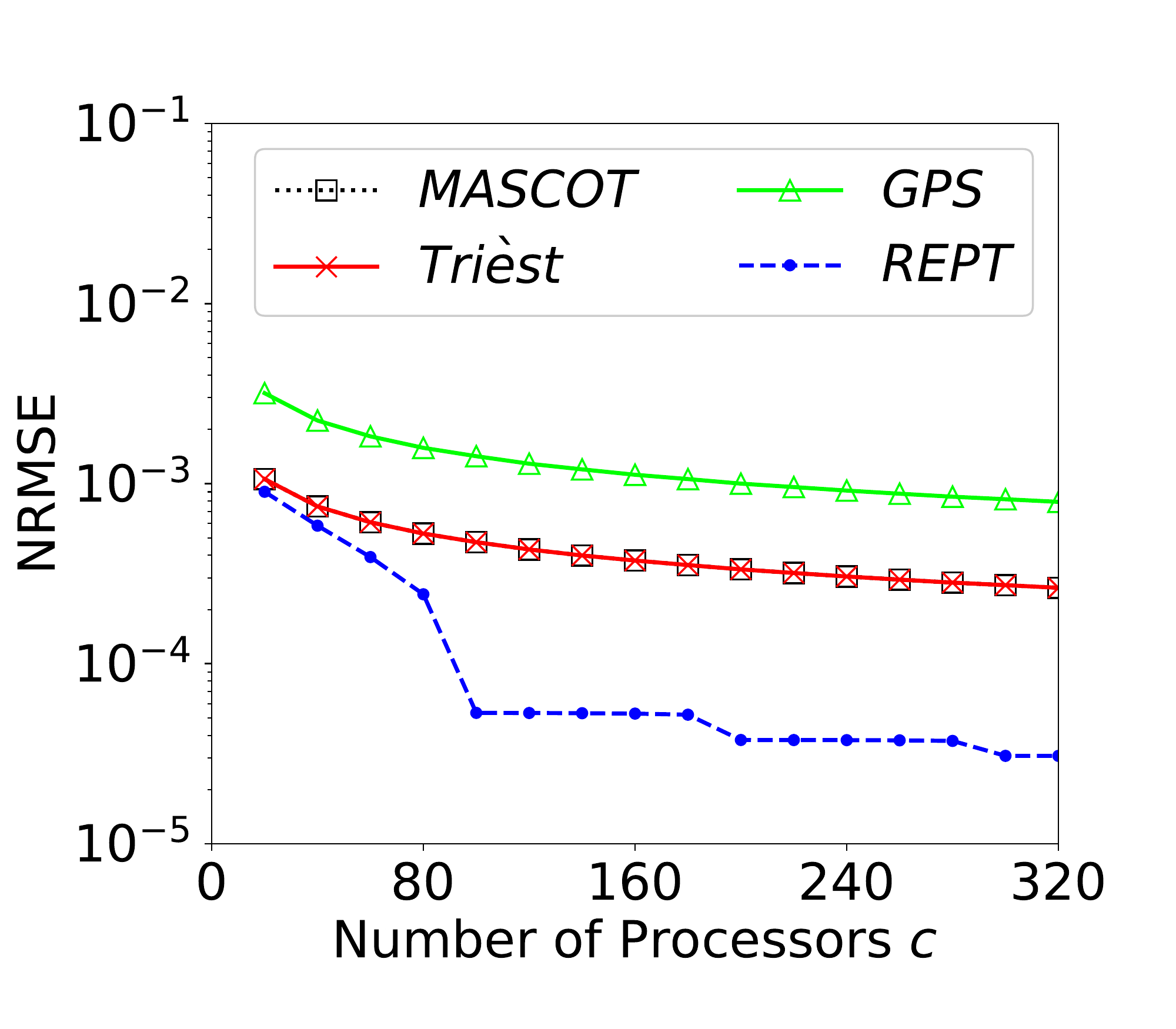}}
\subfigure[com-Orkut]{\includegraphics[width=0.232\textwidth]{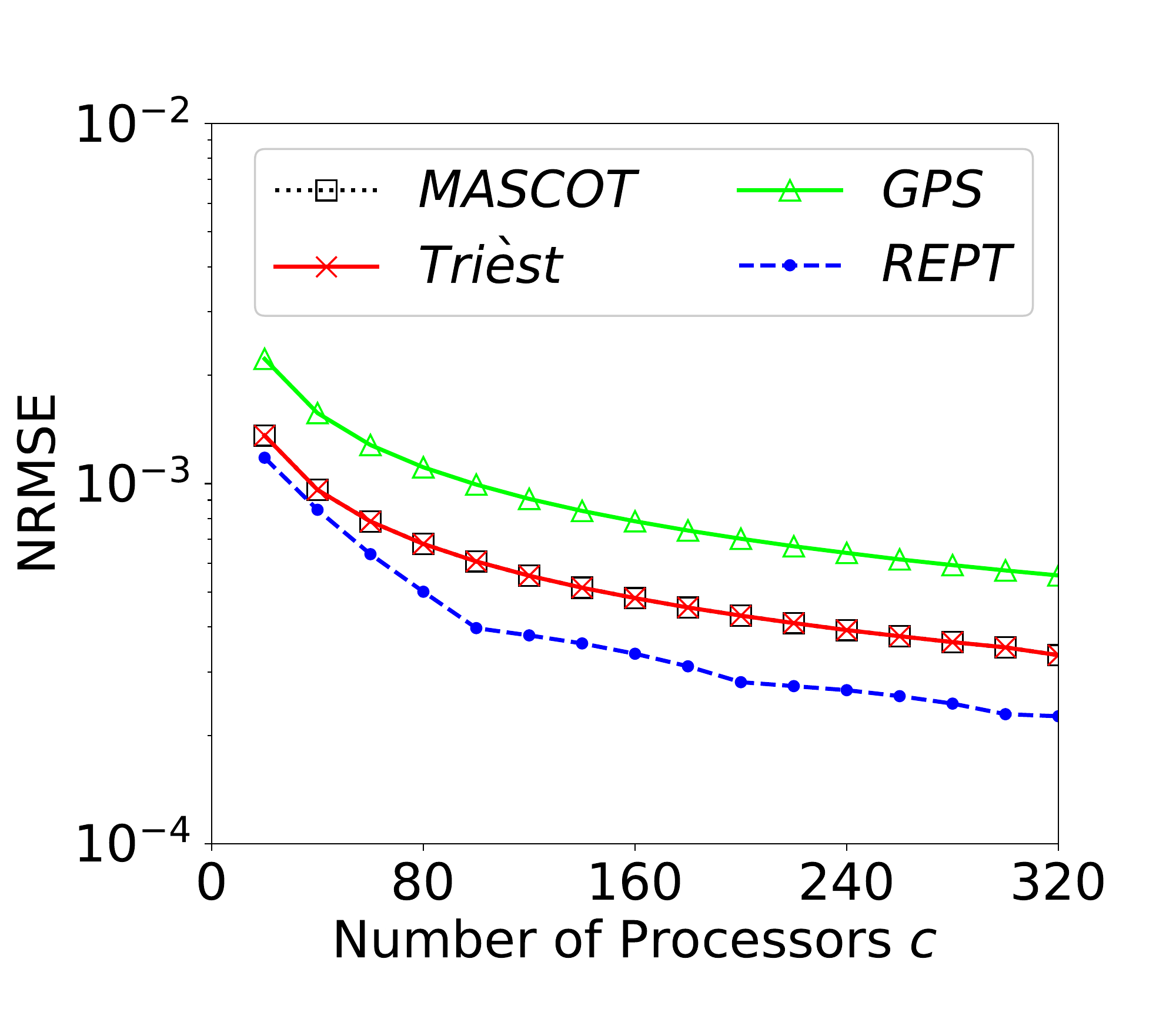}}
\subfigure[LiveJournal]{\includegraphics[width=0.232\textwidth]{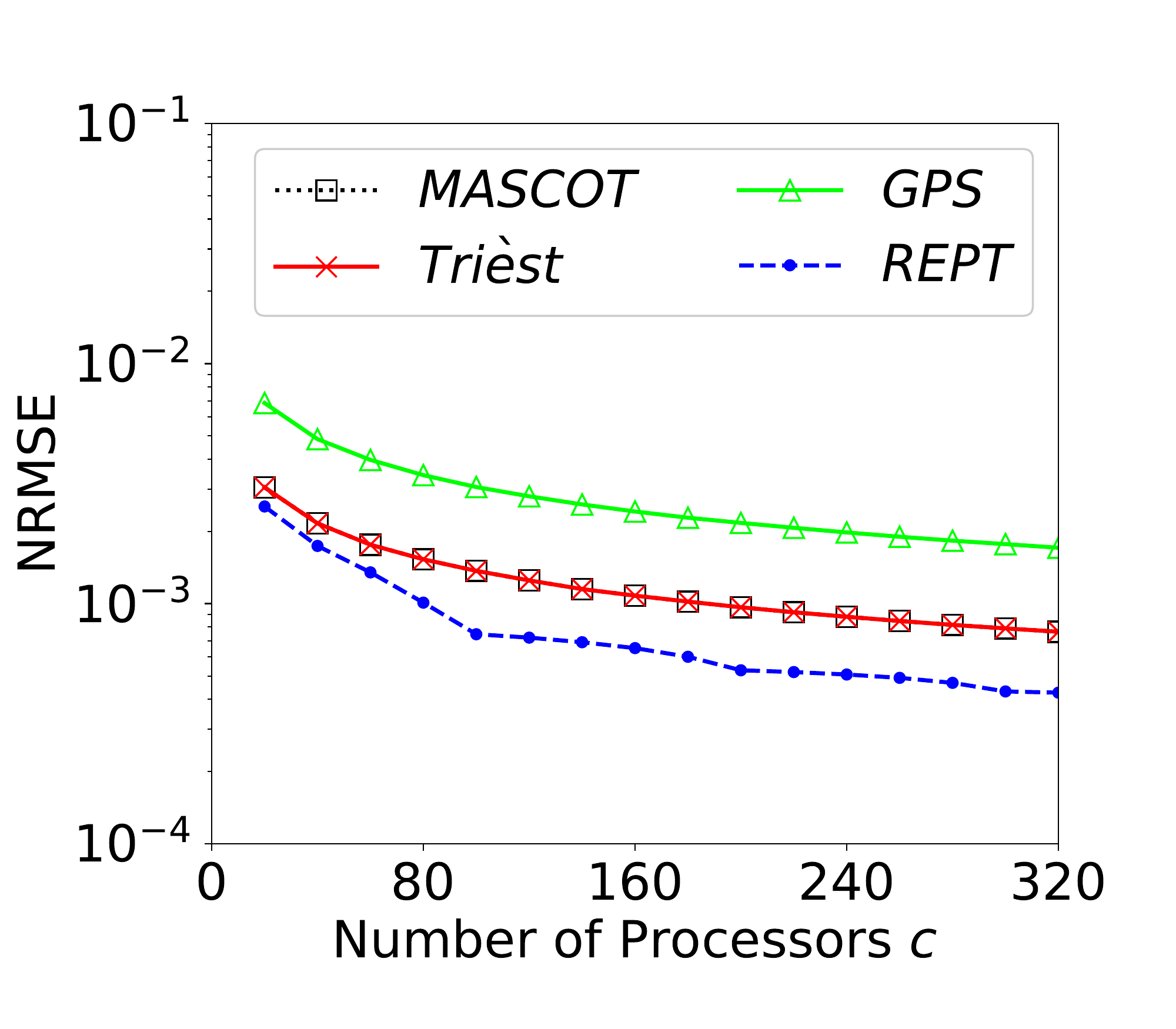}}
\subfigure[Pokec]{\includegraphics[width=0.232\textwidth]{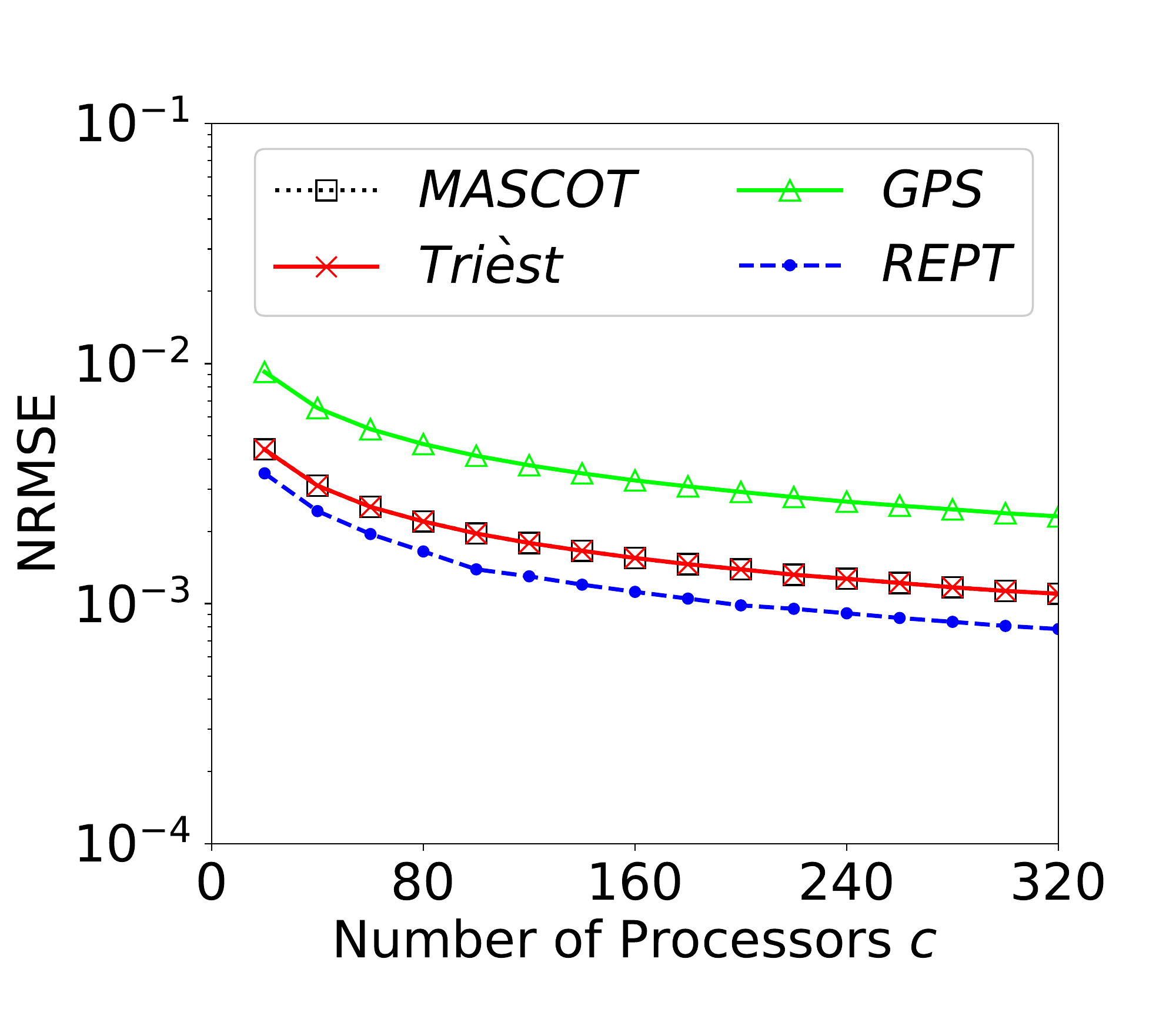}}
\subfigure[Flickr]{\includegraphics[width=0.232\textwidth]{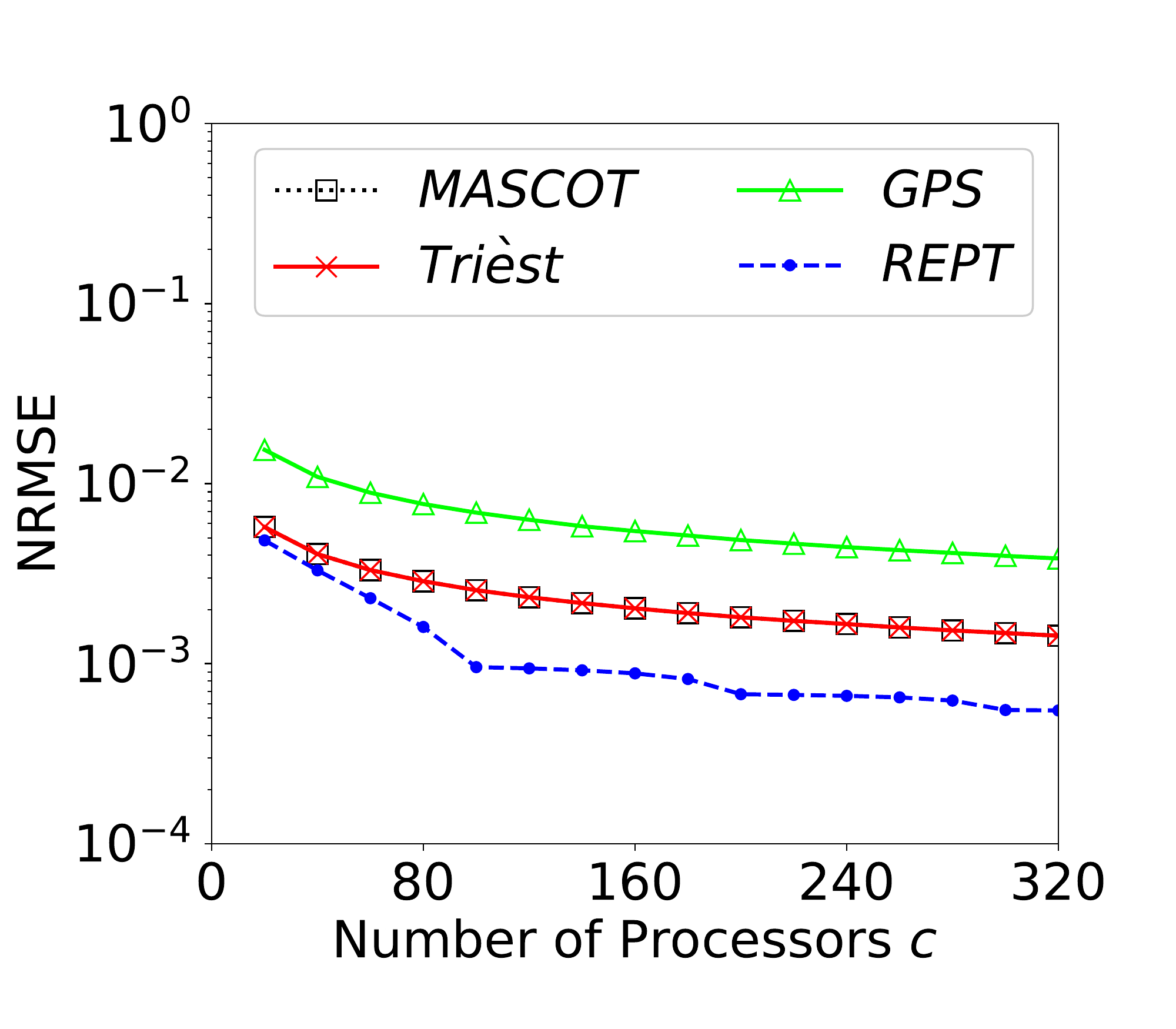}}
\subfigure[Wiki-Talk]{\includegraphics[width=0.232\textwidth]{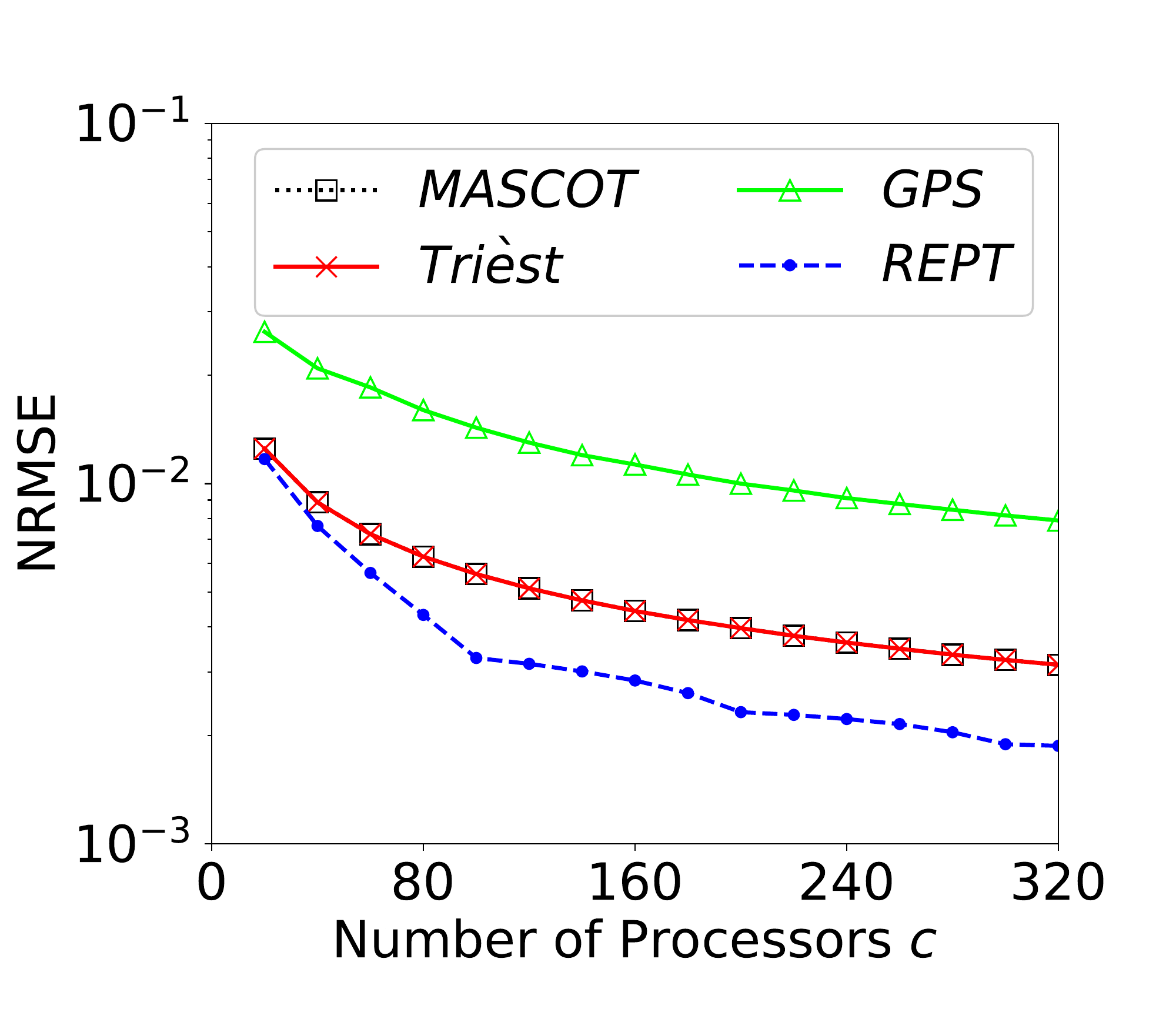}}
\subfigure[Web-Google]{\includegraphics[width=0.232\textwidth]{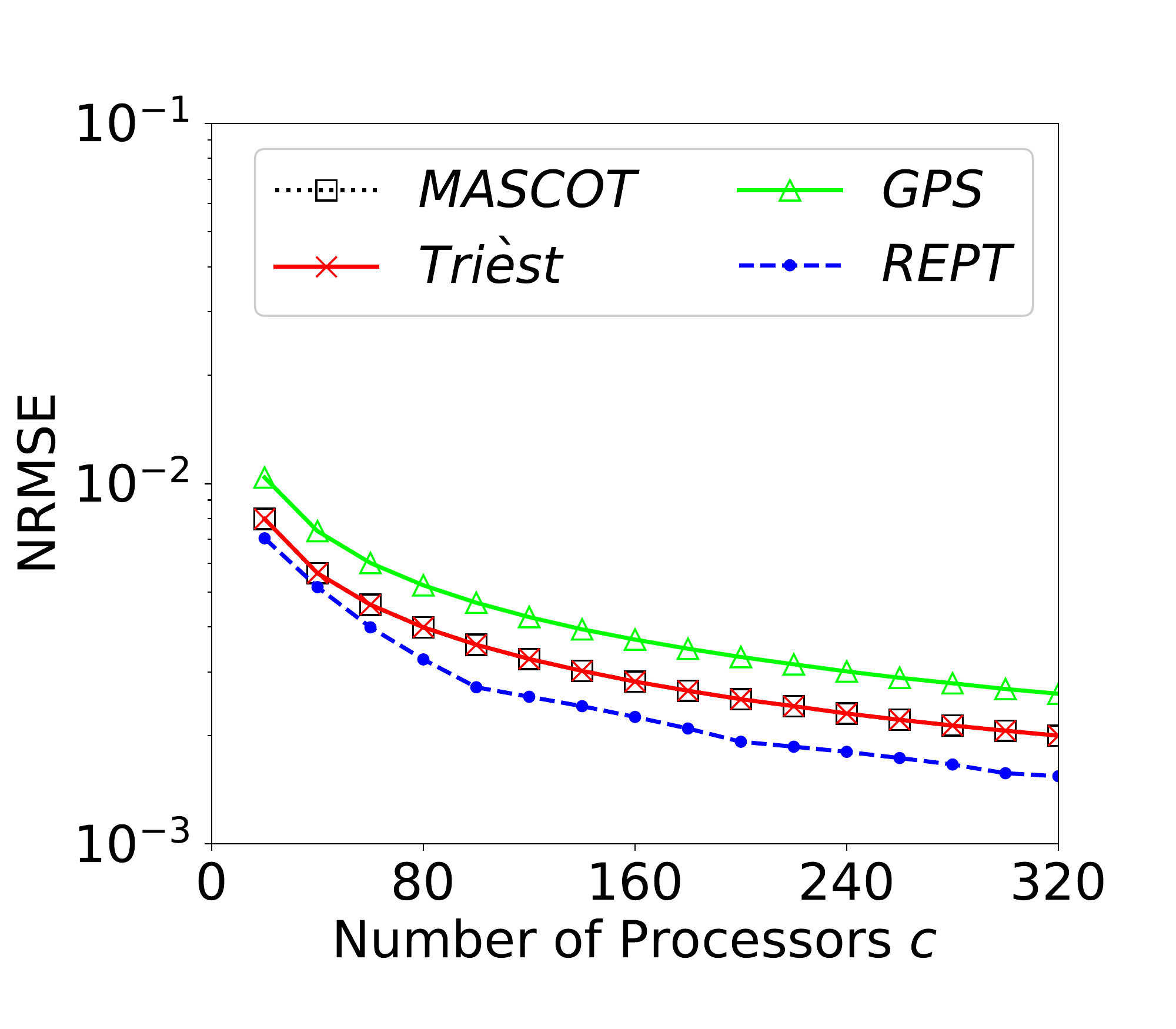}}
\subfigure[YouTube]{\includegraphics[width=0.232\textwidth]{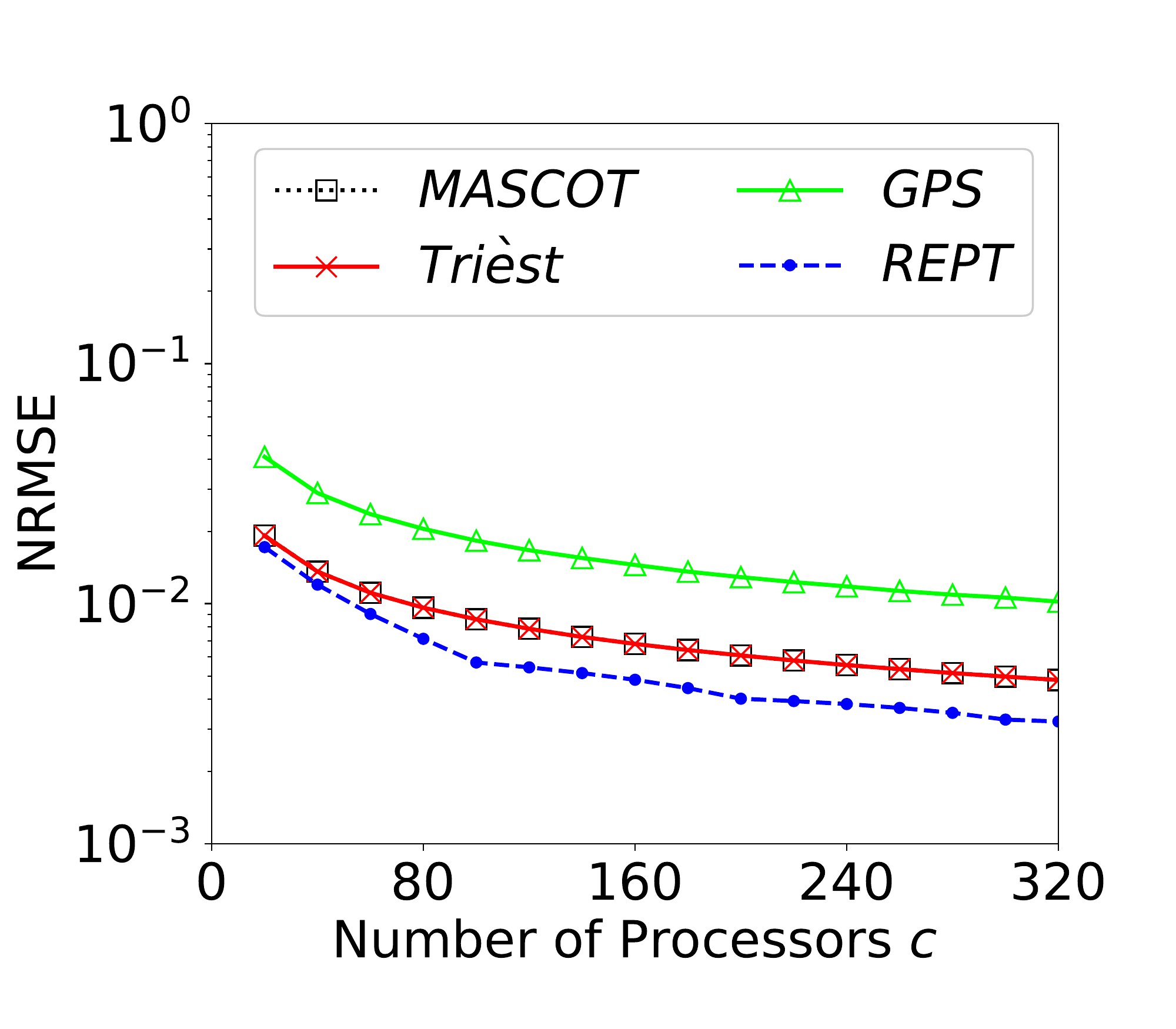}}
\caption{Errors of our method REPT, parallel MASCOT, Tri{\`{e}}st, and GPS for estimating global triangle counts, $p=0.01$.}
\label{fig:Global_m_20}
\end{figure*}

\begin{figure*}[!t]
\centering
\subfigure[Twitter]{\includegraphics[width=0.232\textwidth]{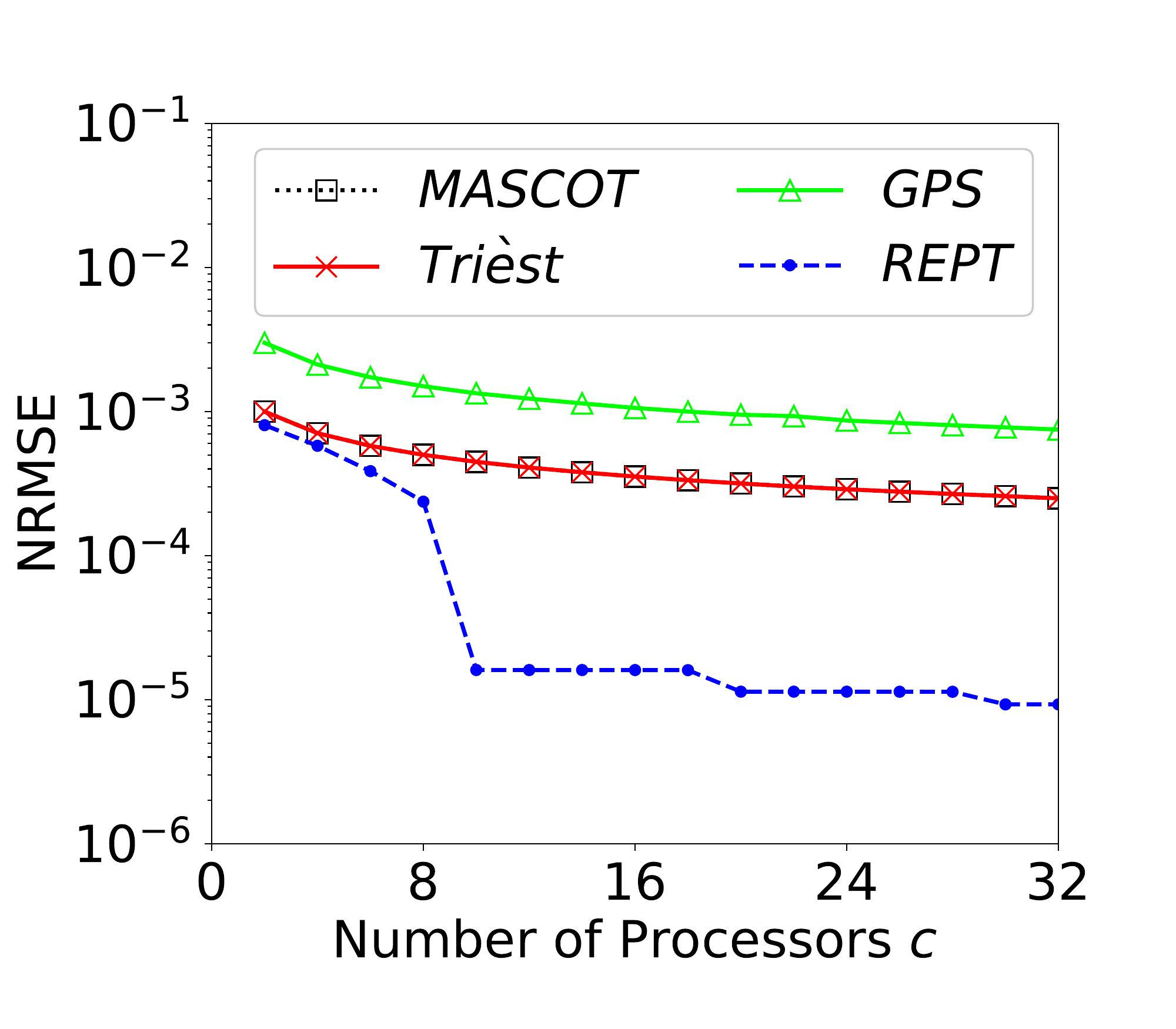}}
\subfigure[com-Orkut]{\includegraphics[width=0.232\textwidth]{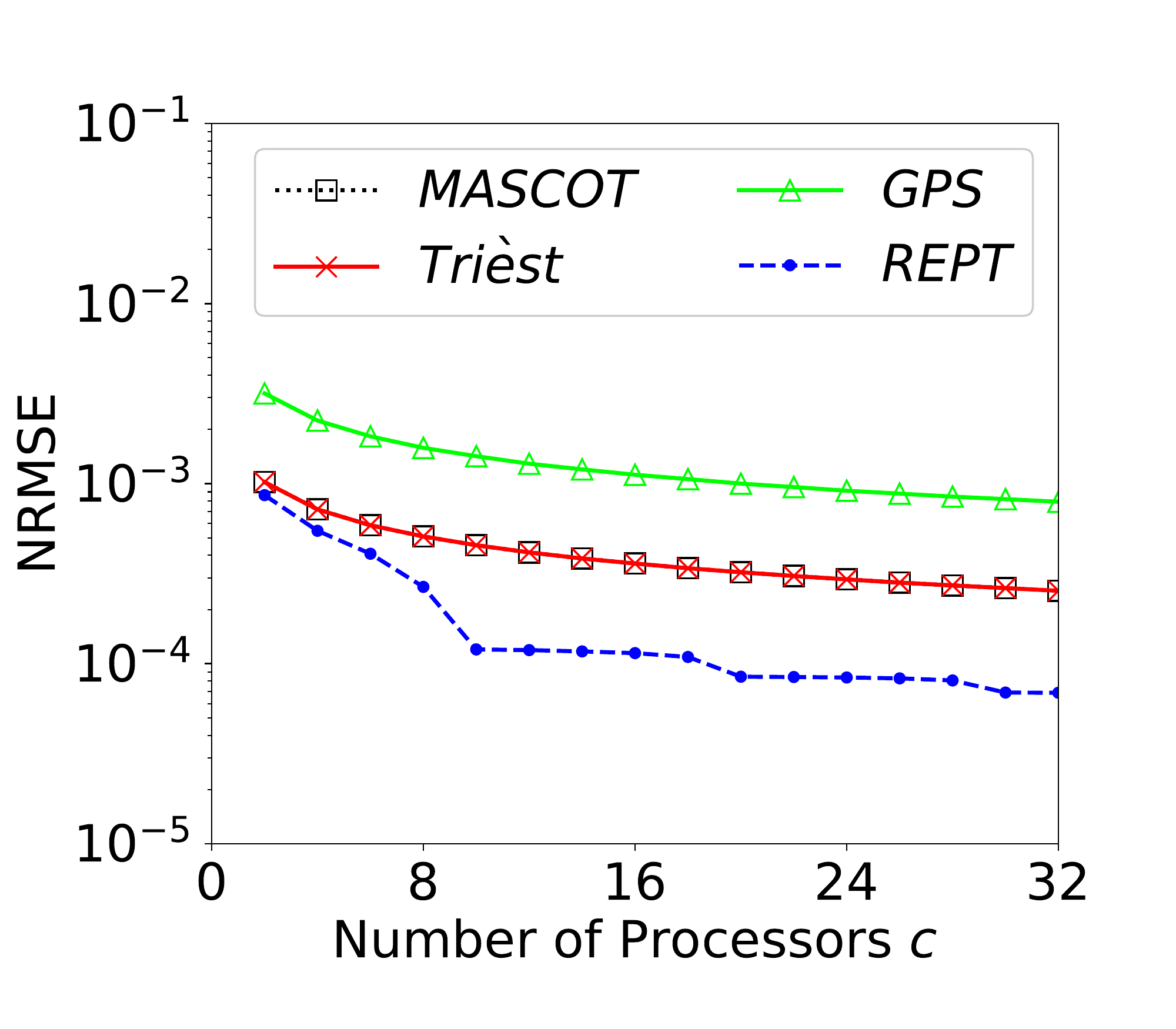}}
\subfigure[LiveJournal]{\includegraphics[width=0.232\textwidth]{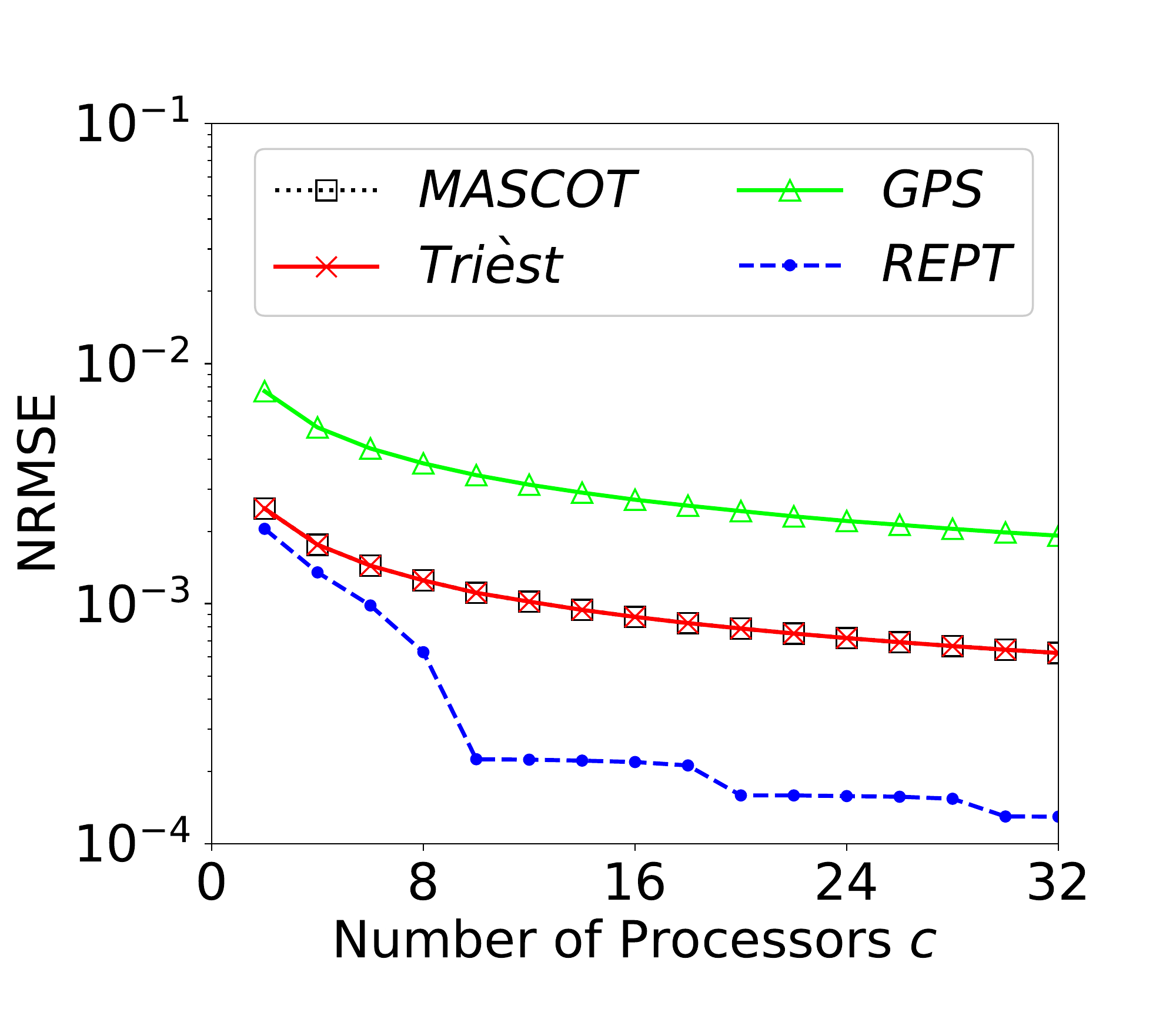}}
\subfigure[Pokec]{\includegraphics[width=0.232\textwidth]{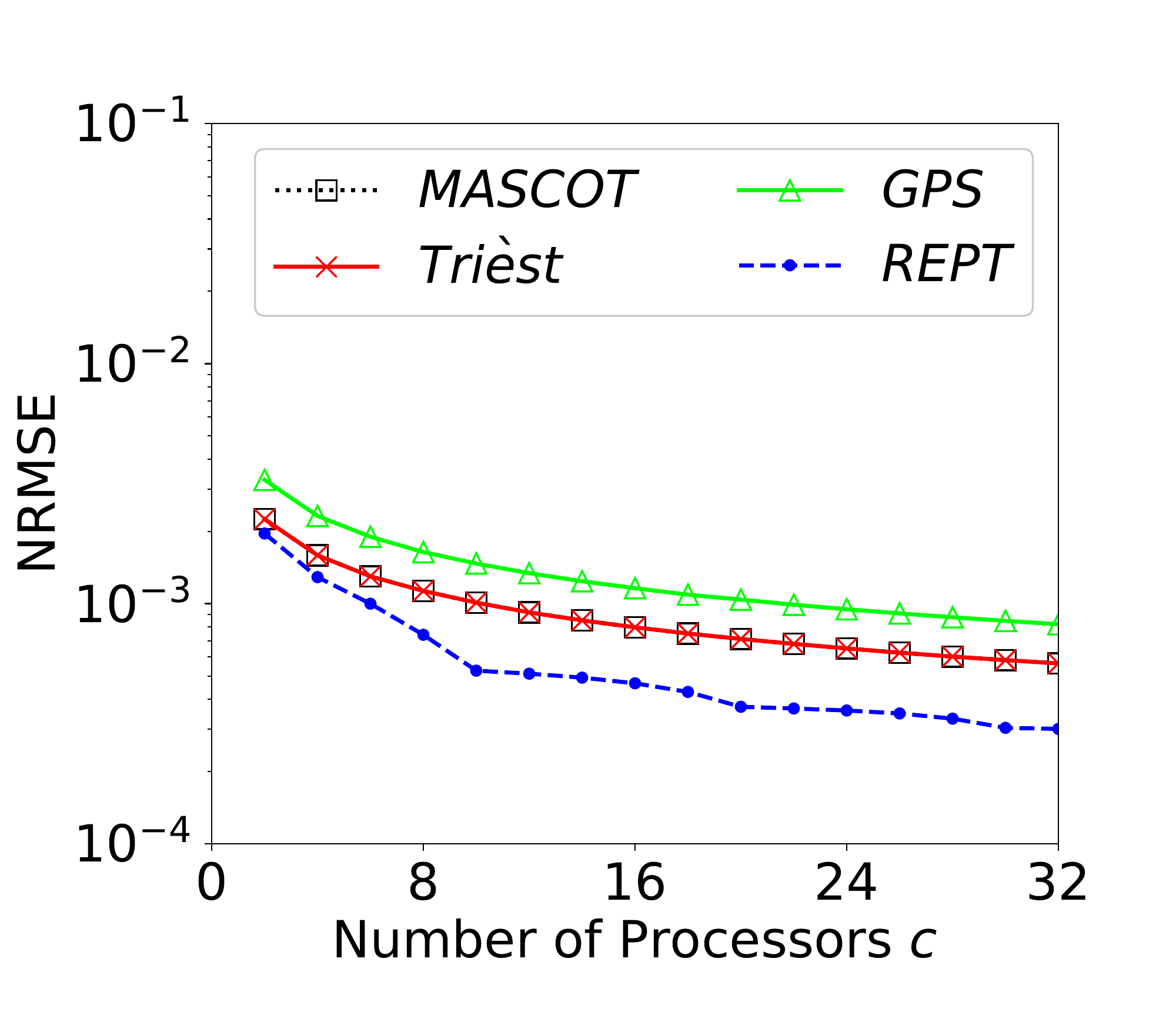}}
\subfigure[Flickr]{\includegraphics[width=0.232\textwidth]{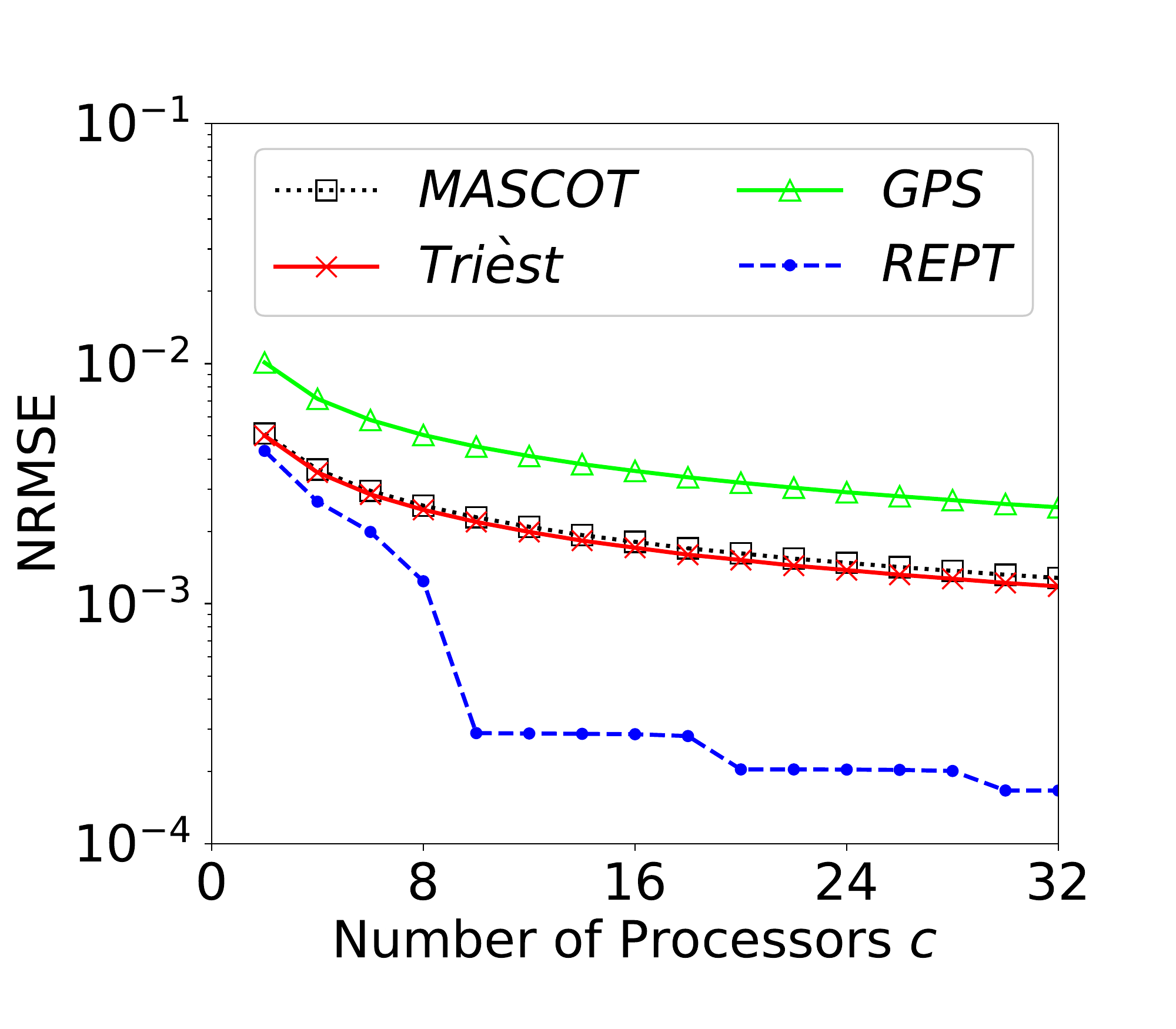}}
\subfigure[Wiki-Talk]{\includegraphics[width=0.232\textwidth]{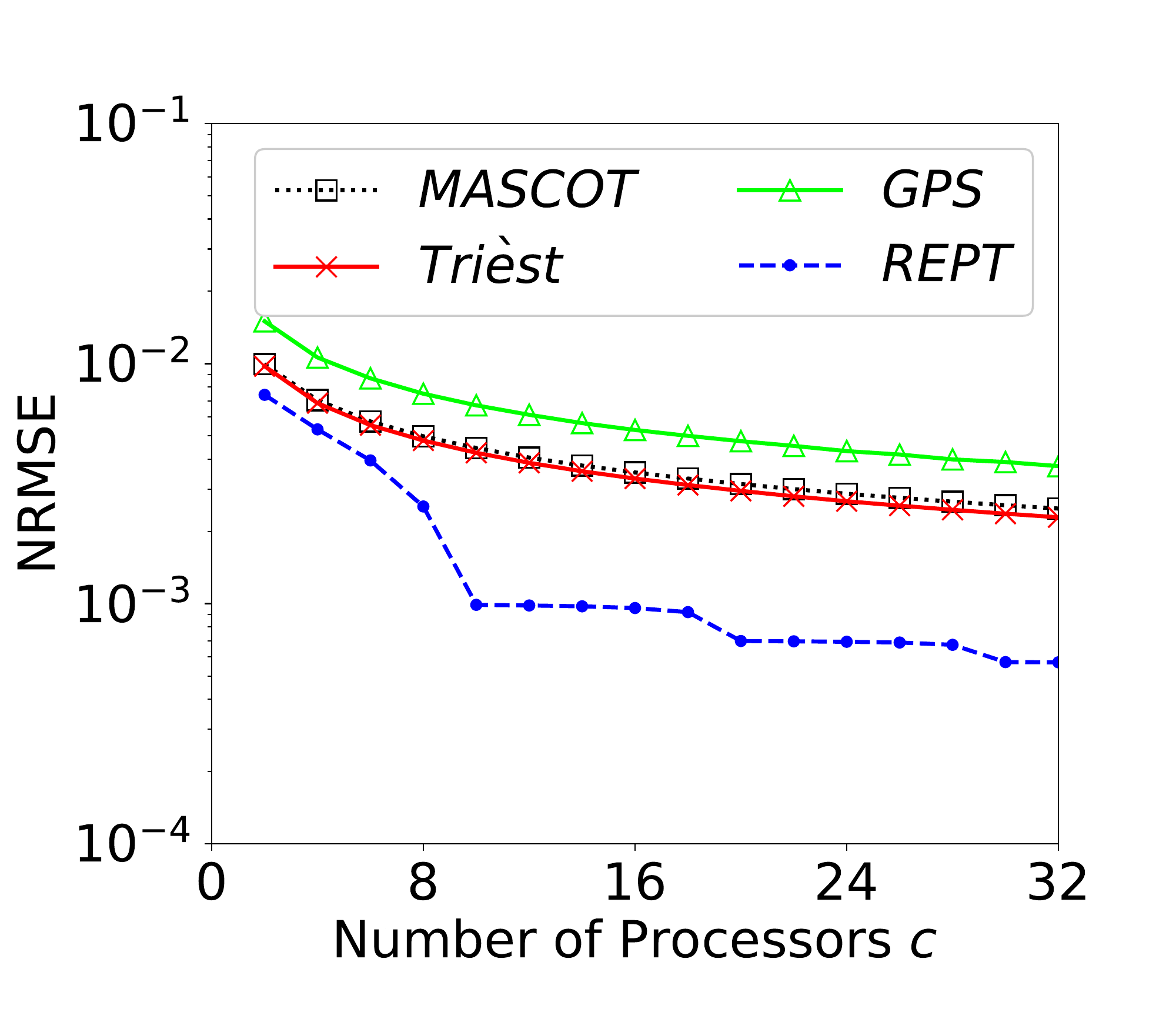}}
\subfigure[Web-Google]{\includegraphics[width=0.232\textwidth]{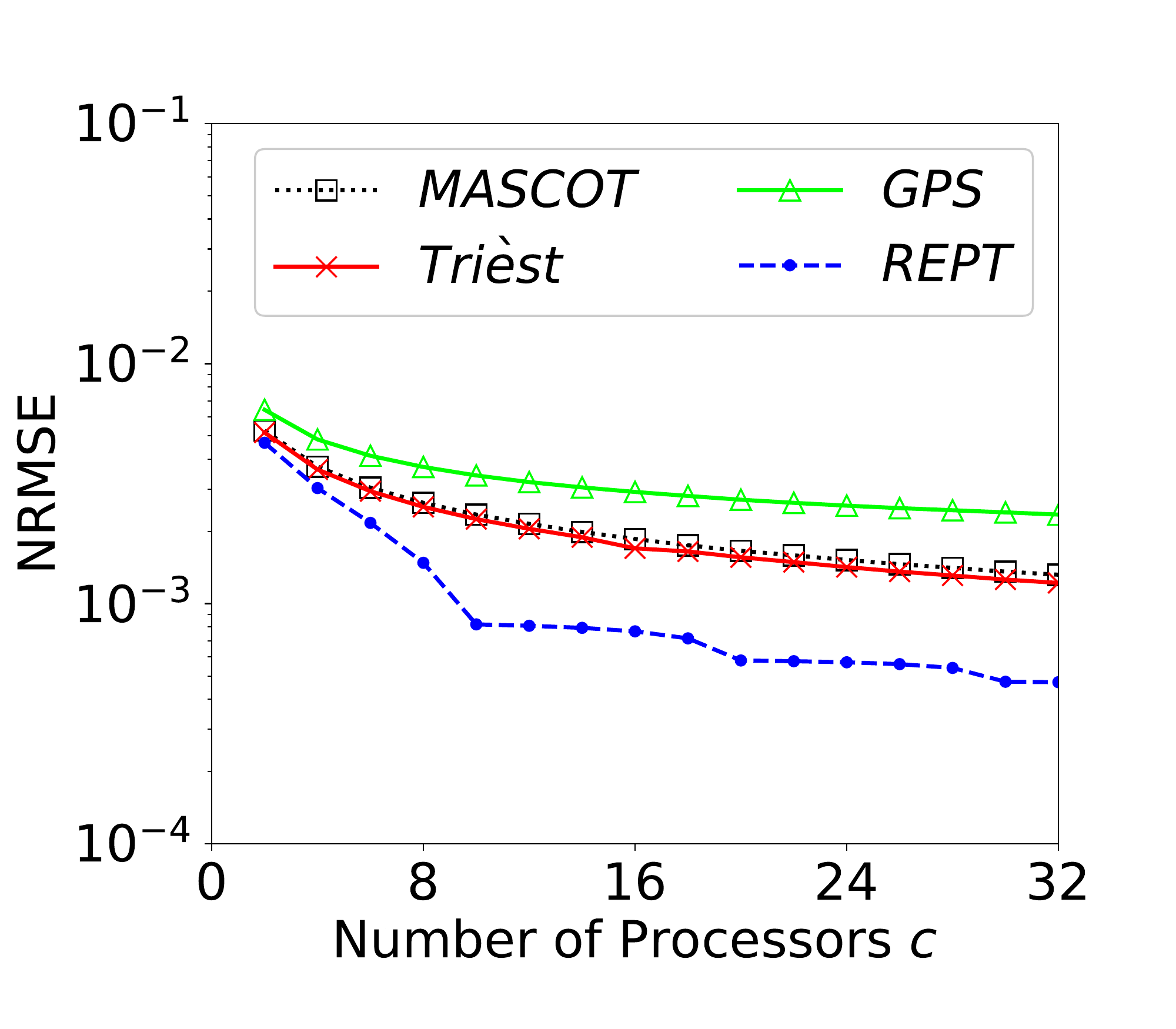}}
\subfigure[YouTube]{\includegraphics[width=0.232\textwidth]{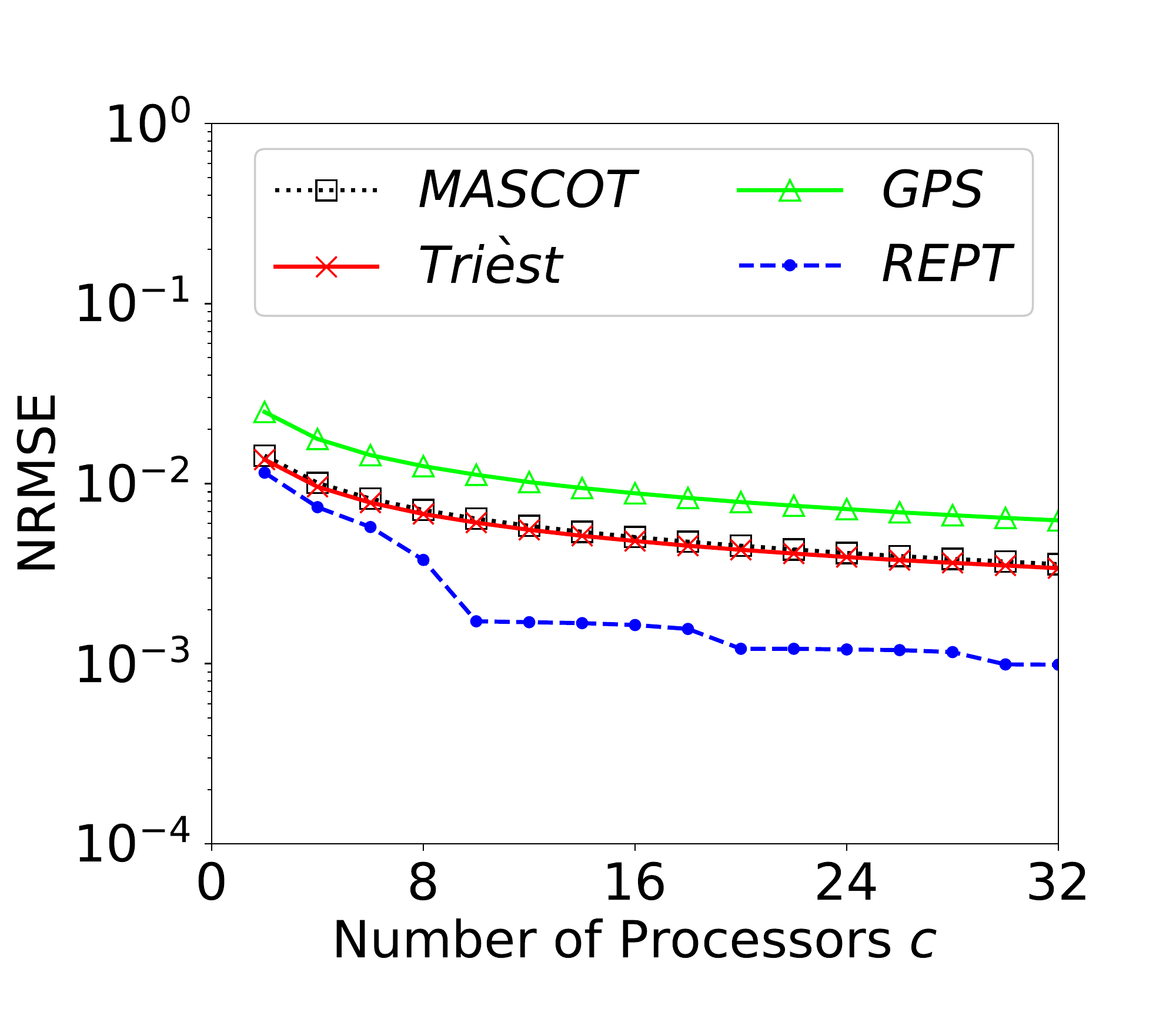}}
\caption{Errors of our method REPT, parallel MASCOT, Tri{\`{e}}st, and GPS for estimating global triangle counts, $p=0.1$.}
\label{fig:Global_m_10}
\end{figure*}
\begin{figure*}[!t]
\centering
\subfigure[Twitter]{\includegraphics[width=0.232\textwidth]{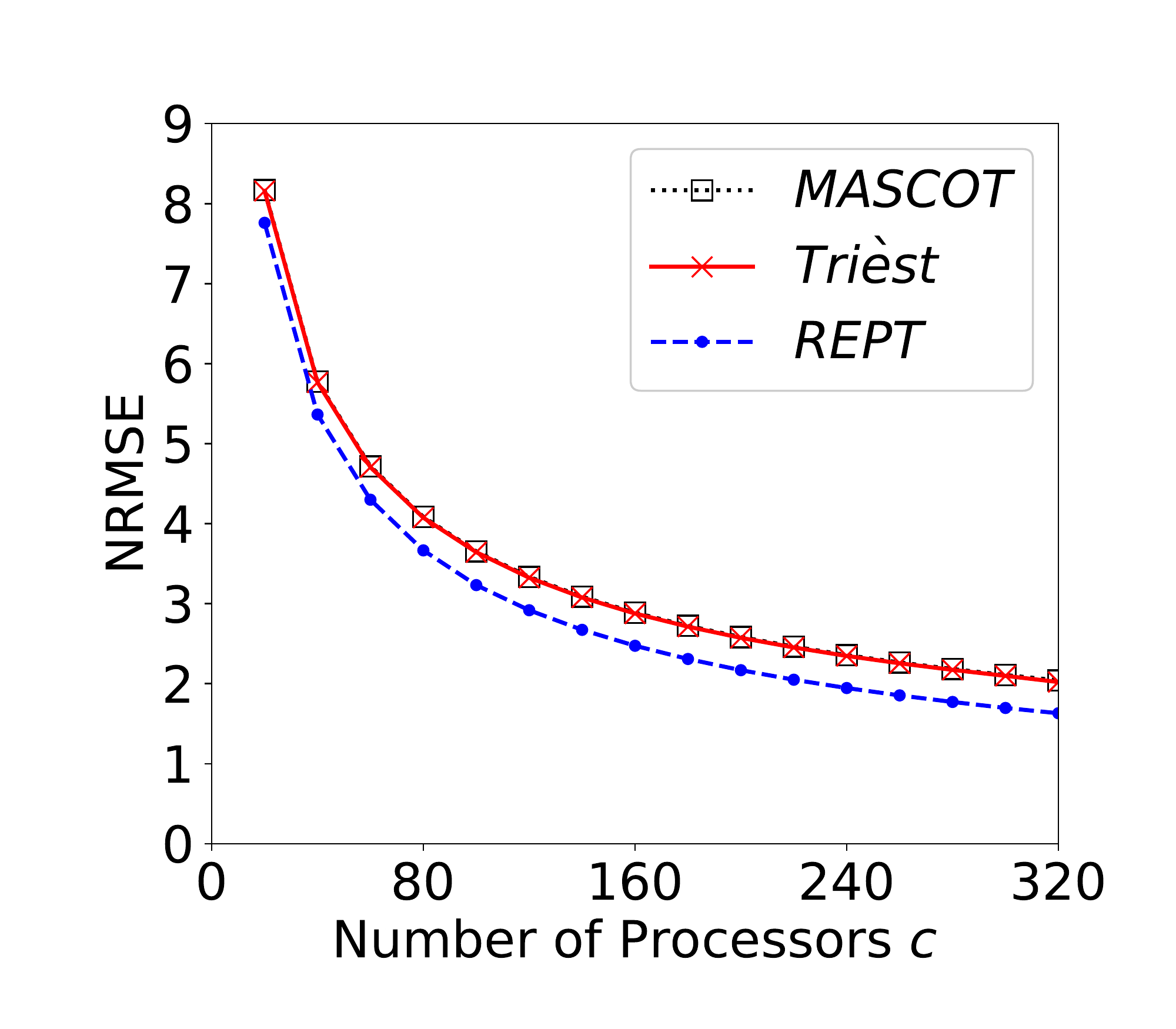}}
\subfigure[com-Orkut]{\includegraphics[width=0.232\textwidth]{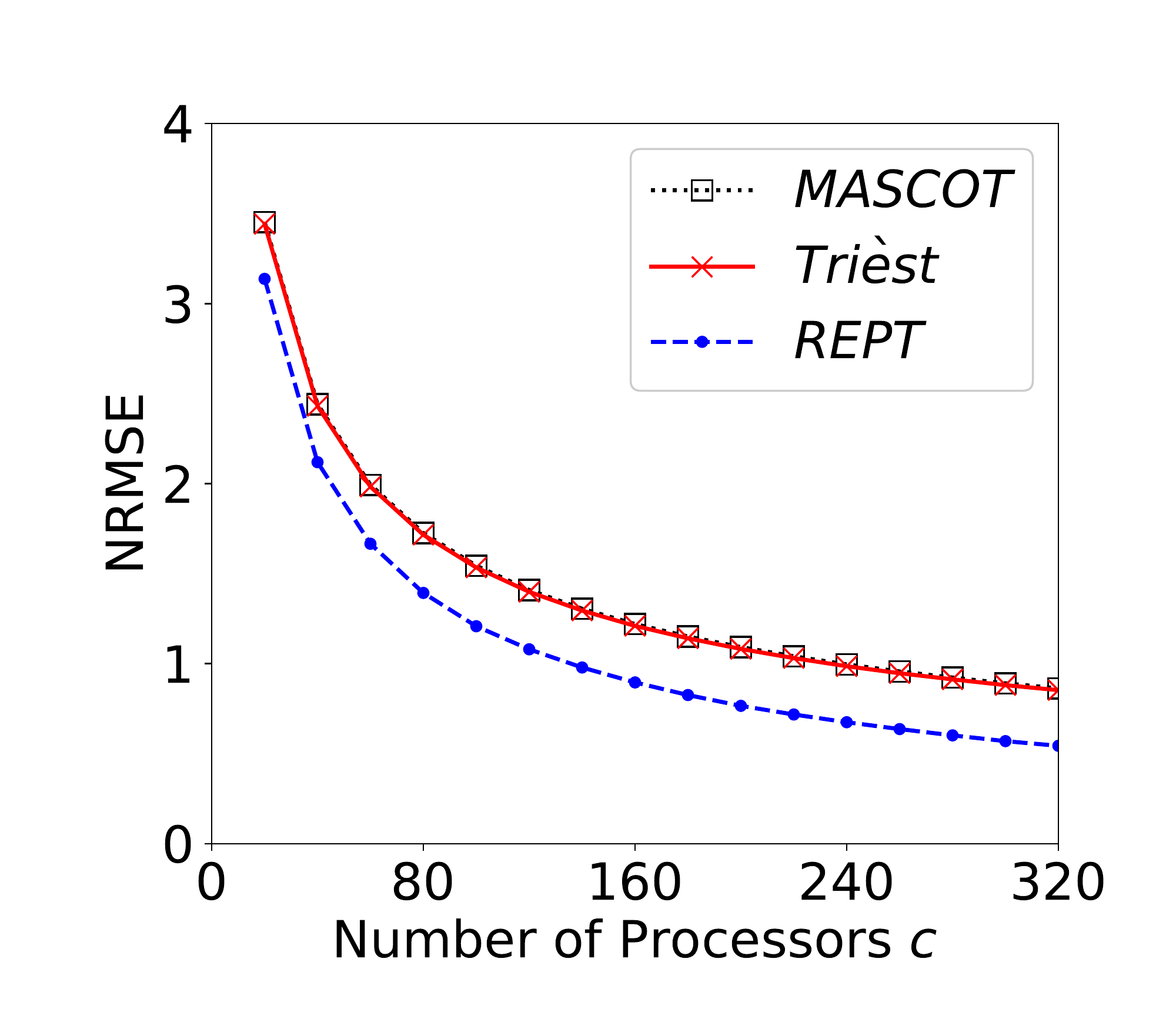}}
\subfigure[LiveJournal]{\includegraphics[width=0.232\textwidth]{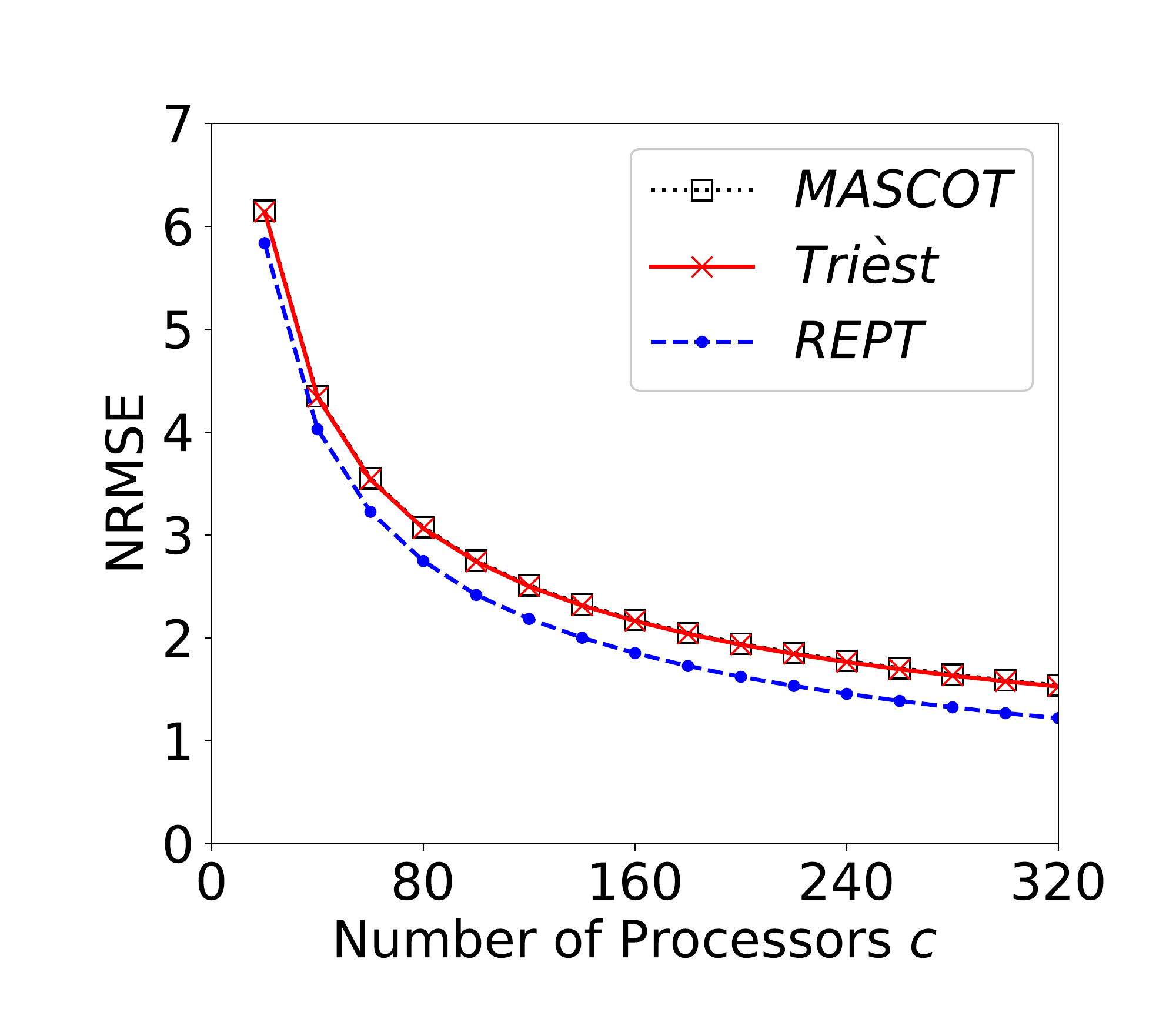}}
\subfigure[Pokec]{\includegraphics[width=0.232\textwidth]{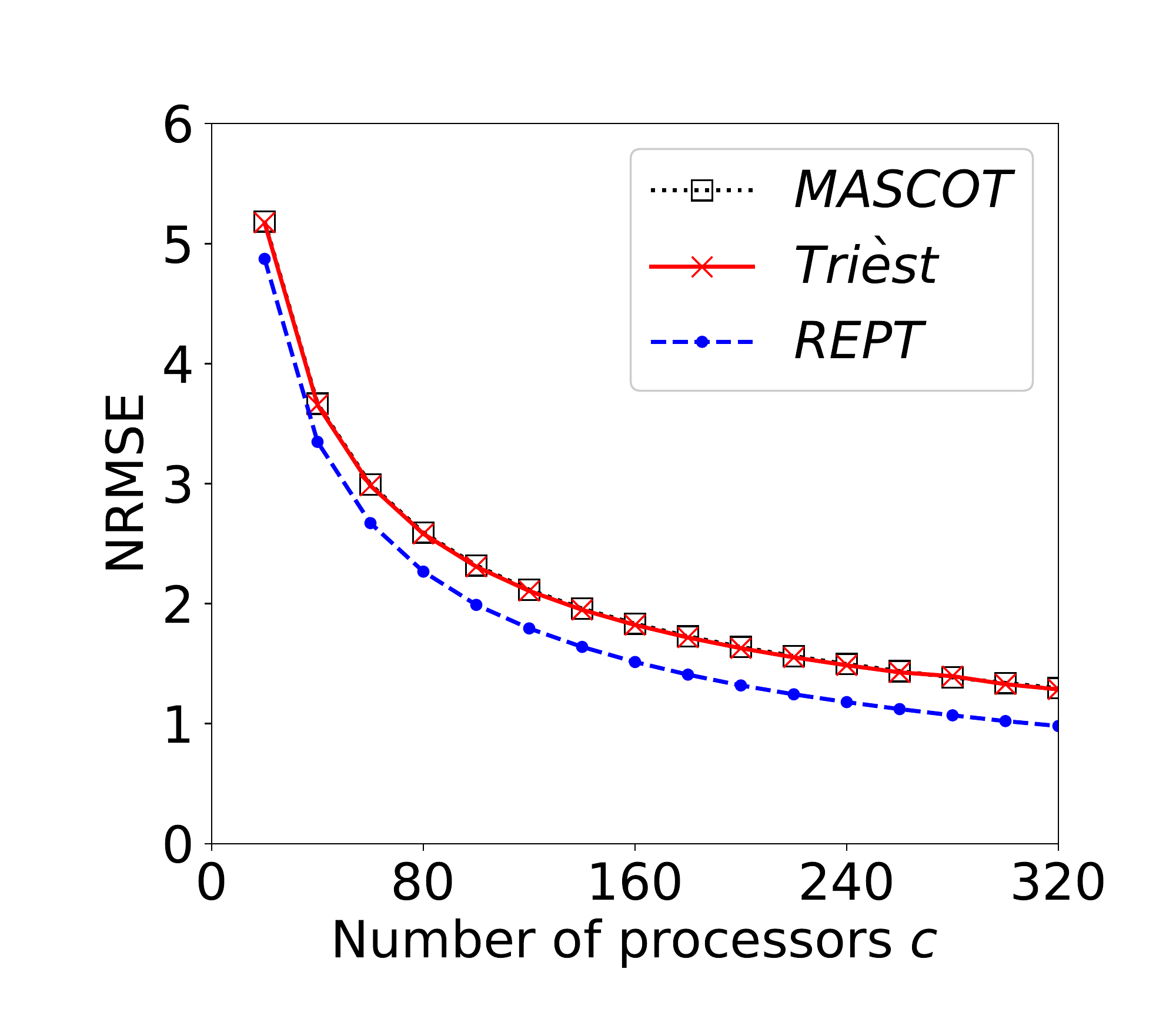}}
\subfigure[Flickr]{\includegraphics[width=0.232\textwidth]{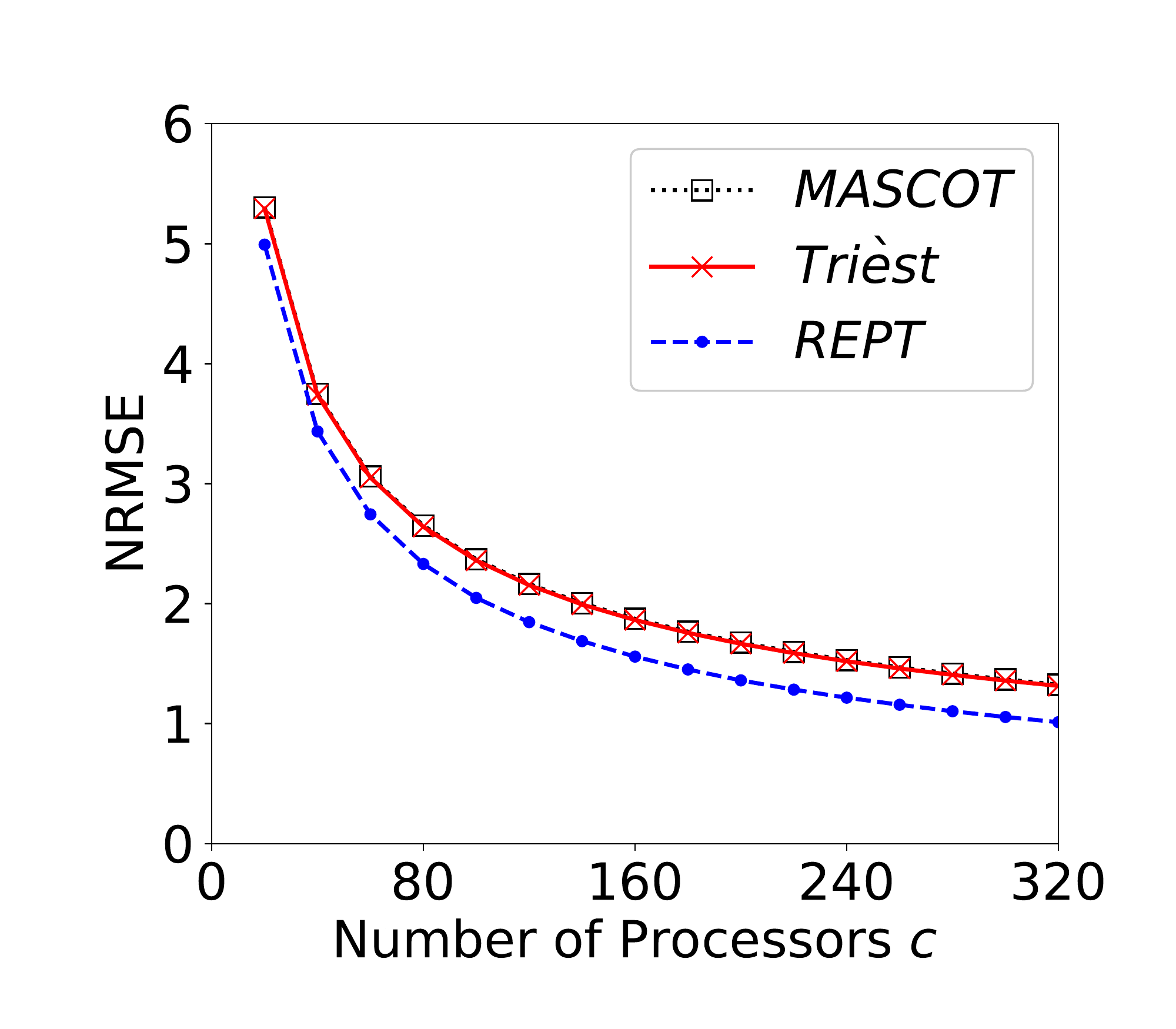}}
\subfigure[Wiki-Talk]{\includegraphics[width=0.232\textwidth]{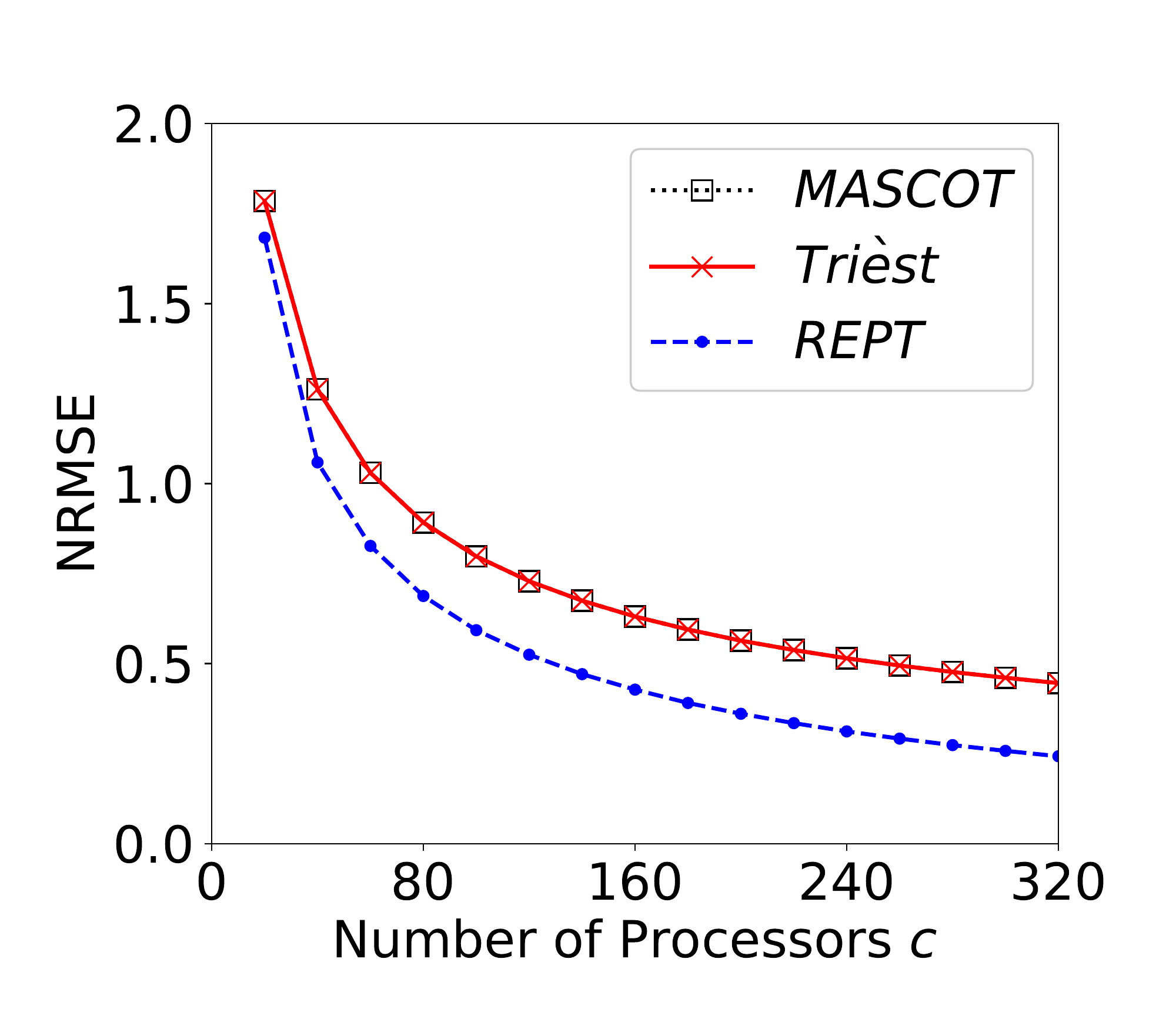}}
\subfigure[Web-Google]{\includegraphics[width=0.232\textwidth]{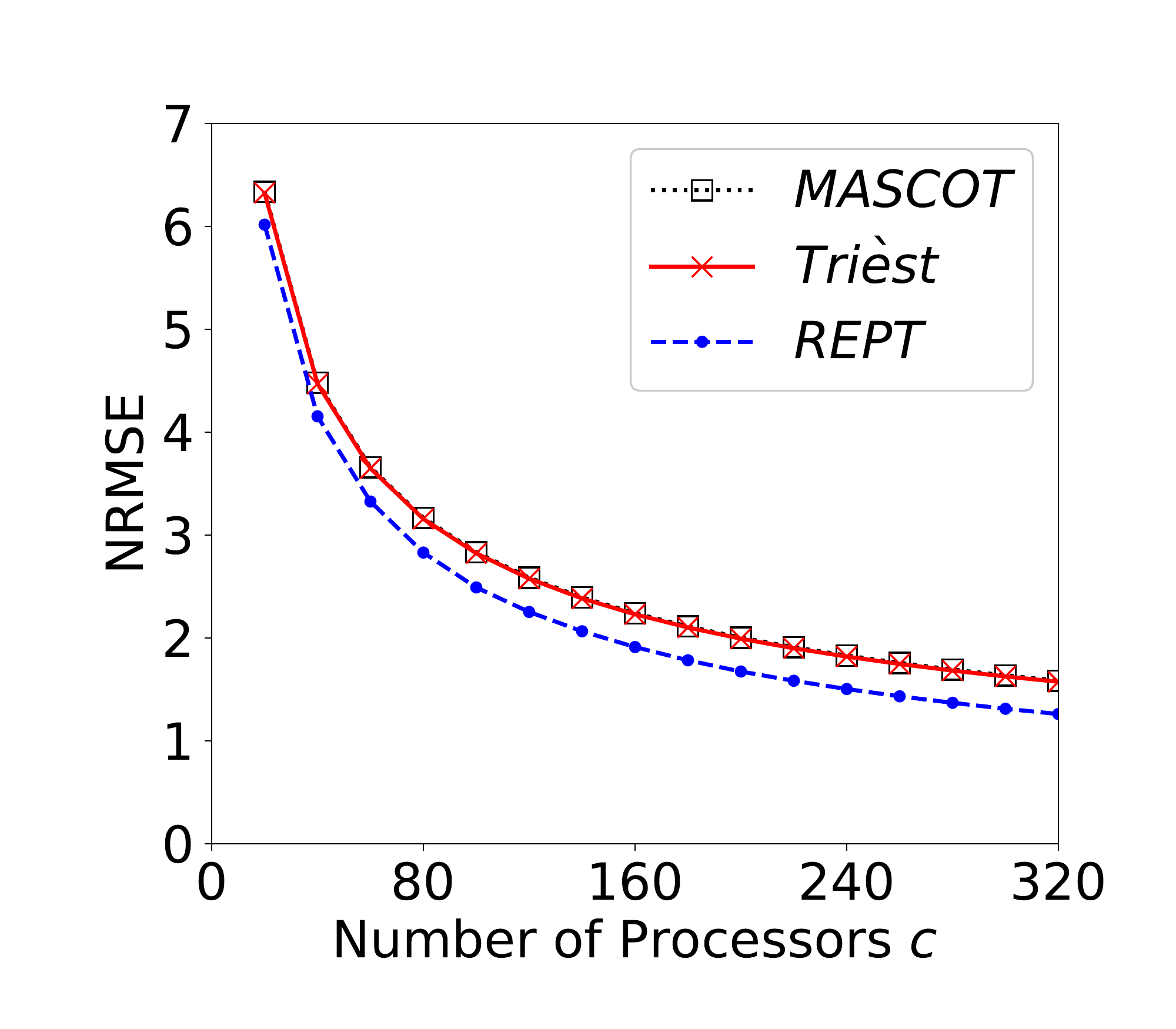}}
\subfigure[YouTube]{\includegraphics[width=0.232\textwidth]{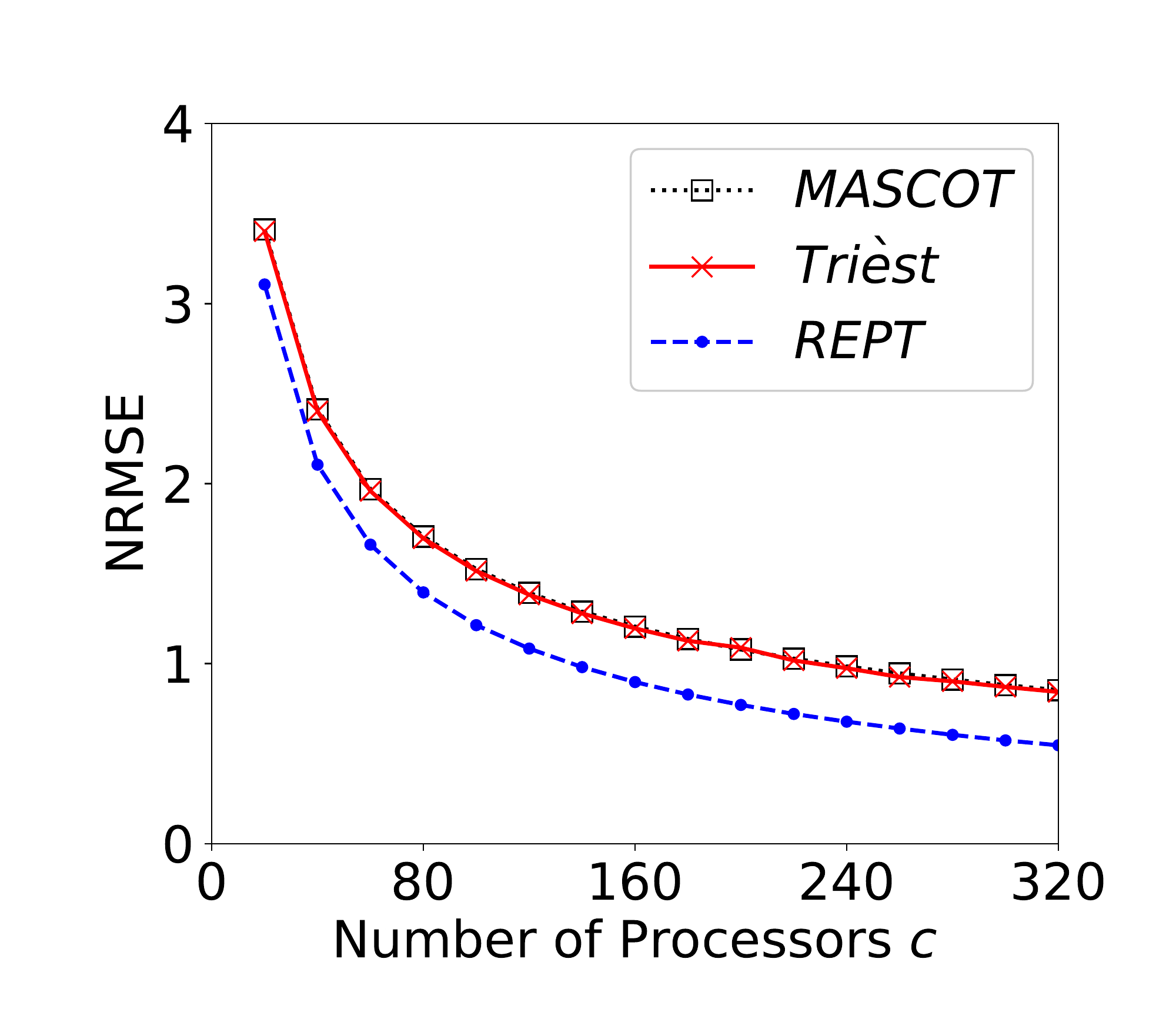}}
\caption{Errors of our method REPT, parallel MASCOT, and parallel Tri{\`{e}}st for estimating local triangle counts, $p=0.01$.}
\label{fig:Local_m_20}
\end{figure*}

\begin{figure*}[!t]
\centering
\subfigure[Twitter]{\includegraphics[width=0.232\textwidth]{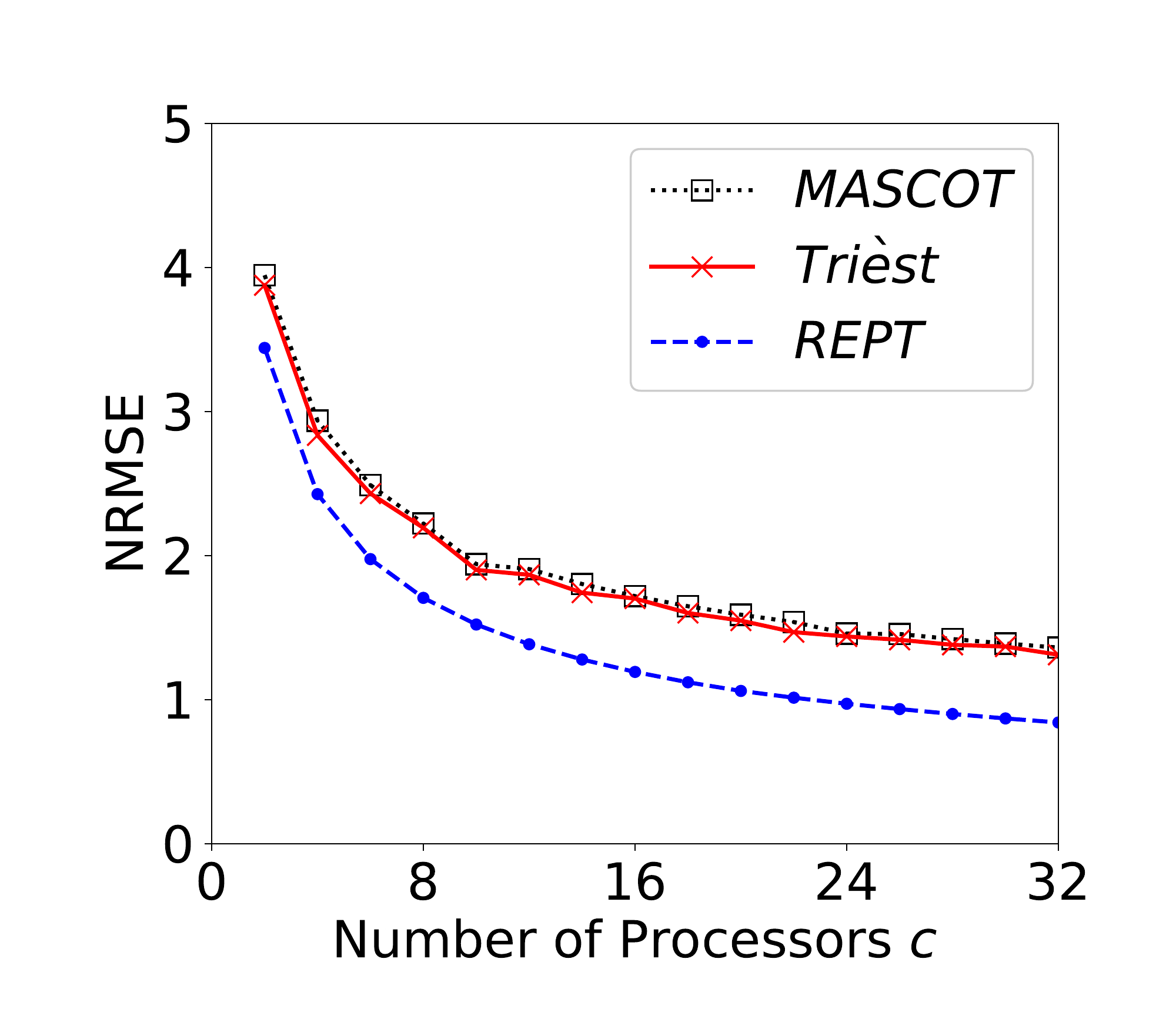}}
\subfigure[com-Orkut]{\includegraphics[width=0.232\textwidth]{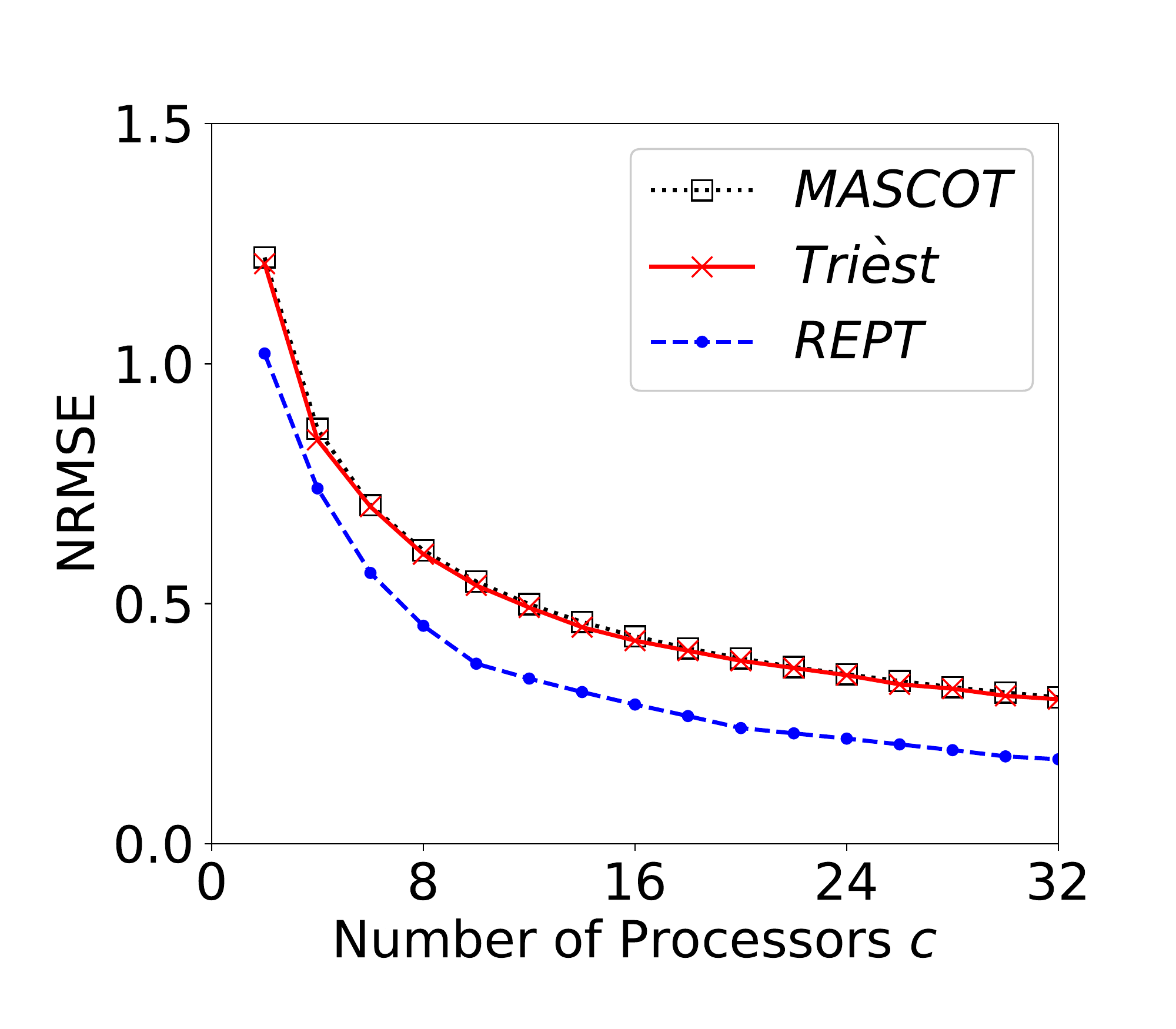}}
\subfigure[LiveJournal]{\includegraphics[width=0.232\textwidth]{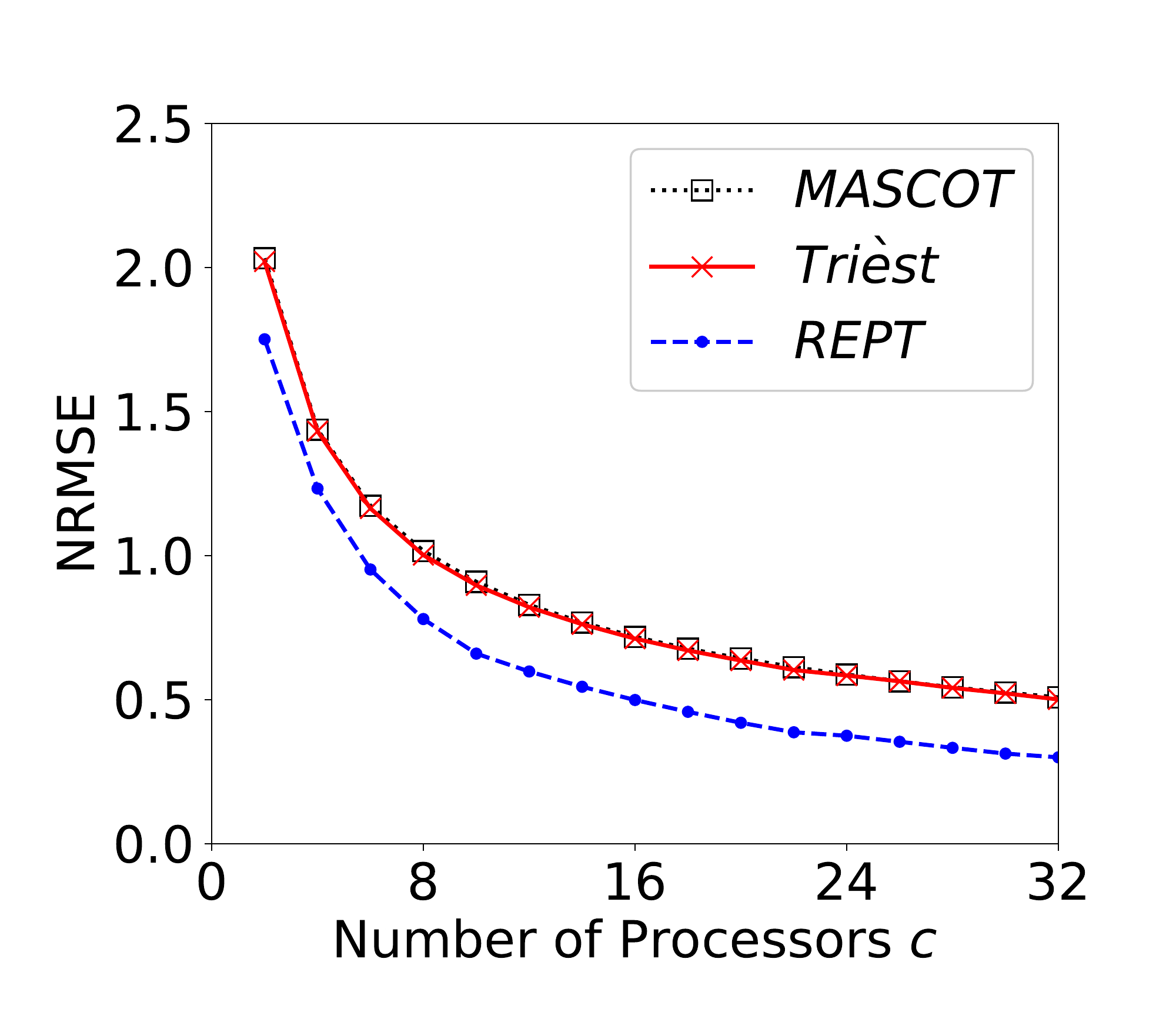}}
\subfigure[Pokec]{\includegraphics[width=0.232\textwidth]{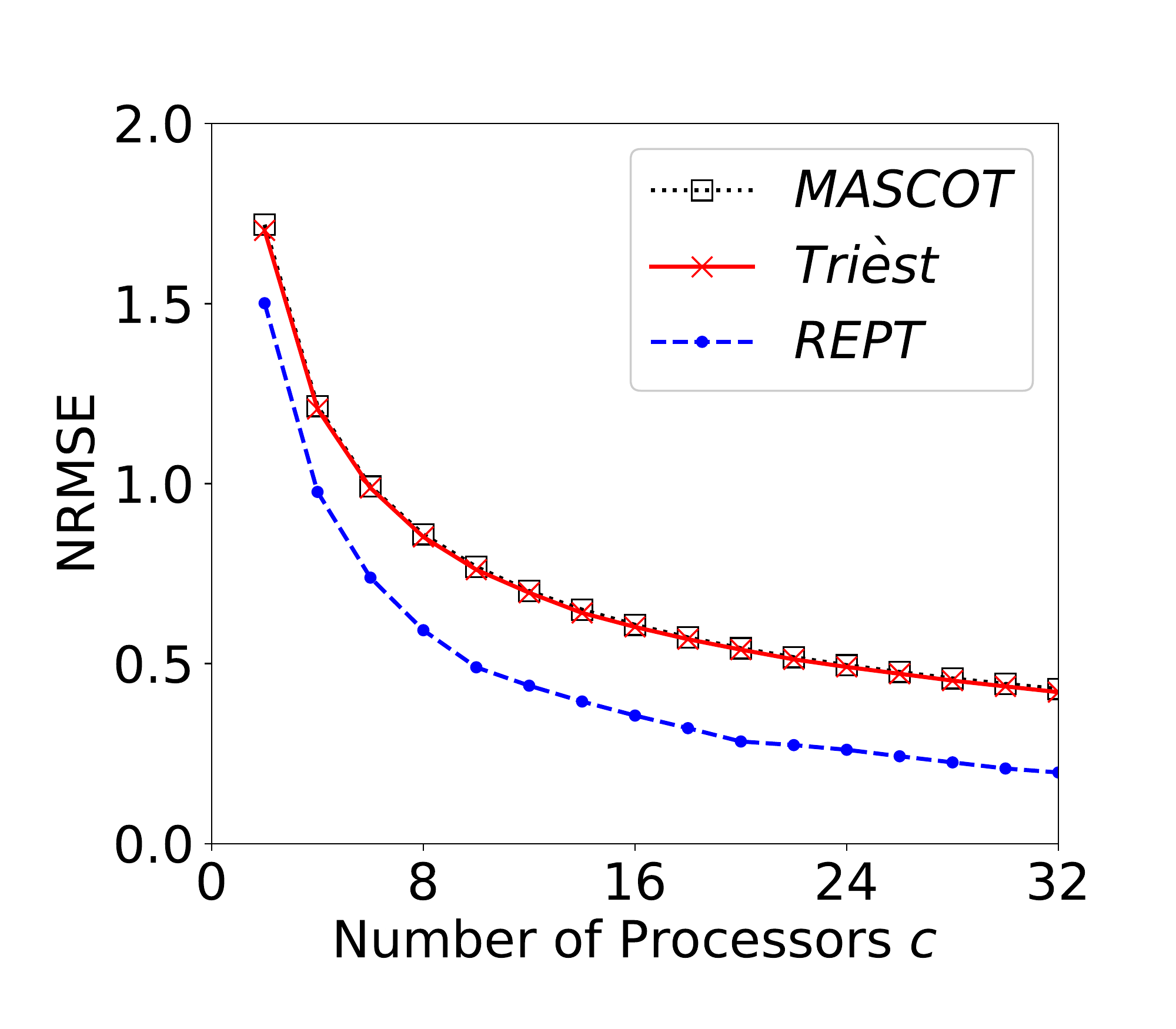}}
\subfigure[Flickr]{\includegraphics[width=0.232\textwidth]{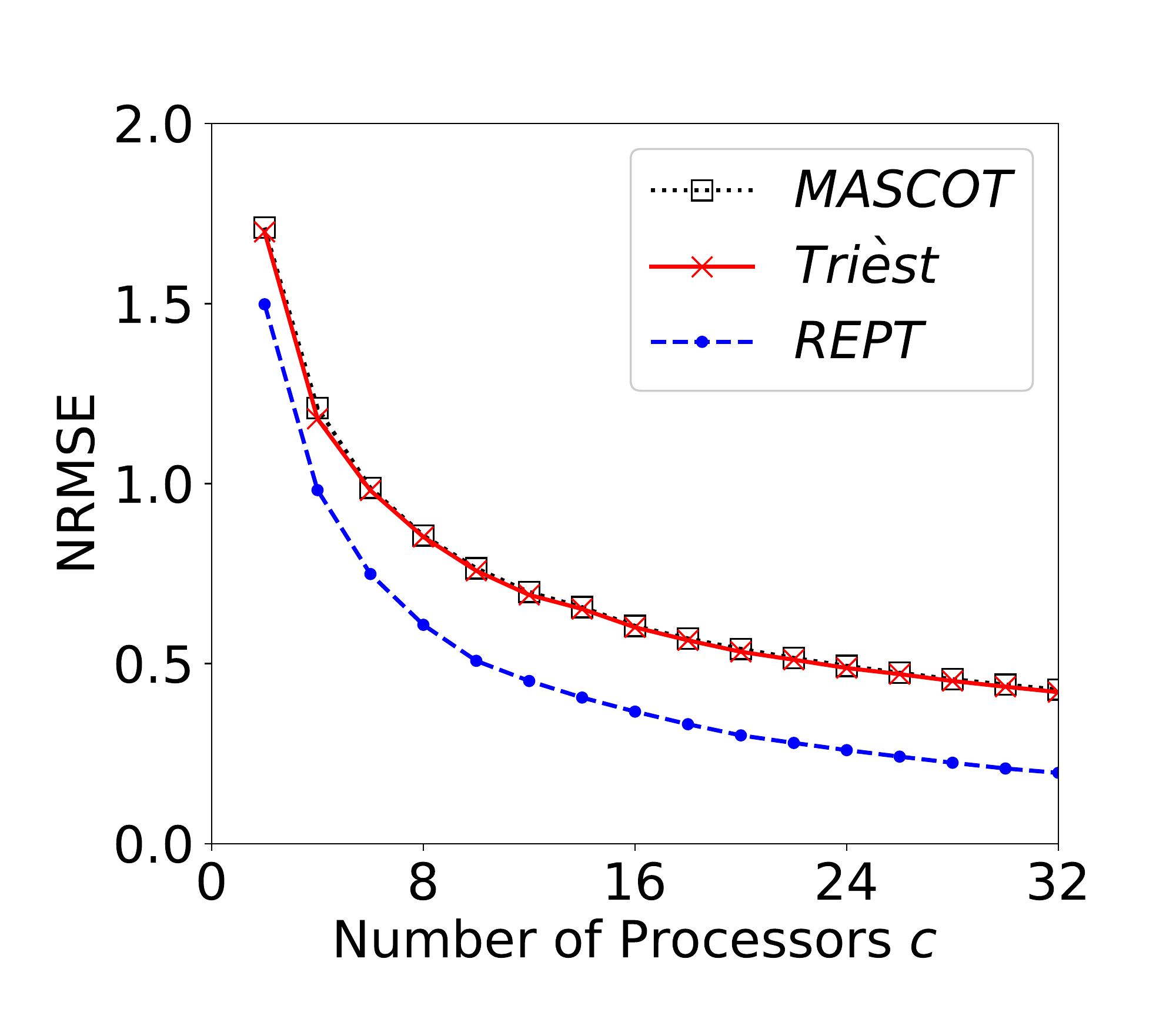}}
\subfigure[Wiki-Talk]{\includegraphics[width=0.232\textwidth]{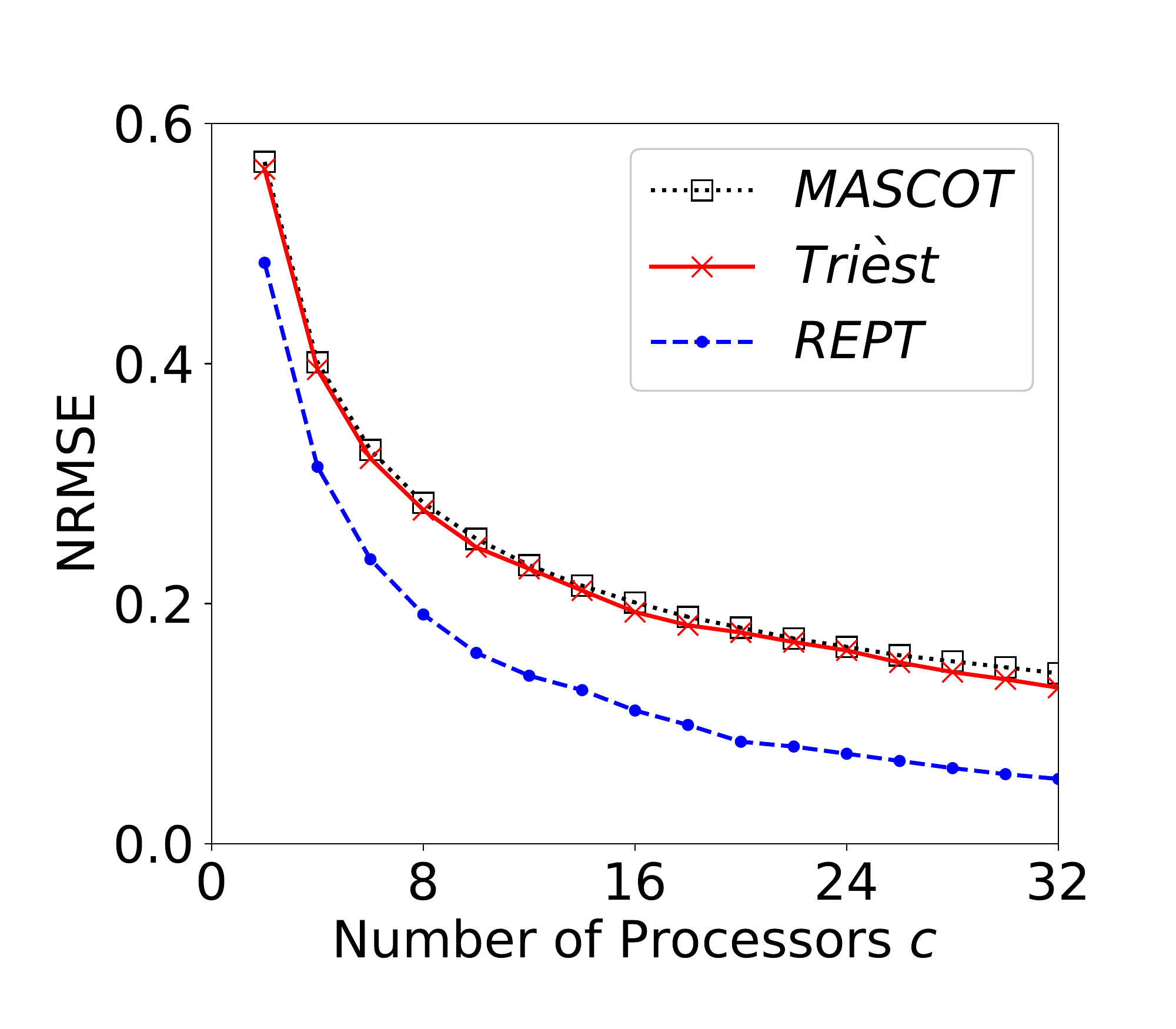}}
\subfigure[Web-Google]{\includegraphics[width=0.232\textwidth]{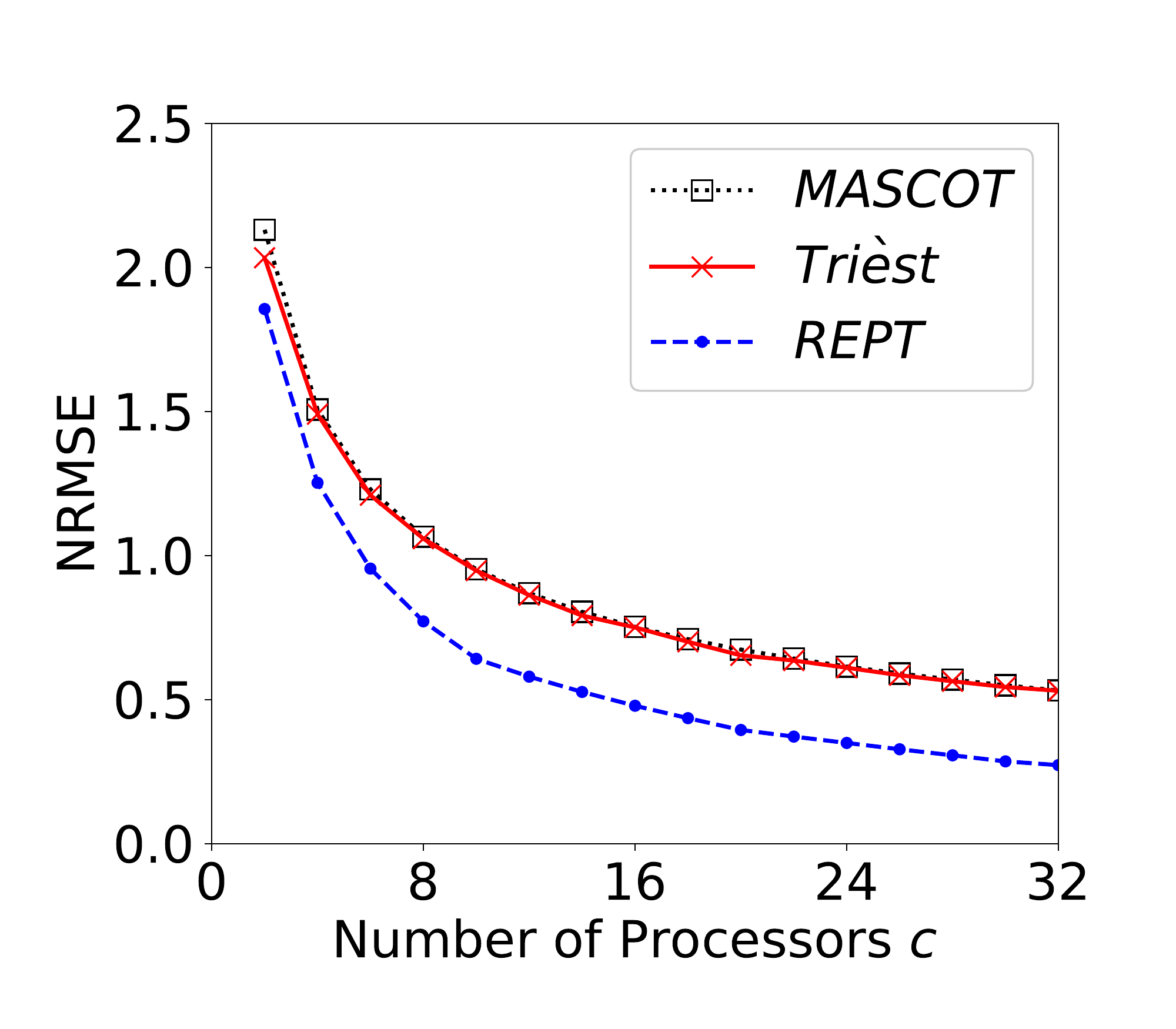}}
\subfigure[YouTube]{\includegraphics[width=0.232\textwidth]{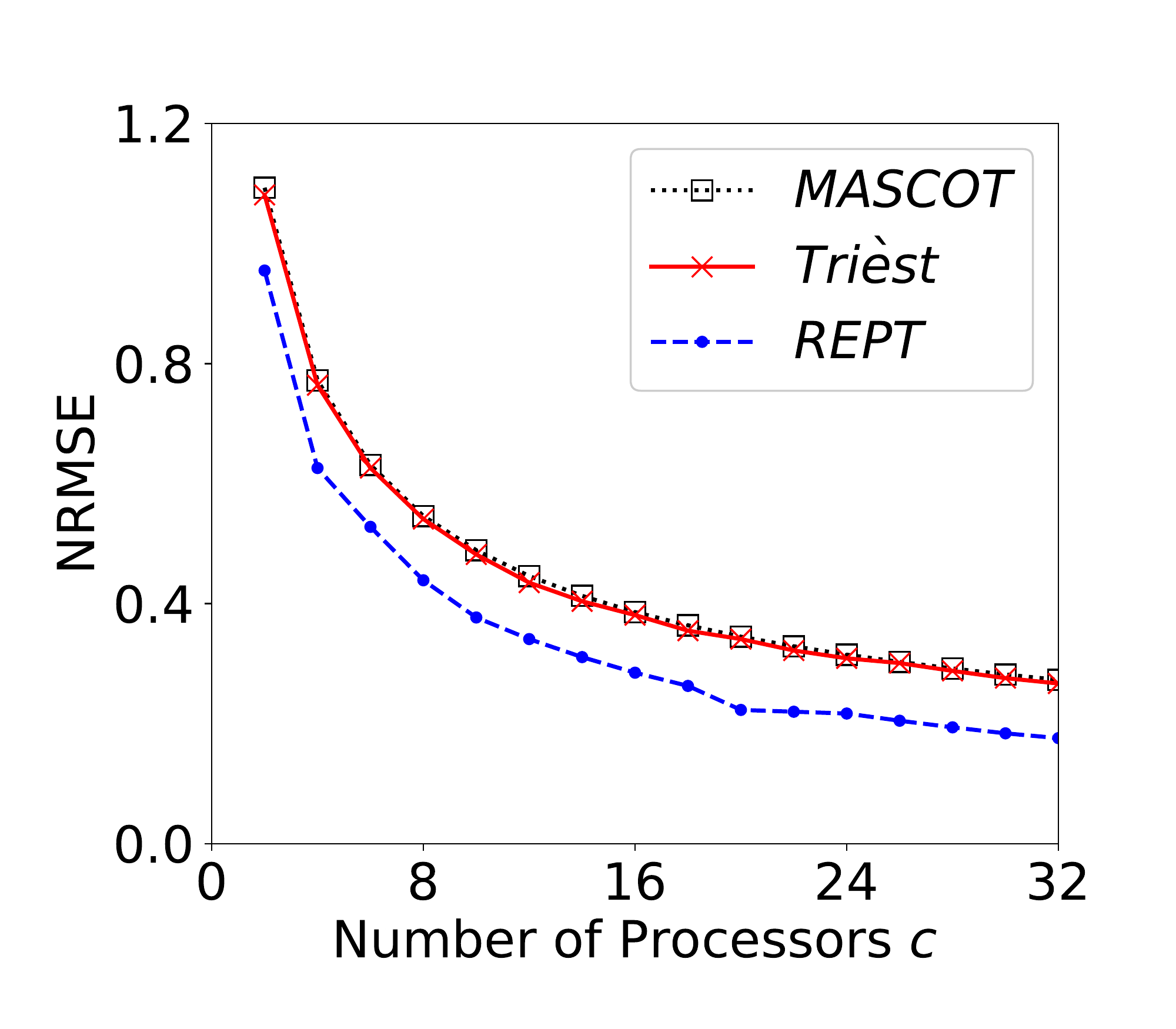}}
\caption{Errors of our method REPT, parallel MASCOT, and parallel Tri{\`{e}}st for estimating local triangle counts, $p=0.1$.}
\label{fig:Local_m_10}
\end{figure*}

\subsection{Performance of REPT vs Parallel Streaming Algorithms}
\noindent \textbf{Accuracy of approximating global triangle counts}. In our experiments, we fix the sampling probability $p=\frac{1}{m}$ as $0.01$ (resp. $0.1$),
and then vary the number of processors $c$ from $20$ to $320$ (resp. $2$ to $32$).
Figures~\ref{fig:Global_m_20} and~\ref{fig:Global_m_10} show the results for $p=0.01$ and $p=0.1$ respectively.
We can see that our method REPT is several times more accurate than parallel MASCOT, Tri{\`{e}}st, and  GPS for different $c$.
For example, the NRMSE of our method REPT on dataset Twitter is about $8.6$ times smaller than parallel MASCOT and Tri{\`{e}}st
and $25.7$ times smaller than parallel GPS when $p=0.01$ and $c=320$,
and is about $26.9$ times smaller than parallel MASCOT and Tri{\`{e}}st,
and $80.8$ times smaller than parallel GPS when $p=0.1$ and $c=32$.
As mentioned in Section~\ref{sec:method},
our method REPT reduces the variance of parallel MASCOT and Tri{\`{e}}st from $\frac{\tau (m^2 - 1) + 2\eta(m - 1)}{c}$
to $\frac{\tau (m^2 - c) + 2\eta(m - c)}{c}$ when $c<m$,
and to $\frac{\tau (m^2 - m)}{c}$ when $c\%m=0$.
Therefore, the error reduction achieved by our method REPT increases as $c$ increases.
It is consistent with the results shown in Figures~\ref{fig:Global_m_20} and~\ref{fig:Global_m_10}.
Although GPS utilizes edges' weights to reduce estimation errors,
it samples a half number of edges less than the other methods under the same memory size.
Therefore, we observe that GPS exhibits the largest estimation errors for all graph datasets.
Compared with parallel MASCOT, Tri{\`{e}}st, and GPS,
our method REPT achieves an error reduction varying for different graphs.
This is because the estimation errors of all these three methods are dominated by the covariance between sampled triangles,
which varies a lot among real-world graphs as shown in Figure~\ref{fig:example}.
From Figures~\ref{fig:Global_m_20} and~\ref{fig:Global_m_10},
we also observe that all four methods' NRMSEs decrease as the sampling probability $p$ increases from $0.01$ to $0.1$ when using the same number of cores.
However, a larger $p$ requires more computations, which will be evaluated in our later experiments.
It is consistent with our analysis in Section~\ref{sec:method}.

\noindent \textbf{Accuracy of approximating local triangle counts.}
Figures~\ref{fig:Local_m_20} and~\ref{fig:Local_m_10} show the errors of local triangle count estimations for $p=0.01$ and $p=0.1$ respectively.
Similar to the results of approximating global triangle counts,
we can see that our method REPT significantly outperforms parallel MASCOT and Tri{\`{e}}st for estimating local triangle counts of all graph datasets,
and the error reduction achieved by REPT increases as $c$ increases.

\begin{figure*}[!t]
\centering
\subfigure[Twitter]{\includegraphics[width=0.232\textwidth]{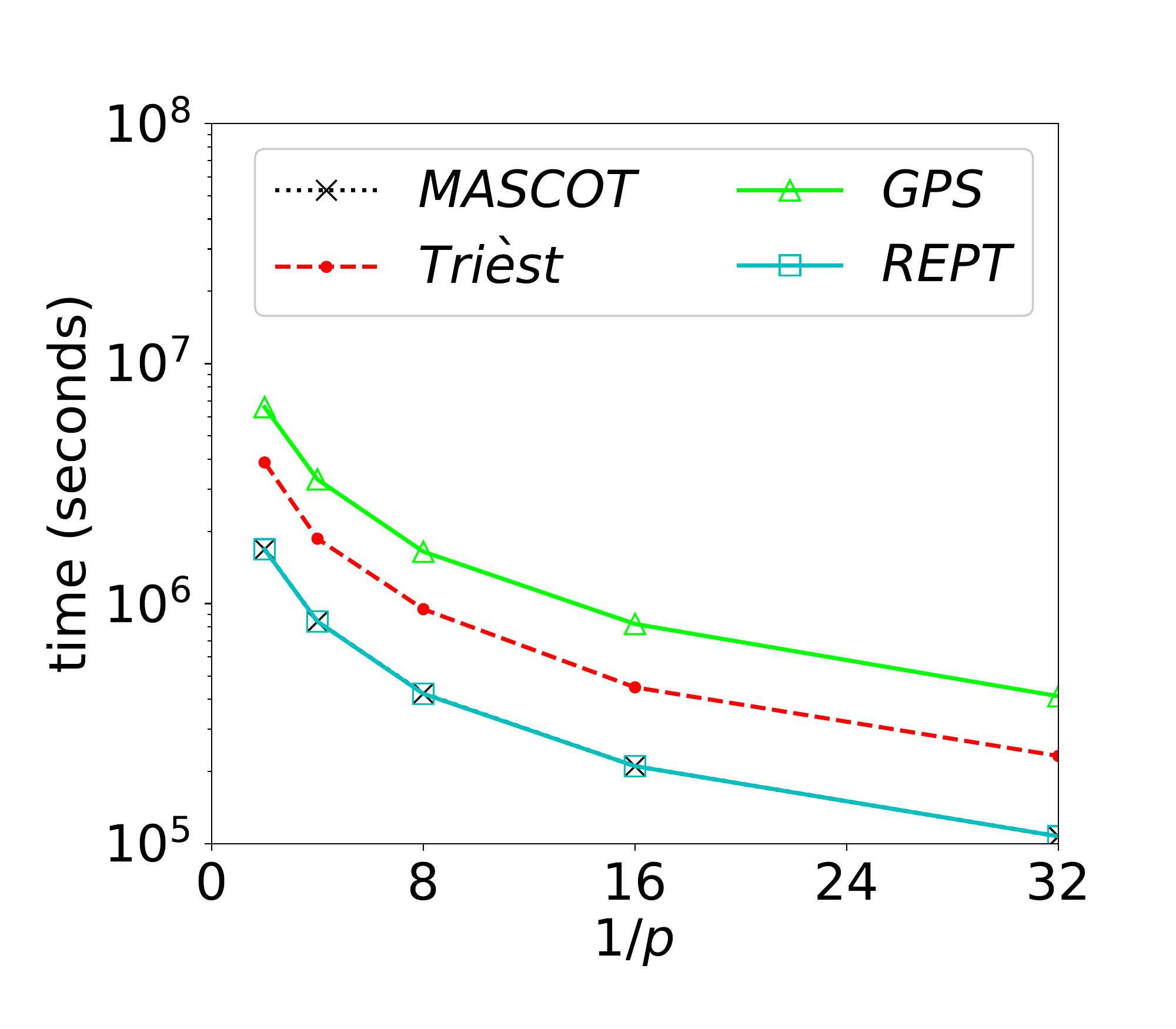}}
\subfigure[com-Orkut]{\includegraphics[width=0.232\textwidth]{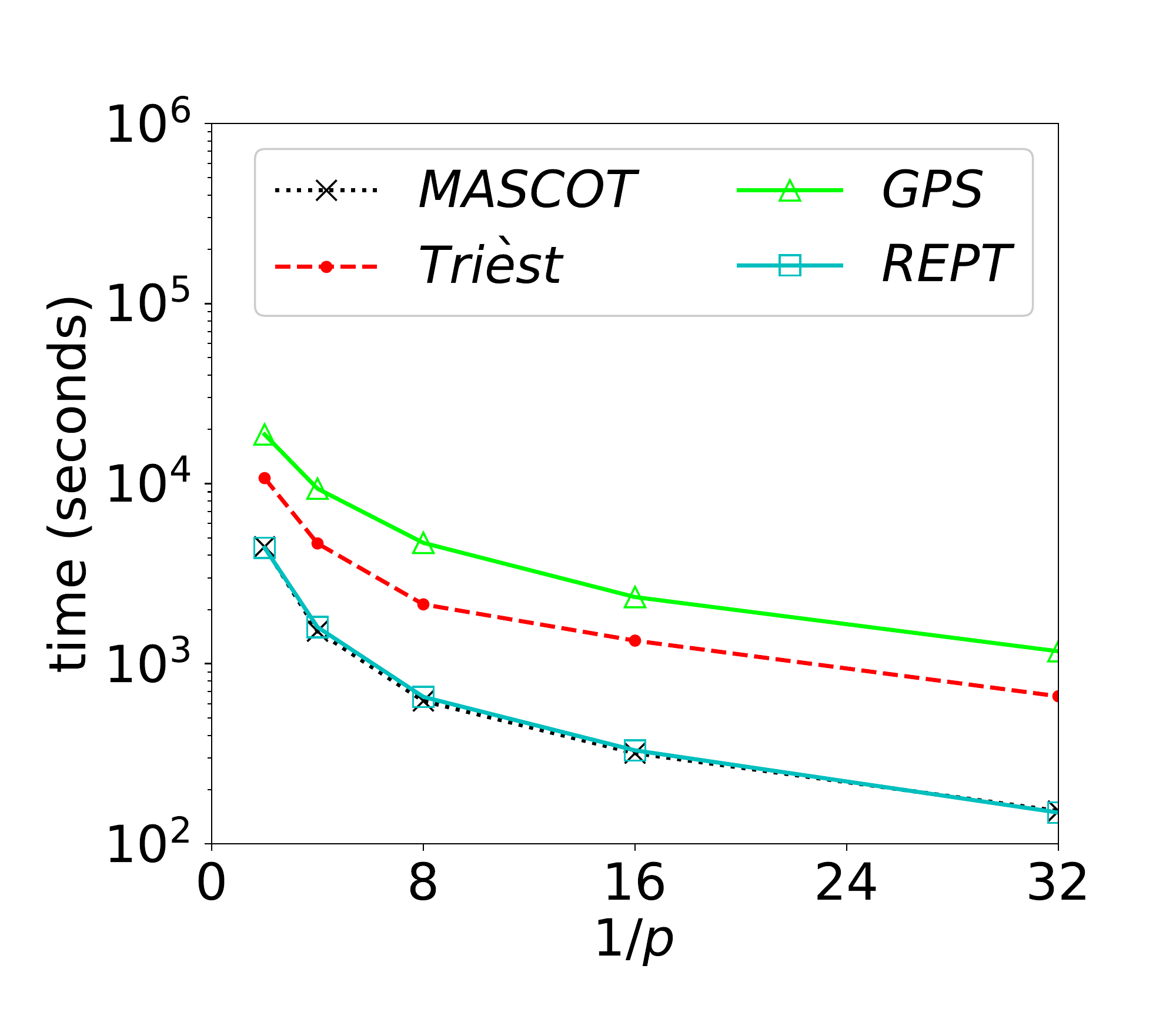}}
\subfigure[LiveJournal]{\includegraphics[width=0.232\textwidth]{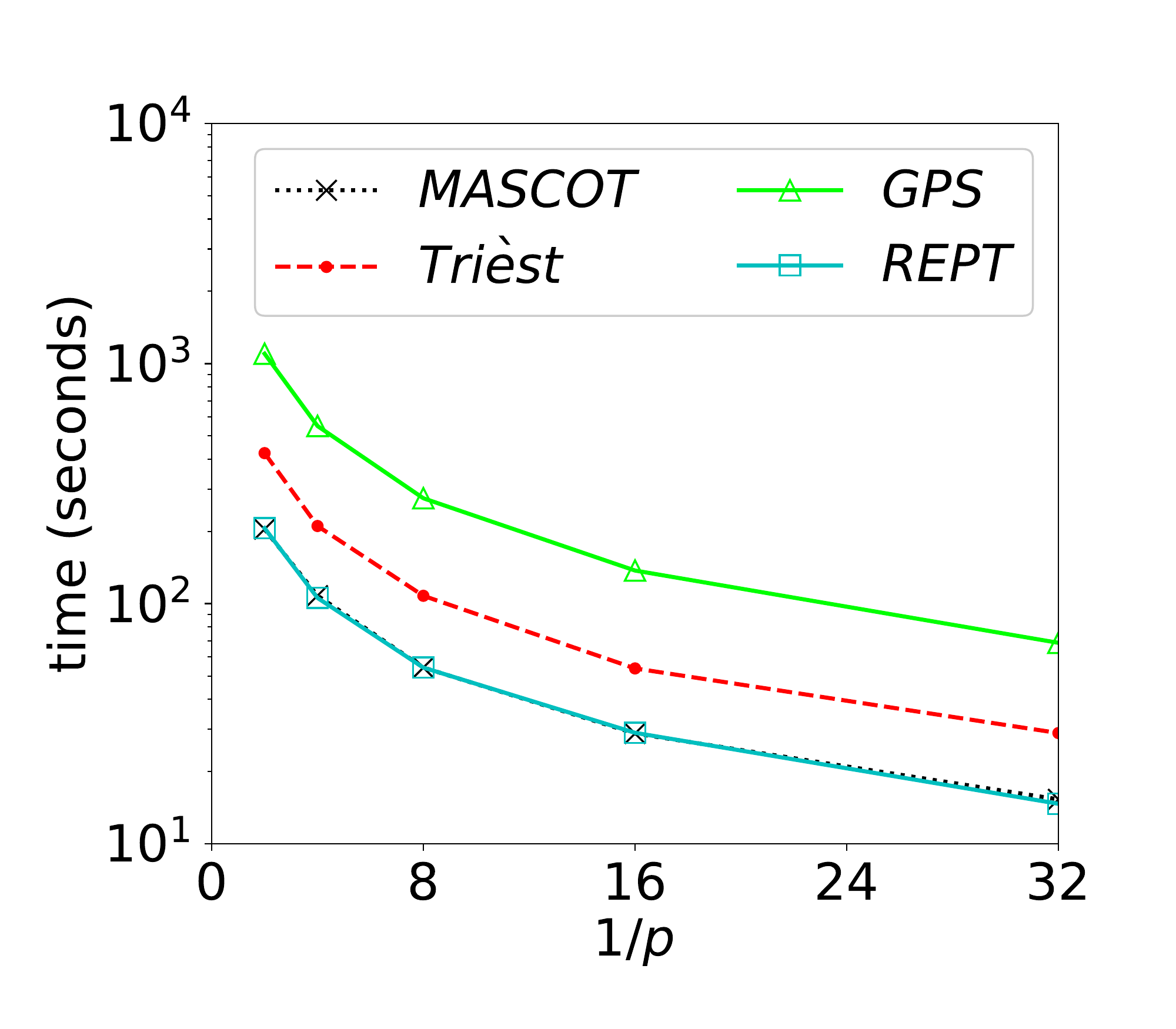}}
\subfigure[Pokec]{\includegraphics[width=0.232\textwidth]{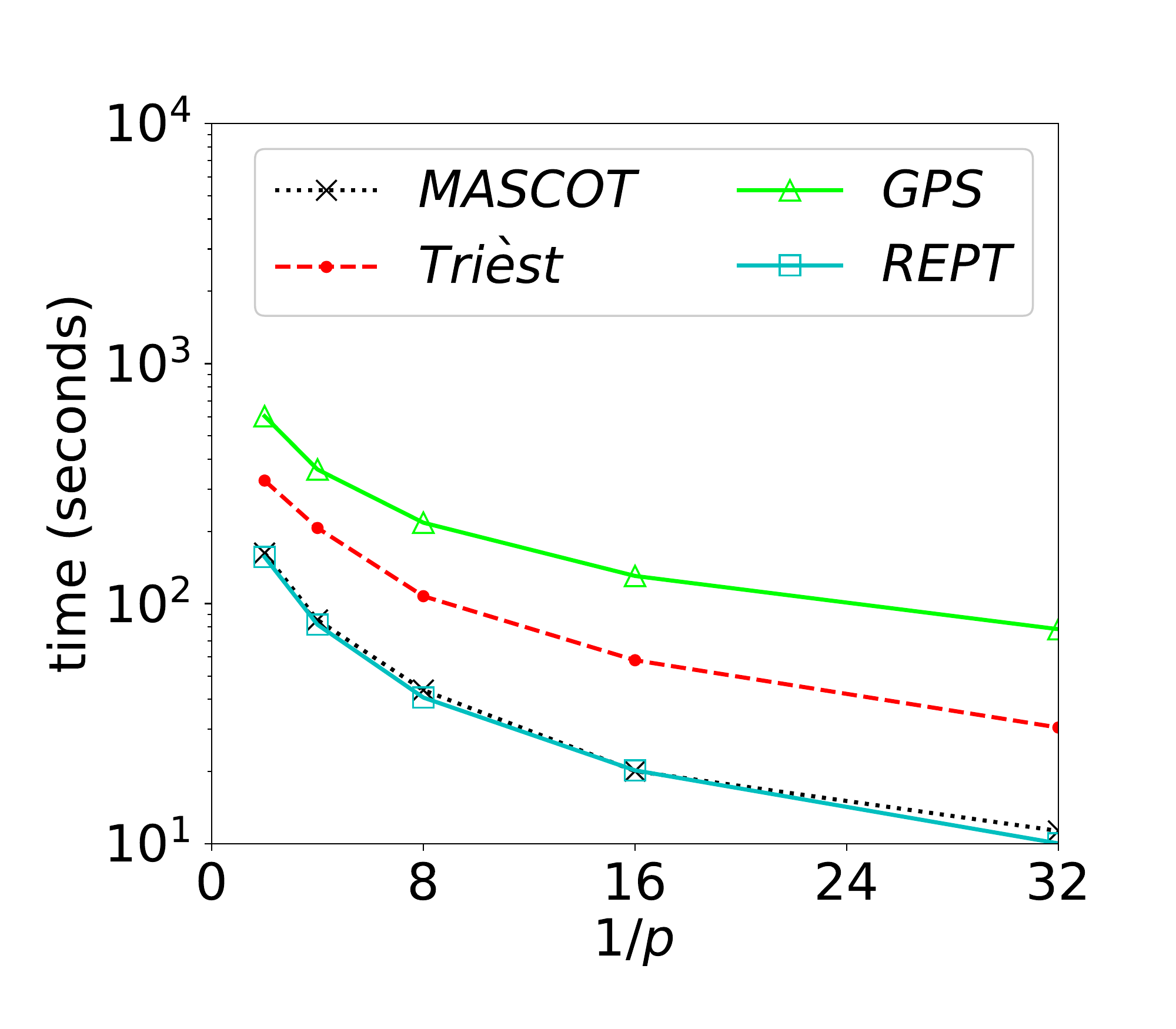}}
\subfigure[Flickr]{\includegraphics[width=0.232\textwidth]{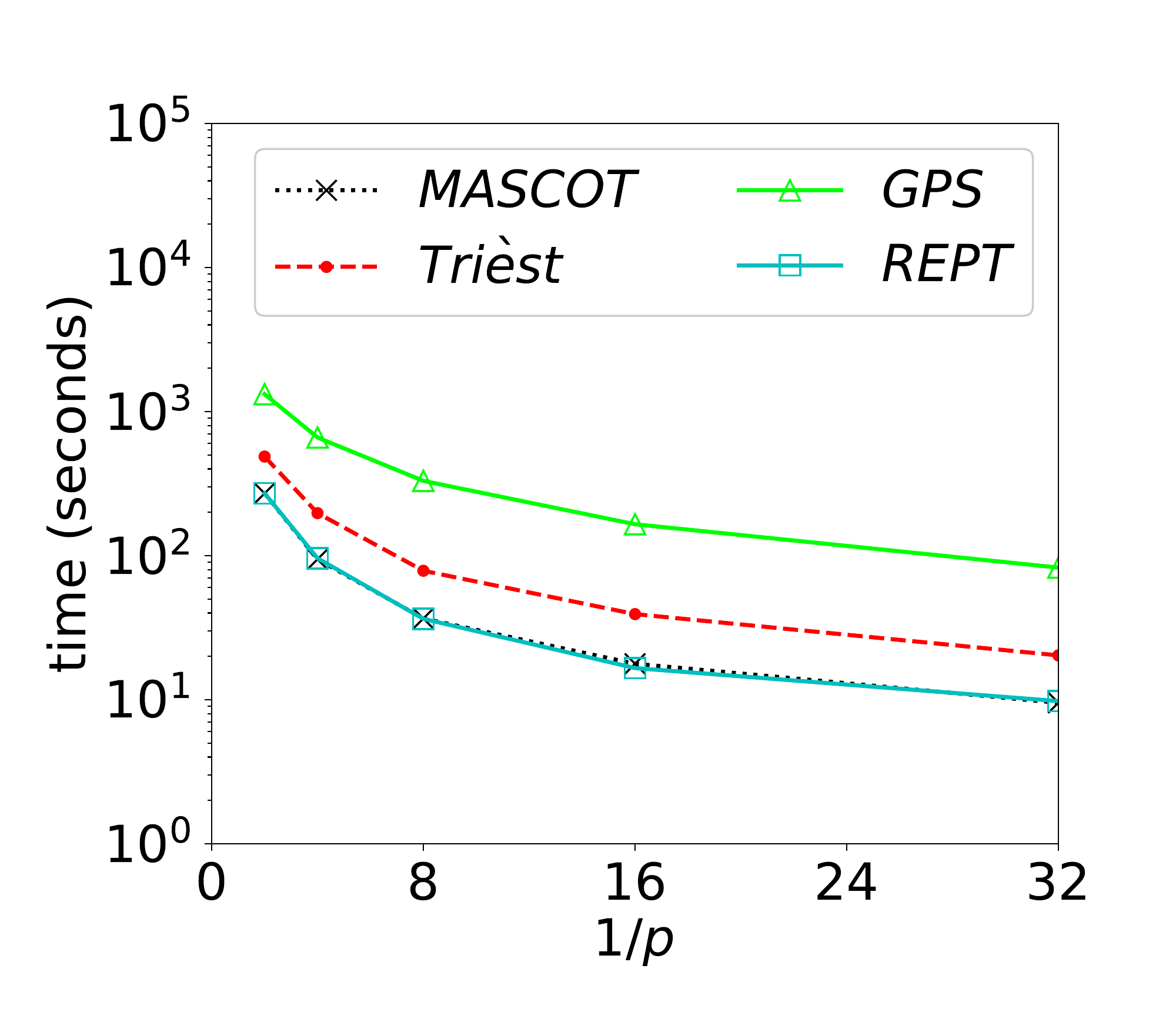}}
\subfigure[Wiki-Talk]{\includegraphics[width=0.232\textwidth]{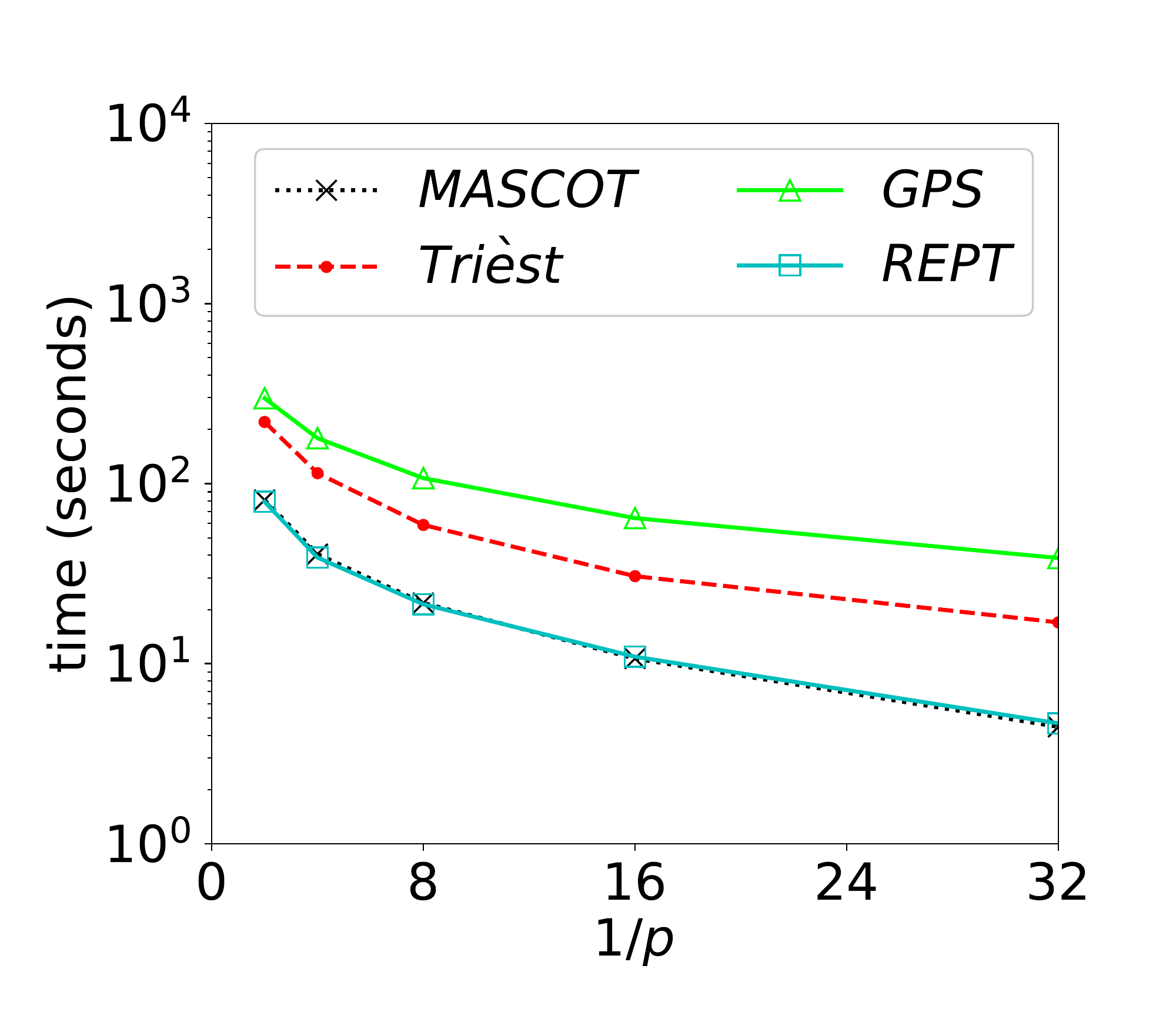}}
\subfigure[Web-Google]{\includegraphics[width=0.232\textwidth]{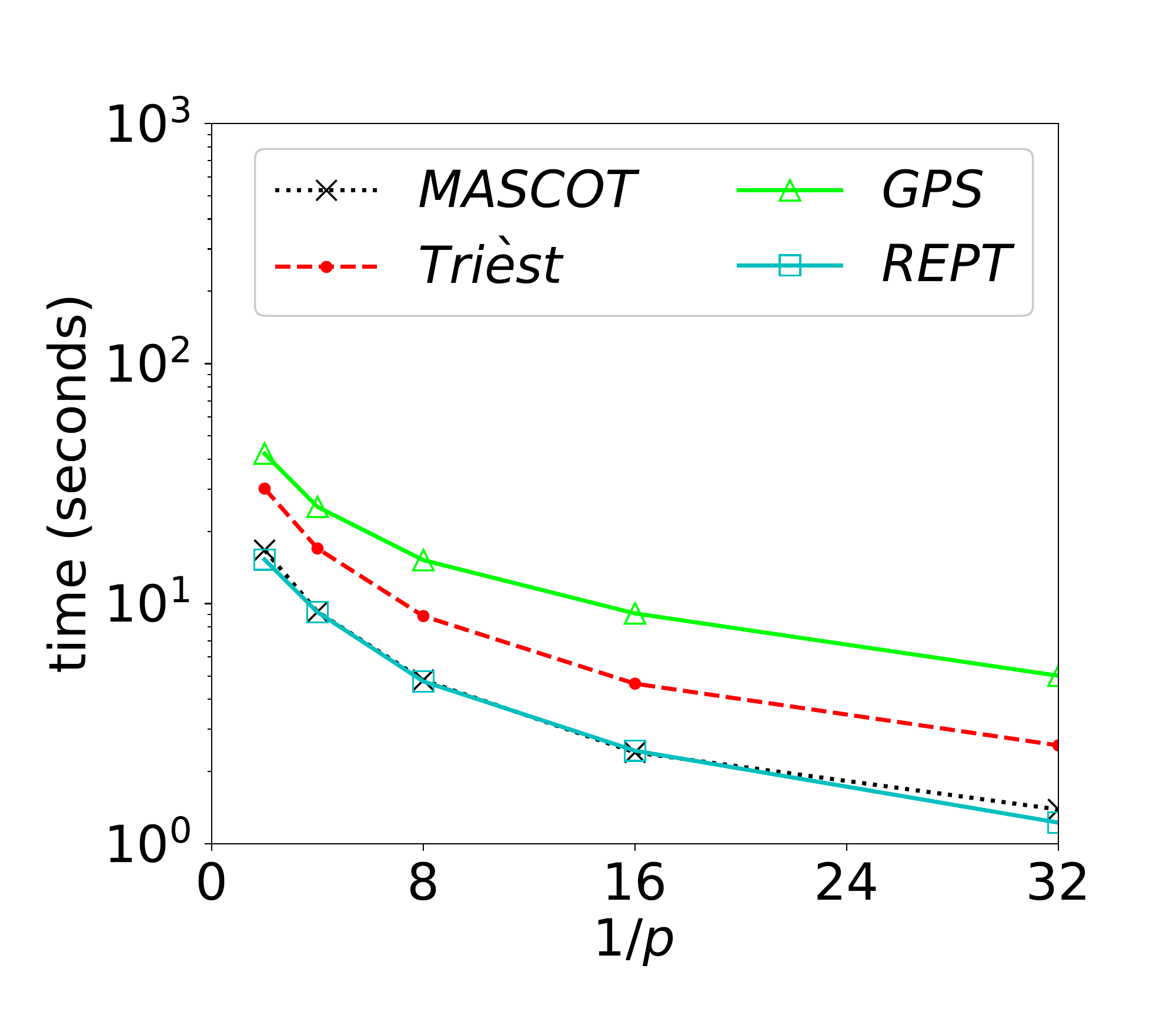}}
\subfigure[YouTube]{\includegraphics[width=0.232\textwidth]{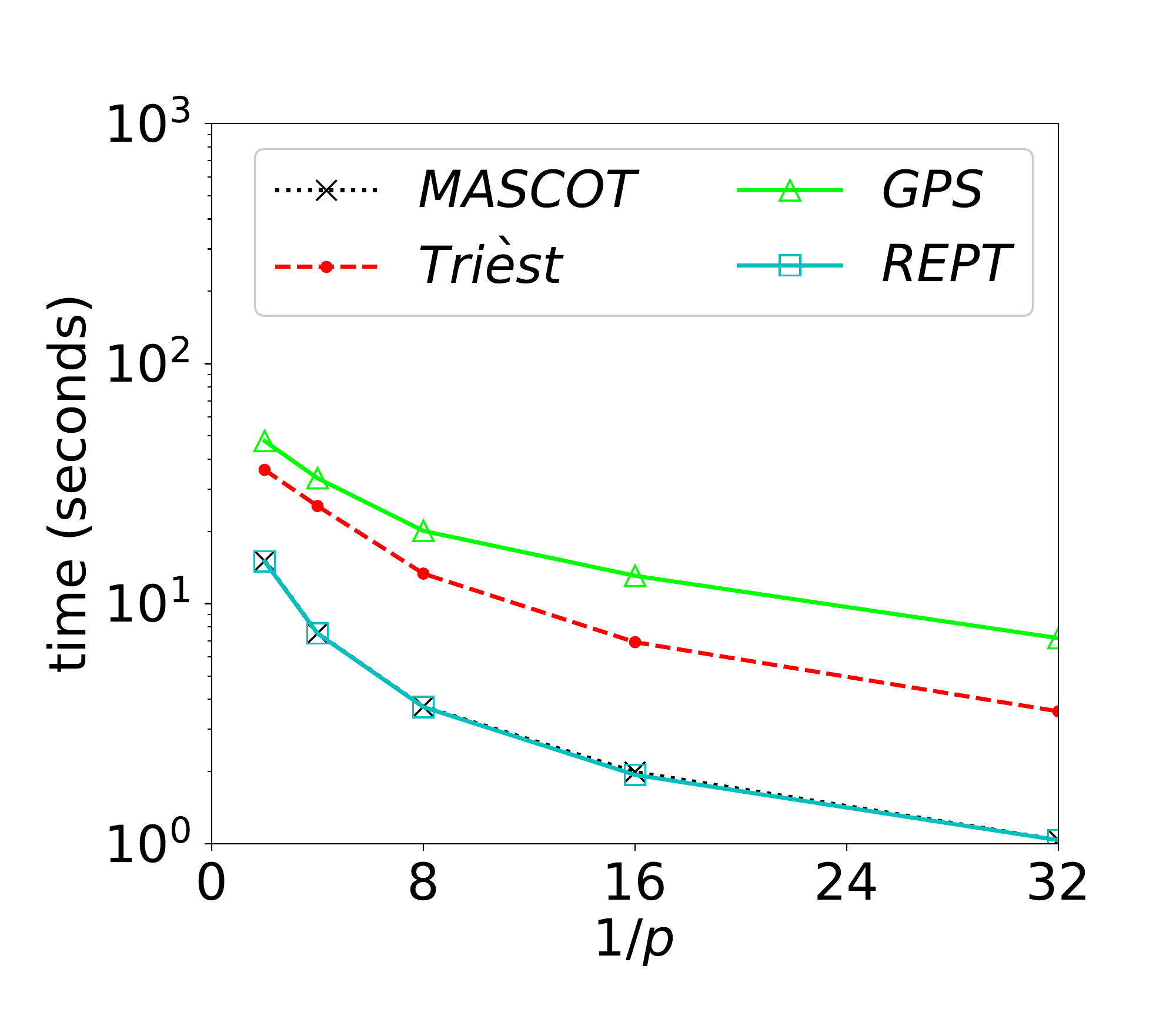}}
\caption{Runtime of our method REPT, parallel MASCOT, and parallel Tri{\`{e}}st for different $p$, where $c=10$.}
\label{fig:Runtime}
\end{figure*}


\noindent \textbf{Runtime.}
For each of the four methods REPT, parallel MASCOT, Tri{\`{e}}st, and GPS,
its running time is mainly determined by the sampling probability $p$,
because each processor samples edges and performs triangle estimation based on $p$.
Therefore, we fix the number of processors $c=10$ and compare the running time of these three methods for different $p$.
The experimental results are shown in Figure~\ref{fig:Runtime}.
We can see that our method REPT is $2$ to $4$ and $4$ to $10$ times faster than parallel Tri{\`{e}}st and GPS respectively,
and almost has the same running time as parallel MASCOT.
This is because all these four methods estimate global and local triangle counts on stream $\Pi$ based on the number of semi-triangles whose first two edges of $\sigma$ on stream $\Pi$ are sampled no matter whether their last edges on stream $\Pi$ are sampled or not.
Also, our method REPT and parallel MASCOT simply sample each edge with a fixed probability $p$ on each processor,
but parallel Tri{\`{e}}st uses the reservoir sampling technique including both edge insertions and deletions during the sampling procedure,
which result in more computation than REPT and parallel MASCOT.
Specially, GPS samples a half number of edges as the other three methods for each processor,
but it is computational intensive to compute the weights of sampled edges.

\subsection{Performance of REPT vs Single-threaded Algorithms}
\begin{figure}[htb]
	\centering
	\subfigure[runtime, $1/p=10$]{\includegraphics[width=0.232\textwidth]{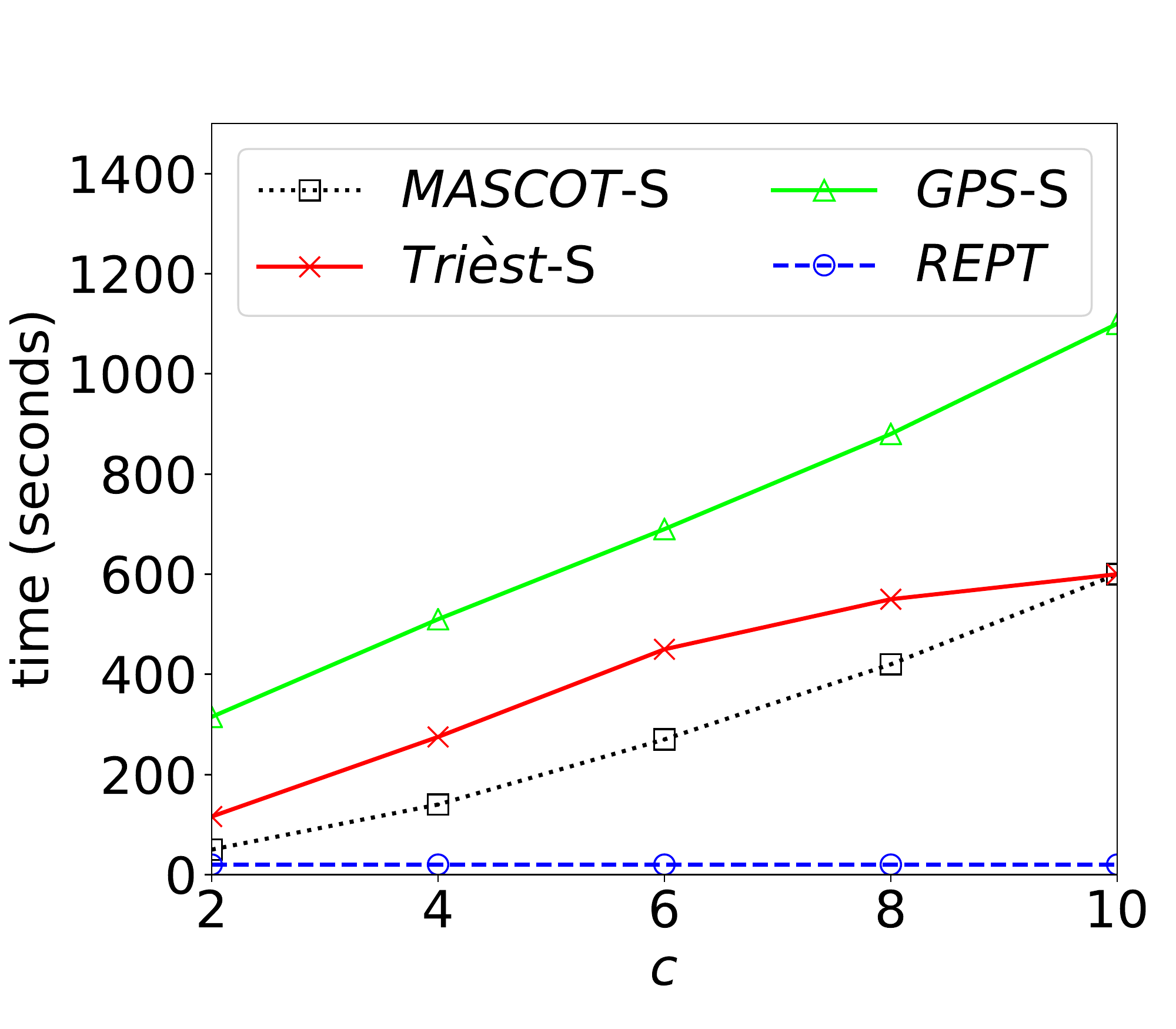}}
	\subfigure[runtime, $1/p=100$]{\includegraphics[width=0.232\textwidth]{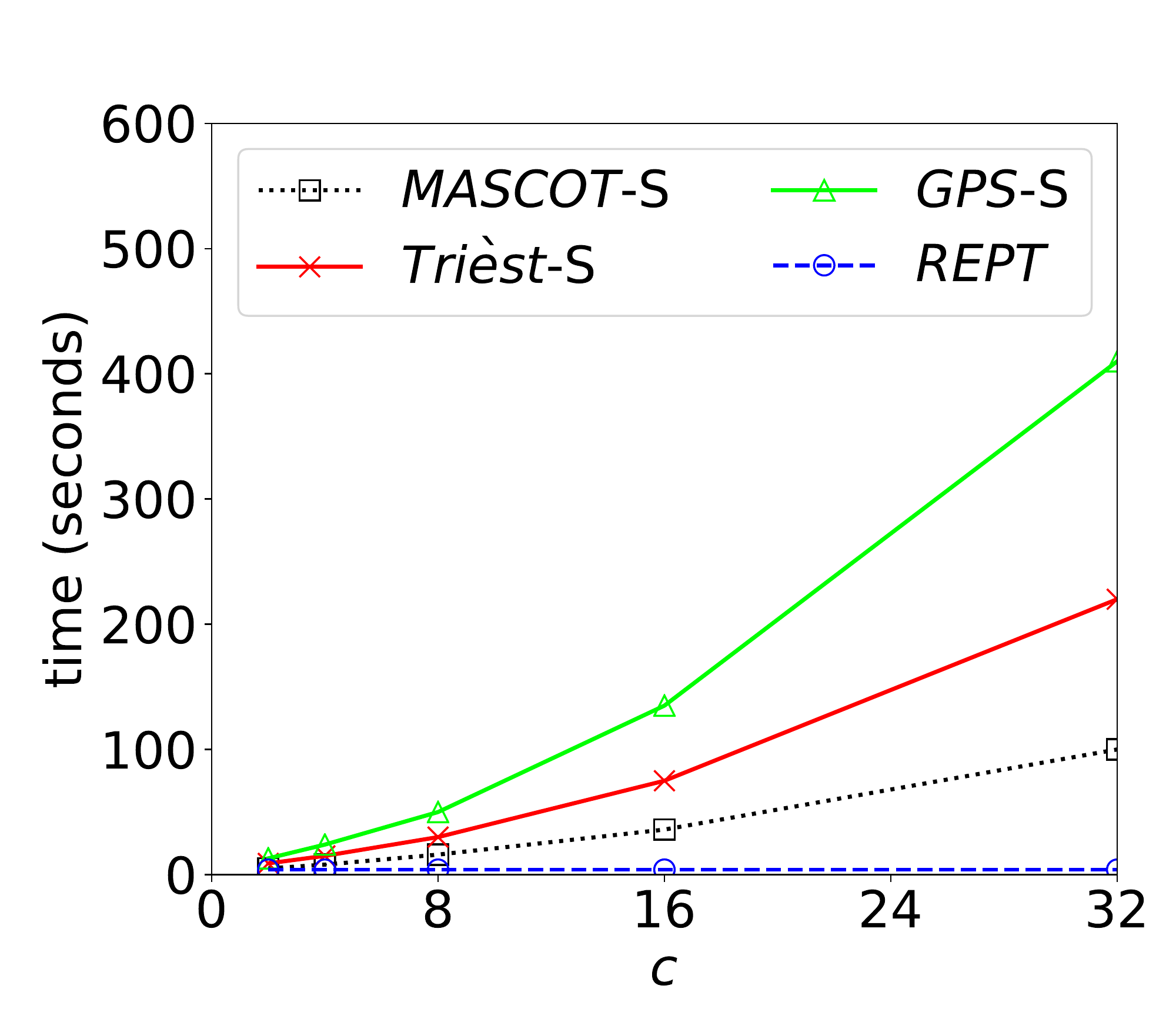}}
	\subfigure[error, $1/p=10$]{\includegraphics[width=0.232\textwidth]{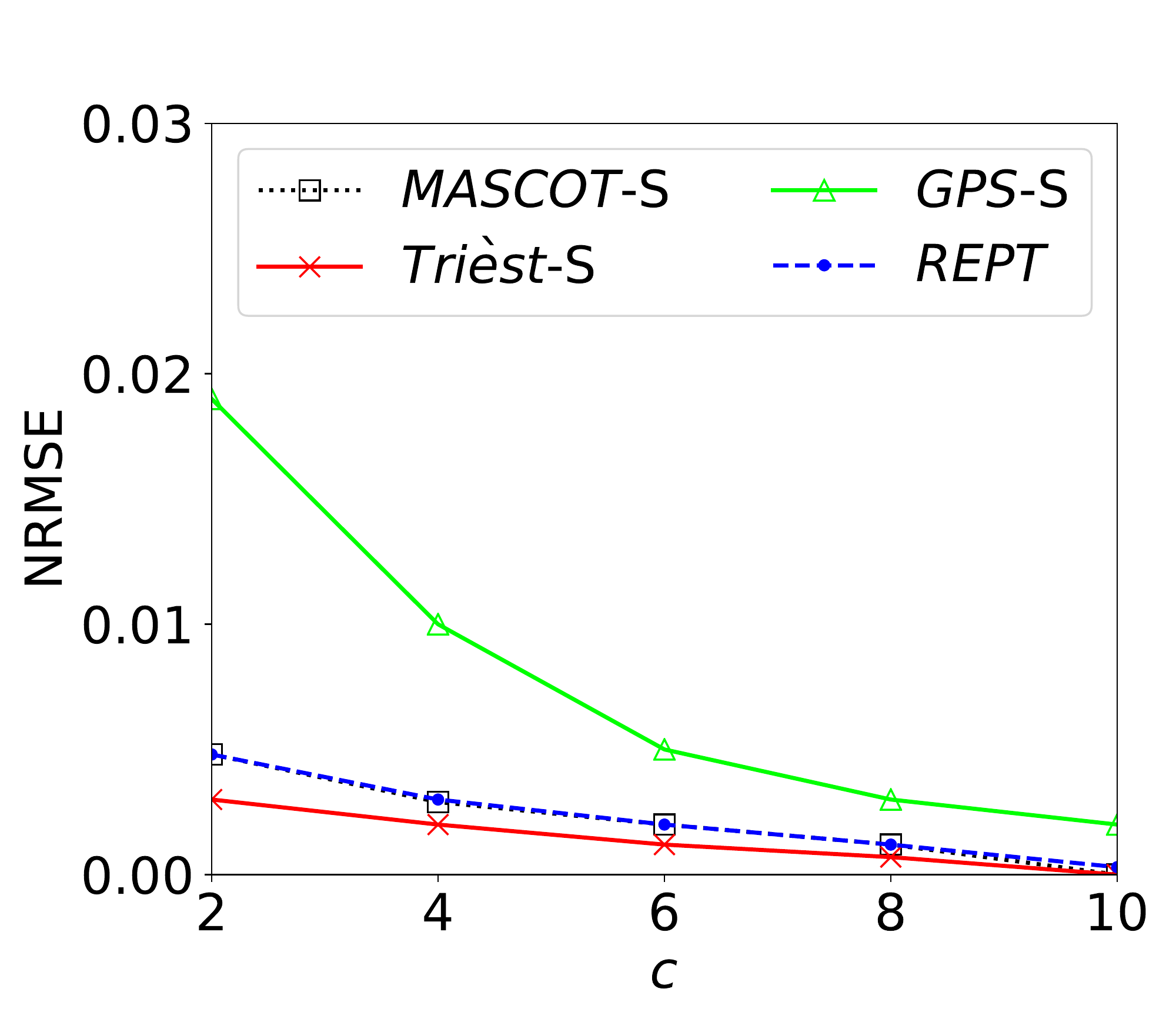}}
	\subfigure[error, $1/p=100$]{\includegraphics[width=0.232\textwidth]{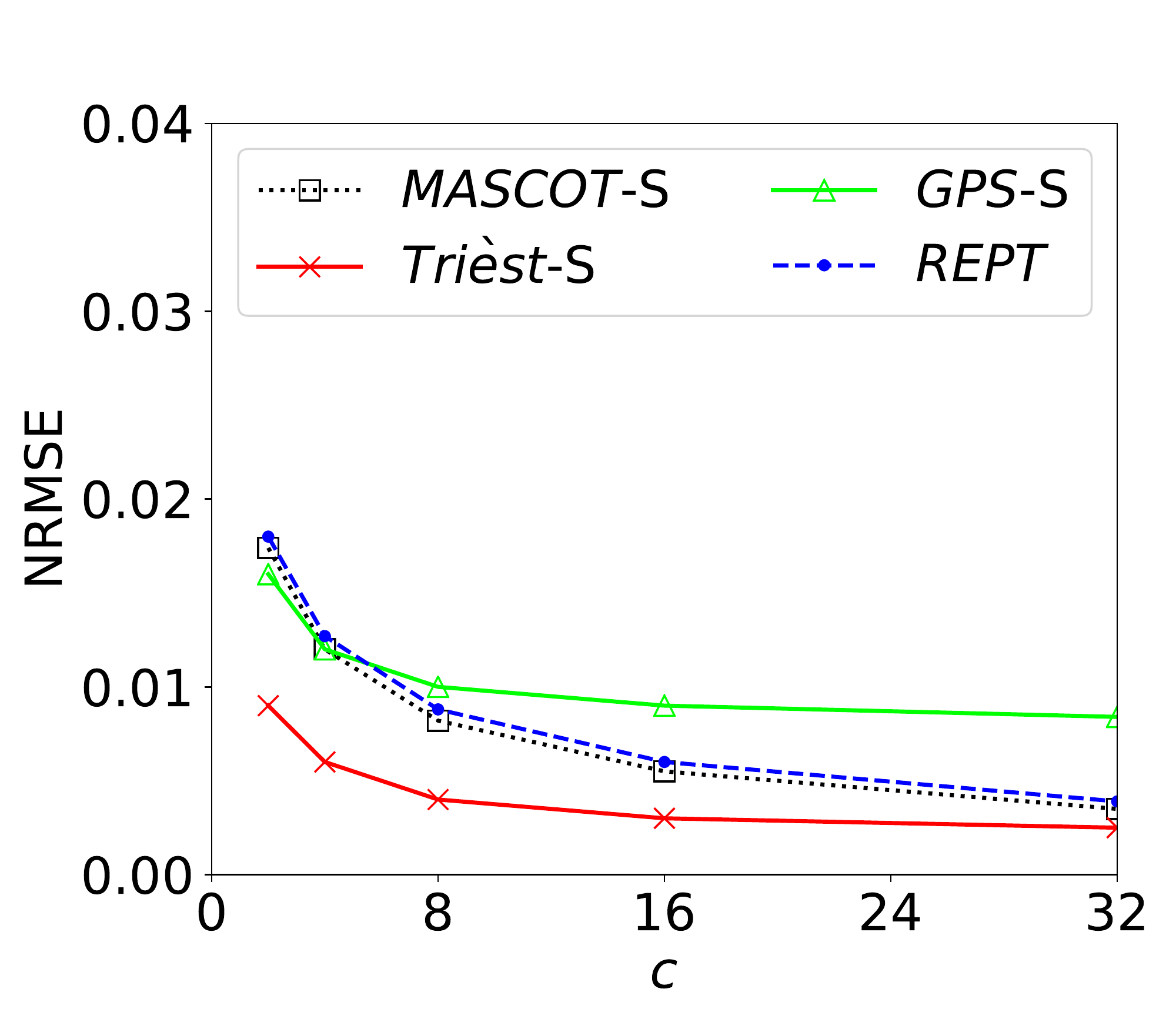}}
	\caption{(Flickr) Runtime and estimation errors of our method REPT, MASCOT-S, Tri{\`{e}}st-S, and GPS-S for different $c$. }
	\label{fig:Single}
\end{figure}

We further compare our method REPT with single-threaded MASCOT, Tri{\`{e}}st, and GPS (in short, MASCOT-S, Tri{\`{e}}st-S, and GPS-S) using the same amount of memory.
We set the sampling probability to $c \times p$ for MASCOT-S,
and sampling budget to $c \times p \times |E|$ for Tri{\`{e}}st-S and GPS-S.
In this experiment, we fix $1/p=10$ and then compare all methods for different $c$.
Due to the limited space, we only show the results of Flickr for $1/p=10$ and $1/p=100$ respectively.
From Figure~\ref{fig:Single}, we can see that our method REPT is up to two orders of magnitude faster than the single-threaded methods while it gives estimations with comparable errors.
To be more specific, when $1/p=100$ and $c=32$, Figure~\ref{fig:Single} (b) shows that REPT is $25$, $50$, and $100$ times faster than MASCOT-S, Tri{\`{e}}st-S, and GPS-S respectively,
while Figure~\ref{fig:Single} (d) reveals that REPT outperforms GPS-S and slightly increases the errors of  MASCOT-S and Tri{\`{e}}st-S.

\section{Related Work} \label{sec:related}
\subsection{Counting Triangles on Just a Machine}
\noindent\textbf{\emph{Exact triangle counting}}. \cite{AlonAlgorithmica1997,SchankWEA2005,LatapyTCS2008} develop fast algorithms using a single machine for listing and counting triangles.
However, these algorithms fail to deal with large graphs due to their high time and space complexities.
To solve this problem,~\cite{HuSIGMOD2013,PaghPODS2014} develop I/O efficient algorithms for listing and counting triangles in a large graph that cannot entirely reside in the main memory.
Kim et al.~\cite{KimSIGMOD2014} present a parallel disk-based triangle enumeration system OPT on a single machine by exploiting the features of solid-state drive (SSD) and multi-core CPU parallelism.
General-purpose disk-based graph computing systems (e.g., GraphChi~\cite{KyrolaOSDI12}, X-Stream~\cite{RoySOSP2013}, TurboGraph~\cite{HanLPL0KY13}, VENUS~\cite{LiuCLL16}, and NXgraph~\cite{ChiDWSLY16}) also provide an implementation for counting triangles.
These algorithms and systems  are not customized for dealing with graph streams because they require that the entire graph of interest is given in advance.\\
\noindent\textbf{\emph{Approximate triangle counting}}. Considerable efforts~\cite{Bar-YossefSODA02,JowhariCOCOON2005,BuriolPODS2006,TsourakakisKDD2009,PavanyVLDB2013,JhaKDD2013,AhmedKDD2014,StefaniKDD16,McGregorPODS2016,Wang2017approximately,Al2018triangle,Mcgregor2014graph,ahmed2017sampling}
have been given to developing one-pass streaming algorithms
for estimating the number of triangles in large graph streams.
Jha et al.~\cite{JhaKDD2013} estimate the triangle count based on a wedge sampling algorithm.
Pavan et al.~\cite{PavanyVLDB2013} develop a neighborhood sampling method to sample and count triangles.
Tsourakakis et at.~\cite{TsourakakisKDD2009} present a triangle
count approximation by sampling each and every edge in the graph stream with a fixed probability.
Ahmed et al.~\cite{AhmedKDD2014} present a general edge sampling based framework
for estimating a variety of graph statistics including the triangle count.
De Stefani et al.~\cite{StefaniKDD16} develop a triangle count estimation method, Tri{\`{e}}st,
which uses the reservoir sampling technique~\cite{VitterTMS1985} to sample edges with fixed memory size.
\cite{ahmed2017sampling} presents a novel weighted edge sampling method, GPS,
which further reduces the estimation error of Tri{\`{e}}st with the same number of sampled edges.
However, GPS requires more memory usage to store the sampling weights of sampled edges,
and more runtime for sampling weights calculation and update.
\cite{Wang2017approximately,JhaACSSC15,JungLLK16} develop one-pass streaming algorithms to deal with large graph streams including edge duplications.
In detail, Wang et al.~\cite{Wang2017approximately} develop PartitionCT for triangle count approximation with a fixed memory usage,
which uses a family of hash functions to uniformly sample distinct edges at a high speed,
and this can reduce the sampling cost per edge to $O(1)$ without additional memory usage.
Jha et al.~\cite{JhaACSSC15} present MG-TRIANGLE algorithm to estimate the triangle counts in multigraph streams.
Jung et al.~\cite{JungLLK16} develop FURL to approximate local triangles for all nodes in multigraph streams.
McGregor et al.~\cite{McGregorPODS2016} present a space efficient one-pass streaming algorithm for counting triangles in adjacency list streams in which all edges incident to the same node appear consecutively
and a two-pass streaming algorithm to further reduce the space complexity of the method in~\cite{JhaKDD2013}.
Wu et al.~\cite{WuTKDE16} theoretically compare the performance of different random sampling algorithms (e.g., subgraph sampling, vertex sampling, triangle sampling and wedge sampling) in adjacency list and edge array streams respectively.
Hasan et al.~\cite{Al2018triangle} present experiments to compare the performance of existing triangle counting approximation methods built under a unified implementation framework.
Also McGregor~\cite{Mcgregor2014graph} give a survey of streaming algorithms for computing graph statistics including the global and local triangle counts.
In addition to global triangle count estimation,
\cite{BecchettiTKDD2010,KutzkovWSDM2013,LimKDD2015} develop methods to compute local (i.e., incident to each node) counts of triangles in a large graph.
Besides these streaming algorithms,~\cite{SeshadhriSADM2014} presents triangle count approximation algorithms for large static graphs.

\subsection{Counting Triangles on a Cluster of Machines}
Cohen~\cite{Cohen2009} develops the first MapReduce algorithm for listing triangles in a large graph.
Suri and Vassilvitskii~\cite{SuriWWW2011} give another MapReduce based algorithm \emph{Graph Partition} (GP) using a graph partitioning technique to count the number of triangles.
\cite{ParkCIKM2013,ParkCIKM2014,ParkKDD16} further reduce a large amount of intermediate data (e.g., shuffled data) generated by GP that causes network congestion and increases the processing time.
Arifuzzaman et al.~\cite{ArifuzzamanCIKM2013} develop a distributed-memory algorithm based on Message Passing Interface (MPI),
which divides the graph into overlapping subgraphs and enumerates triangles in each subgraph in parallel.
\cite{Pavan2013parallel,Tangwongsan2013parallel} develop parallel cache-oblivious algorithms for global triangle counting estimation on both multi-core machines and distributed systems based on the neighbor sampling technique~\cite{PavanyVLDB2013}.
PDTL~\cite{GiechaskielICPP2015} is a distributed extension of the I/O efficient triangle enumeration algorithm in~\cite{HuSIGMOD2013}.
General-purpose distributed graph computing systems (e.g., GraphLab~\cite{LowPVLDB2012}, PowerGraph~\cite{GonzalezLGBG12}, and GraphX~\cite{GonzalezXDCFS14}) also provide an implementation for counting triangles.
Shun et al.~\cite{ShunICDE2015} present a shared-memory parallel triangle counting algorithm for multi-core machines,
which is designed in the dynamic multithreading framework to take full advantage of multi-cores.
\cite{ArifuzzamanCIKM2013, ShunICDE2015} further improve the computational cost by directly combining their algorithms with sampling techniques.
The above algorithms are customized for handling static graphs (i.e., the entire graph of interest is given in advance) but not graph streams.

\section{Conclusions} \label{sec:conclusions}
In this paper, we observe that state-of-the-art triangle count estimation algorithms' errors are significantly dominated by the covariance between sampled triangles.
To solve this problem, we develop a parallel method REPT to significantly reduce the covariance or even completely eliminate the covariance for some cases.
We theoretically prove that REPT is more accurate than parallelizing existing approximate triangle counting algorithms such as MASCOT and Tri{\`{e}}st in a direct manner.
In addition, we also conduct extensive experiments on a variety of real-world graphs,
and the experimental results demonstrate that our method REPT is several times more accurate than state-of-the-art triangle count estimation methods
with the same computational cost.
In future, we plan to extend our algorithm to distributed platforms to estimate triangle counts in parallel.

\section*{Acknowledgment}
The research presented in this paper is supported in part by National Key R\&D Program of China (2018YFC0830500), National Natural Science Foundation of China (U1301254, 61603290, 61602371), the Ministry of Education\&China Mobile Research Fund (MCM20160311), the Natural Science Foundation of Jiangsu Province (SBK2014021758), 111 International Collaboration Program of China, the Prospective Joint Research of Industry-Academia-Research Joint Innovation Funding of Jiangsu Province (BY2014074), Shenzhen Basic Research Grant (JCYJ20160229195940462, JCYJ20170816100819428), China Postdoctoral Science Foundation (2015M582663), Natural Science Basic Research Plan in Shaanxi Province of China (2016JQ6034).

\balance
\bibliographystyle{IEEEtran}

\end{document}